\documentclass[letterpaper,twoside,english,iop,appendixfloats,numberedappendix,twocolappendix]{emulateapj}
\usepackage[LGR,T1]{fontenc}
\usepackage[active]{srcltx}
\usepackage{color}
\usepackage{amsbsy}
\usepackage{amstext}
\usepackage{graphicx}

\makeatletter

\DeclareRobustCommand{\greektext}{%
  \fontencoding{LGR}\selectfont\def\encodingdefault{LGR}}
\DeclareRobustCommand{\textgreek}[1]{\leavevmode{\greektext #1}}
\ProvideTextCommand{\~}{LGR}[1]{\char126#1}

\providecommand{\tabularnewline}{\\}

\usepackage{apjfonts}
\usepackage[hyperfootnotes=false,colorlinks=true,citecolor=blue]{hyperref}

\usepackage{apjfonts}
\usepackage{amsmath}
\usepackage[none]{hyphenat}
\usepackage[hyperfootnotes=false]{hyperref}
\hypersetup{colorlinks=true,citecolor=blue,urlcolor=blue,linkcolor=blue}
\shorttitle{Ly$\alpha$ Polarization}
\shortauthors{SEON, SONG, \& CHANG}

\DeclareSymbolFont{CMletters}{OML}{cmm}{m}{it}
\DeclareMathSymbol{\nu}{\mathord}{CMletters}{23}
\DeclareMathSymbol{v}{\mathord}{CMletters}{`v}
\DeclareMathSymbol{w}{\mathord}{CMletters}{`w}
\DeclareMathSymbol{g}{\mathord}{CMletters}{`g}

\makeatother

\usepackage{babel}
\begin{document}

\title{Ly$\alpha$ Radiative Transfer: A Stokes Vector Approach to Ly$\alpha$
Polarization}

\author{Kwang-il Seon\altaffilmark{1,2}, Hyunmi Song\altaffilmark{3,4}
and Seok-Jun Chang\altaffilmark{1,5}}

\altaffiltext{1}{Korea Astronomy \& Space Science Institute, 776 Daedeokdae-ro, Yuseong-gu, Daejeon 34055, Republic of Korea; kiseon@kasi.re.kr}
\altaffiltext{2}{Astronomy and Space Science Major, University of Science and Technology, 217, Gajeong-ro, Yuseong-gu, Daejeon 34113, Republic of Korea}
\altaffiltext{3}{Department of Astronomy, Yonsei University, 50 Yonsei-ro, Seodaemun-gu, Seoul 03722, Republic of Korea}
\altaffiltext{4}{Department of Astronomy and Space Science, Chungnam National University, 99 Daehak-ro, Yuseong-gu, Daejeon, 34134, Republic of Korea}
\altaffiltext{5}{Department of Physics and Astronomy, Sejong University, 209 Neungdong-ro, Gwangjin-gu, Seoul, 05006, Republic of Korea}
\begin{abstract}
Ly$\alpha$ emitting galaxies and giant Ly$\alpha$ blobs (LABs) have
been extensively observed to study the formation history of galaxies.
However, the origin of their extended Ly$\alpha$ emission, especially
of LABs, remains controversial. Polarization signals from some LABs
have been discovered, and this is commonly interpreted as strong evidence
supporting that the extended Ly$\alpha$ emission originates from
the resonance scattering. The Monte Carlo Ly$\alpha$ radiative transfer
code LaRT is updated to investigate the polarization of Ly$\alpha$
using the Stokes vector formalism. We apply LaRT to a few models to
explore the fundamental polarization properties of Ly$\alpha$. Interestingly,
individual Ly$\alpha$ photon packets are found to be almost completely
polarized by a sufficient number of scatterings ($N_{{\rm scatt}}\gtrsim10^{4}-10^{5}$
in a static medium) or Doppler shifts induced by gas motion, even
starting from unpolarized light. It is also found that the polarization
pattern can exhibit a non-monotonically increasing pattern in some
cases, besides the commonly-known trend that the polarization monotonically
increases with radius. The polarization properties are primarily determined
by the degree of polarization of individual photon packets and the
anisotropy of the Ly$\alpha$ radiation field, which are eventually
controlled by the medium's optical depth and velocity field. If once
Ly$\alpha$ photon packets achieve $\sim$100\% polarization, the
radial profile of polarization appears to correlate with the surface
brightness profile. A steep surface brightness profile tends to yield
a rapid increase of the linear polarization near the Ly$\alpha$ source
location. In contrast, a shallow surface brightness profile gives
rise to a slowly increasing polarization pattern.
\end{abstract}


\section{INTRODUCTION}

Ly$\alpha$ is one of the most powerful tracers to investigate star-forming
galaxies and circumgalactic/intergalactic media (CGM/IGM) in the universe.
After the prediction by \citet{1967ApJ...147..868P} that star-forming
galaxies should be strong Ly$\alpha$ emitters (LAEs), many photometric
and spectroscopic surveys have been performed to detect Ly$\alpha$
emission from nearby or high-redshift galaxies \citep{1981ApJ...246L.109M,1998AJ....115.1319C,2000ApJ...545L..85R,2014ApJ...797...11O,2015A&A...575A..75B,2018PASJ...70S..13O,2020ARA&A..58..617O}.
The observed Ly$\alpha$ spectra and/or surface brightness profiles
of LAEs can be well understood by resonance scattering of Ly$\alpha$
in simplified galactic outflow models. The most well-known one is
a thin shell model, in which a central Ly\textgreek{a} source is surrounded
by a constantly expanding spherical thin shell of atomic hydrogen
gas. This shell model has surprisingly well reproduced diverse Ly$\alpha$
line profiles \citep{2004ApJ...601L..25A,2008A&A...480..369S,2008A&A...491...89V,2011A&A...531A..12S,2015ApJ...812..123G,2016ApJ...820..130Y,2017A&A...608A.139G,2017A&A...599A..28K},
even though it is questionable whether the model can reproduce the
observed surface brightness profiles as well. Another model designed
to better represent real galaxy halos was able to successfully reproduce
both the spectra and surface brightness profiles of LAEs \citep{2020ApJ...901...41S}.
In this halo model, the distributions of Ly\textgreek{a} source and
gas are described by exponential functions of radius. The outflowing
velocity of the halo is characterized by a piecewise linear function
of radius.

The extended Ly$\alpha$ nebulae (also known as Ly$\alpha$ blobs
or LABs) observed at $z=2-6$ can provide clues to galaxy formation
in the early universe \citep{1999AJ....118.2547K,2000ApJ...532..170S,2004AJ....128..569M,2009ApJ...693.1579Y,2010ApJ...719.1654Y,2012ApJ...752...86P,2017ApJ...845..172B}.
They are often found in association with high-density regions of LAEs;
the connection of LABs with the overdensities of LAEs suggests that
they are associated with matter density peaks in the universe and
thus likely to evolve into the present-day groups and clusters of
galaxies. However, the mechanism powering the extended Ly$\alpha$
emission in LABs remains debated. Many possible mechanisms have been
proposed to explain LABs, including (1) rapid cooling of the accreting
gas that is heated by galactic outflows during powerful starbursts
\citep{2000ApJ...532L..13T,2004ApJ...613L..97M,2005MNRAS.363.1398G}
or by the dissipation of gavitational energy as gas falls toward galaxies
\citep{2000ApJ...537L...5H,2001ApJ...562..605F,2009MNRAS.400.1109D},
(2) photoionization by luminous active galactic nuclei (AGNs), young
stars, and/or the intergalactic ultraviolet background \citep{2001ApJ...556...87H,2006Natur.440..501J,1996ApJ...468..462G},
and (3) resonant scattering of Ly$\alpha$ photons produced by star
forming galaxies and/or AGNs hosted within the nebulae \citep{2011Natur.476..304H,2011ApJ...736..160S,2016ApJ...818..138B}.
In the former two scenarios, Ly$\alpha$ photons are produced in situ
inside the nebulae, whereas they are scattered light by neutral hydrogen
gas surrounding the central Ly$\alpha$ source(s) in the last scenario.

It is challenging to discern which scenario among the above is better
suited using only the surface brightness profile and spectrum. With
this regards, polarization signal can provide critical, additional
information about the nature of the Ly$\alpha$ nebulae or the diffuse
Ly$\alpha$ halos around high-redshift galaxies, which cannot be offered
by the photometric and spectroscopic measurements alone; the resonant
scattering of Ly$\alpha$ gives rise to polarization, but other mechanisms
would not. Therefore, the polarization phenomenon of Ly$\alpha$ has
attracted attention in both observational and theoretical studies
to reveal the nature of LABs. The first attempt to detect any polarization
signal of the Ly$\alpha$ emission from LABd05 was not sensitive enough
\citep{2011ApJ...730L..25P}. Later, subsequent observations have
detected clear polarization signatures in Ly$\alpha$ nebulae \citep{2011Natur.476..304H,2013ApJ...768L...3H,2016ApJ...818..138B,2017ApJ...834..182Y,2020ApJ...894...33K},
which qualitatively accord with the theoretical prediction that the
polarization level of the scattered Ly$\alpha$ tends to increase
with distance from the central source. The detection of polarization
signals that follow the predicted trend is interpreted as strong evidence
supporting that the LABs are caused primarily by the resonance scattering
of Ly$\alpha$ originating from star-forming galaxies or AGNs. Moreover,
\citet{2021MNRAS.502.2389L} showed that the observed Ly$\alpha$/H$\beta$
ratios in LAB2 support the resonant scattering scenario, and the observed
Ly$\alpha$ spectra can be well reproduced using an RT model in a
clumpy medium. In contrast, \citet{2016A&A...593A.122T} demonstrated
that the radial profile of polarization could be well explained by
the scenario in which Ly$\alpha$ photons are produced in the cooling
gas surrounding galaxies and then self-scattered by the gas. Their
results likely suggest that the polarimetric observations alone are
unable to disentangle between the two scenarios. Recently, \citet{2020A&A...642A..55H}
found that ionizing radiation and mechanical heating would be dominant
near the embedded galaxies in LAB1, based on the observation of \ion{He}{2}
$\lambda$1640 and non-detection of \ion{C}{4} $\lambda\lambda$1548,
1550, while Ly$\alpha$ scattering contributes more significantly
at larger distances.

To interpret the above observations, modeling of Ly$\alpha$ radiative
transfer (RT) process is essential. This is because Ly$\alpha$ is
a strong resonance line and thus undergoes a large number of scatterings
by neutral hydrogen atoms in the interstellar medium (ISM) within
galaxies, in the CGM that immediately surrounds them, and in the IGM
between galaxies. Monte Carlo simulation is the most flexible and
straightforward approach in calculating the Ly$\alpha$ RT effects
in astrophysical systems. The Monte Carlo simulation technique for
Ly$\alpha$ RT has now become almost a standard, and thus many Monte
Carlo Ly$\alpha$ RT codes have been developed in many different contexts
\citep{1968ApJ...152..493A,2000JKAS...33...29A,2002ApJ...578...33Z,2005ApJ...628...61C,2006ApJ...645..792T,2006ApJ...649...14D,2006A&A...460..397V,2007A&A...474..365S,2007ApJ...657L..69L,2009ApJ...696..853L,2009MNRAS.393..872P,2010MNRAS.403..870B,2011MNRAS.415.3666F,2012MNRAS.424..884Y,2012MNRAS.425...87O,2013A&A...556A...5B,2014MNRAS.444.1095G,2015MNRAS.449.4336S,2020A&A...635A.154M,2020ApJS..250....9S}.
Most of them have focused on predicting the emergent Ly$\alpha$ spectrum
and surface brightness, except \citet{2007A&A...474..365S} and \citet{2020ApJS..250....9S},
who investigated the Wouthuysen-Field (WF) effect of Ly$\alpha$ radiation
on 21 cm emission.

In addition to predicting the spectra and surface brightness profiles,
RT simulations for the Ly$\alpha$ polarization have also been developed
\citep{2002ApJ...567..922A,2008MNRAS.386..492D,2012MNRAS.424.1672D,2015JKAS...48..195A,2016A&A...593A.122T,2017MNRAS.464.5018C,2018ApJ...856..156E,2020PhRvD.101h3032M}.
\citet{1998ApJ...504L..61L} proposed, for the first time, that the
polarimetry of Ly$\alpha$ emission can be a useful probe about the
geometry and kinematics of the neutral hydrogen gas in nearby starburst
galaxies and primeval galaxies. It was also shown that the scattered
Ly$\alpha$ radiation would gain a high level of polarization for
a wide range of galactic outflows in high-redshift galaxies \citep{2008MNRAS.386..492D}.
Prior to the epoch of cosmic reionization, the Ly$\alpha$ photons
emitted by early galaxies would be resonantly scattered by the neutral
IGM in their vicinity, leading to a fairly compact Ly$\alpha$ halo
\citep{1999ApJ...524..527L}. The scattered Ly$\alpha$ light would
be highly polarized, and the polarized, intergalactic Ly$\alpha$
halos may then provide a unique tool for probing the neutral IGM before
and during the epoch of reionization \citep{1999ApJ...520L..79R}.
These calculations showed that the degree of polarization increases,
even up to $\sim40-60$\%, with increasing radial distance from the
central source.

To date, two methods have been developed to implement polarization
in the Ly$\alpha$ RT calculations. The first approach deals with
individual ``monochromatic'' photons that are 100\% linearly polarized
\citep{1999ApJ...520L..79R,2008MNRAS.386..492D,2016A&A...593A.122T}.
The second approach employs the density matrix to represent the polarization
state of a partially-polarized ``quasi-monochromatic'' ensemble of
photons \citep{1994MNRAS.267..303L,1998ApJ...504L..61L,2002ApJ...567..922A,2015JKAS...48..195A,2017MNRAS.464.5018C}.
However, neither approaches are complete, at least from a theoretical
point of view. First, instead of adopting a scattering phase function
that constantly varies as the Ly$\alpha$ frequency changes, they
use two distinct phase functions selected according to the frequency
regimes. Second, the change in polarization state due to dust scattering
was not taken into account properly. Third, no circular polarization,
which can naturally arise by dust grains in some circumstances, was
included in the formulae.

The present study was mainly motivated in two respects: (1) incomplete
treatment of Ly$\alpha$ polarization in previous theoretical studies,
and (2) the controversy over the interpretation of the polarization
signals observed in LABs. For this purpose, we developed a new, third
approach using the Stokes parameters. The method is superb in dealing
with quantum-mechanically-derived phase function compared to the methods
mentioned above. It is also able to take the dust effect into account
appropriately when dealing with the Ly$\alpha$ polarization. In our
first paper \citep{2020ApJS..250....9S}, we introduced the state-of-the-art
Monte Carlo Ly$\alpha$ RT code LaRT\footnote{LaRT is publicly available via https://doi.org/10.5281/zenodo.5618511
and https://github.com/seoncafe/LaRT.} to study Ly$\alpha$-related phenomena, including the WF effect.
LaRT was utilized to simultaneously analyze the Ly$\alpha$ spectra
and surface brightness profiles of the LAEs at $z=3-6$ \citep{2020ApJ...901...41S}.
This paper, the third in this series, extends the code to investigate
the Ly$\alpha$ polarization. We also present the results from various
simple models to demonstrate the ability of LaRT and discuss intriguing
properties on the Ly$\alpha$ polarization obtained from the models.

This paper is organized as follows. In Section \ref{sec:2}, we present
algorithms developed for LaRT, which were not explained in the first
paper. Section \ref{section:3} explains two main factors, the polarization
of individual photon packets and the degree of isotropy (or anisotropy)
in the radiation field, that determine the ensemble average of polarization
signals. These two concepts are essential to understanding the results
of this paper. Section \ref{section:4} describes the results obtained
from four types of models. We discuss a few topics that are relevant
to our results and Ly$\alpha$ RT in Section \ref{sec:5}. In Section
\ref{sec:6}, we summarize the main results. In Appendices, we derive
the scattering matrix for the Ly$\alpha$ scattering by hydrogen atoms
(\ref{sec:scattering_matrix_for_hydrogen}), and provide approximate
formulae for the scattering matrix elements for the scattering by
dust (\ref{sec:scattering_matrix_for_dust_scattering}). We also discuss
the algorithm to randomly draw scattering angles from the phase function
using the inversion method (\ref{sec:App_scattering_angle}). In addition,
we obtain generalized formulae for the density matrix method from
our formulae (\ref{sec:density_matrix}) and discuss the technique
utilizing 100\% polarized photons (\ref{sec:App_100=000025-Polarized}).

\section{Monte Carlo Radiative Transfer Methods}

\label{sec:2}

The basic Ly$\alpha$ RT algorithms of LaRT, except those for polarization
and the ``peeling-off'' technique, were described in the first paper
of this series \citep{2020ApJS..250....9S}. In the following, we
describe only the basic definitions and the contents relevant to polarization
and peeling-off. We use the polarization RT algorithm that is similar
to those of \citet{1996ApJ...465..127B} and \citet{2017A&A...601A..92P},
which were developed for the dust RT. We assume that a newly emitted
photon packet is unpolarized, unless otherwise stated. The dust grains
are assumed to be spherical and thus no dichroic attenuation due to
elongated dust grains is considered (see \citealt{2018ApJ...862...87S}
for the polarization caused by the dichroic attenuation). The polarization
algorithm described in this paper was already exploited to study the
polarization of continuum radiation in the dusty halo of edge-on galaxies
\citep{2018ApJ...862...87S}; but, no detailed description of the
algorithm was published.

The overall sequence of the RT algorithm is as follows. First, we
setup the temperature, density, and bulk velocity fields in a three
dimensional grid system. Second, an initial position, a propagation
direction, and a frequency of the individual photon packet are randomly
generated according to appropriate probability distribution functions.
Third, an optical depth $\tau$ that the photon will propagate before
it interacts with a hydrogen atom or a dust grain is randomly drawn
from the exponential distribution function $P(\tau)=e^{-\tau}$. The
physical distance is then calculated by numerically inverting the
optical depth integral given as a function of distance. Fourth, we
choose a scattering agent, either a hydrogen atom or dust grain, from
the relative ratio between the optical depths due to hydrogen atoms
and dust grains. We then find the velocity vector of an atom that
scatters the photon if it is scattered by a hydrogen atom. If the
photon is assumed to interact with dust, it is either absorbed or
scattered by a dust grain; the decision is made by comparing a uniform
random number with the dust albedo $a$. The absorbed photon can be
totally removed from the system or its weight is reduced by a factor
of $1-a$, depending on the simulation mode. Unless the photon is
removed from the system by a dust grain, we proceed further RT processes
for the photon packet. Fifth, we draw the scattering angles $(\theta,\phi)$
of the photon from an appropriate scattering phase function, and update
the Stokes vector and propagation direction (Sections \ref{subsec:2.5}-\ref{section2.8}).
A new frequency is also obtained, if the scatterer is a hydrogen atom,
from the velocity of the scattering atom and the old and new direction
vectors of the photon. If a dust grain is the scatterer, the frequency
is not altered. Sixth, to predict observational spectropolarimetric
data, the peeling-off procedure is performed for a predefined virtual
detector plane at the observer's location (Section \ref{section2.9}).
The processes are repeated until the photon either escapes the system
or is removed by dust. The whole procedures are performed for all
photon packets and the resulting outputs are converted into physical
units.

\subsection{Scattering Cross Section}

The fine structure of the $n=2$ quantum state of hydrogen should
be considered to take the Ly$\alpha$ polarization into account properly.
We refer the transition $^{2}S_{1/2}\leftrightarrow{}^{2}P_{1/2}^{{\rm o}}$
(corresponding to the lower frequency) to as ``H'' and the transition
$^{2}S_{1/2}\leftrightarrow{}^{2}P_{3/2}^{{\rm o}}$ (higher frequency)
to as ``K,'' as in the \ion{Ca}{2} $\lambda\lambda3933,3968$ doublet
line. The frequencies for the H and K transitions are denoted by $\nu_{{\rm H}}$
and $\nu_{{\rm K}}$, respectively. The central frequency of Ly$\alpha$
is $\nu_{\alpha}=2.466\times10^{15}$ Hz and the frequency difference
between the fine structure levels is $\Delta\nu_{{\rm HK}}=\nu_{{\rm K}}-\nu_{{\rm H}}=1.08\times10^{10}$
Hz, which is equivalent to a Doppler shift of 1.34 km s$^{-1}$. The
scattering cross-section of a Ly$\alpha$ photon in the rest frame
of a hydrogen atom is derived in Appendix \ref{sec:scattering_matrix_for_hydrogen}:
\begin{equation}
\sigma_{\nu}^{{\rm rest}}=\chi_{0}\left[\frac{(1/3)\Gamma/4\pi^{2}}{(\nu-\nu_{{\rm H}})^{2}+(\Gamma/4\pi)^{2}}+\frac{(2/3)\Gamma/4\pi^{2}}{(\nu-\nu_{{\rm K}})^{2}+(\Gamma/4\pi)^{2}}\right],\label{eq1}
\end{equation}
where $\chi_{0}=f_{\alpha}\pi e^{2}/m_{e}c$. Here, $f_{\alpha}=0.4162$
is the oscillator strength, and $\Gamma=A_{\alpha}=6.265\times10^{8}$
s$^{-1}$ the damping constant (the Einstein A coefficient) of the
Ly$\alpha$ transition. The multiplication factor 2 for the K transition
is due to the difference in the statistical weights of $2J+1$ between
the two transitions. Integrating over the one-dimensional Maxwellian
velocity distribution of the hydrogen gas at temperature $T$, the
cross section in a reference frame comoving with the gas fluid results
in
\begin{equation}
\sigma_{\nu}=\frac{\chi_{0}}{\sqrt{\pi}\Delta\nu_{{\rm D}}}\left[\frac{1}{3}H(x_{{\rm H}},a)+\frac{2}{3}H(x_{{\rm K}},a)\right],\label{eq2}
\end{equation}
where $H(x,a)$ is the Voigt-Hjerting function given by
\begin{align}
H(x,a) & =\frac{a}{\pi}\int_{-\infty}^{\infty}\frac{e^{-y^{2}}}{(x-y)^{2}+a^{2}}dy.\label{eq3}
\end{align}
We define $x_{{\rm H}}$, $x_{{\rm K}}$ and $x$ as the relative
frequencies of the photon normalized to the thermal Doppler width
$\Delta\nu_{{\rm D}}=\nu_{\alpha}(v_{{\rm th}}/c)$:
\begin{align}
x_{{\rm H}} & =(\nu-\nu_{{\rm H}})/\Delta\nu_{{\rm D}}=x+\Delta\nu_{{\rm HK}}/\left(2\Delta\nu_{{\rm D}}\right)\nonumber \\
x_{{\rm K}} & =(\nu-\nu_{{\rm K}})/\Delta\nu_{{\rm D}}=x-\Delta\nu_{{\rm HK}}/\left(2\Delta\nu_{{\rm D}}\right)\nonumber \\
x & =\left[\nu-(\nu_{{\rm H}}+\nu_{{\rm K}})/2\right]/\Delta\nu_{{\rm D}}=(x_{{\rm H}}+x_{{\rm K}})/2,\label{eq:4}
\end{align}
Here, $v_{{\rm th}}=(2k_{{\rm B}}T/m_{p})^{1/2}$ and $a=\Gamma/(4\pi\Delta\nu_{{\rm D}})$
are the thermal speed of hydrogen gas and the natural width parameter
of $H(x,a)$, respectively. Note that $\Delta\nu_{{\rm HK}}/\Delta\nu_{{\rm D}}\simeq(T/100\ {\rm K})^{-1/2}$;
thus, the fine structure splitting is negligible if $T\gg10^{2}$
K.

The optical depth $\tau_{\nu}(s)$ of a photon with frequency $\nu$
traveling along a path length $s$ is given by 
\begin{equation}
\tau_{\nu}(s)=\int_{0}^{s}\int_{-\infty}^{\infty}n(v_{\parallel})\sigma_{\nu}dv_{\parallel}d\ell,\label{eq:5}
\end{equation}
where $n(v_{\parallel})$ is the number density of the neutral hydrogen
atom with the velocity component $v_{\parallel}$ parallel to the
photon's propagation direction. In this paper, the total amount of
gas is measured using the column density $N_{{\rm HI}}$ of neutral
hydrogen or the optical depth $\tau_{0}$ defined by
\begin{equation}
\tau_{0}\equiv\left(\chi_{0}/\Delta\nu_{{\rm D}}\right)N_{{\rm HI}}\phi_{x}(0)\simeq\left(\chi_{0}/\sqrt{\pi}\Delta\nu_{{\rm D}}\right)N_{{\rm HI}},\label{eq:6}
\end{equation}
where $\phi_{x}=H(x,a)/\sqrt{\pi}$ is the normalized Voigt profile.
This definition of $\tau_{0}$ is consistent with that of the monochromatic
optical depth measured at the line center ($x=0$) when the fine structure
is ignored.

The parallel component $u_{\parallel}=v_{\parallel}/v_{{\rm th}}$
of the scattering hydrogen atom is obtained from the following composite
distribution function:
\begin{equation}
f_{{\rm FS}}(u_{\parallel}|x)=\mathcal{P}_{{\rm H}}f(u_{\parallel}|x_{{\rm H}})+(1-\mathcal{P}_{{\rm H}})f(u_{\parallel}|x_{{\rm K}}),\label{eq:7}
\end{equation}
where
\begin{align}
\mathcal{P}_{{\rm H}} & =\frac{H(x_{{\rm H}},a)}{H(x_{{\rm H}},a)+2H(x_{{\rm K}},a)},\label{eq:8}\\
f(u_{\parallel}|x) & =\frac{a}{\pi H(x,a)}\frac{e^{-u_{\parallel}^{2}}}{a^{2}+(x-u_{\parallel})^{2}}.\label{eq:9}
\end{align}
If a uniform random number $\xi$ ($0\le\xi\le1$) is choosen to be
smaller than $\mathcal{P}_{{\rm H}}$, the photon is scattered via
the H transition, otherwise, via the K transition. A random parallel
velocity component of the scattering atom, given a transition, is
obtained by the algorithm developed in \citet{2020ApJS..250....9S}.

When considering the fine-structure splitting, we need to compute
$H(x,a)$ more than twice as much as the case ignoring the splitting.
It is, therefore, essential to use a fast algorithm to evaluate $H(x,a)$.
We developed a fast yet accurate function, which is inspired by the
approximation formula of \citet{1948ApJ...108..112H} and the function
``voigt\_king'' in the VPFIT program \citep{2014ascl.soft08015C}.
It was found that the routine is accurate over the domain of $a\lesssim0.4$
($T\gtrsim$ 1 K for the hydrogen gas) and the whole range of $x$,
with a relative error lower than $10^{-4}$. This routine was already
utilized in \citet{2020ApJS..250....9S}. We also implemented two
more Fortran subroutines translated from the Matlab routine of \citet{2015JMR.7.163},
which are very accurate with an average accuracy of $10^{-14}$ over
a broad domain of $(a,x)$.

\begin{figure}[t]
\begin{centering}
\includegraphics[clip,scale=0.6]{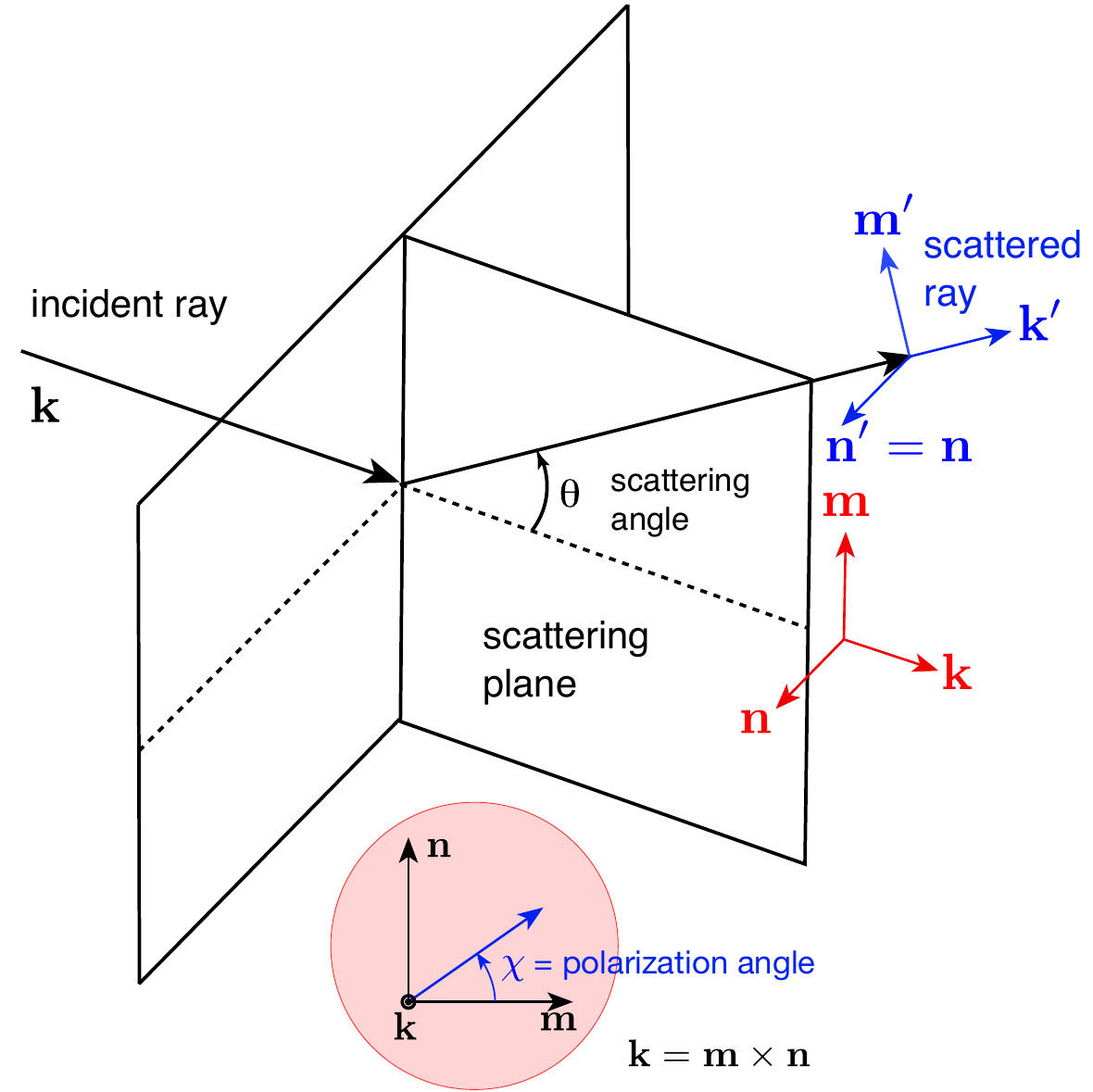}
\par\end{centering}
\begin{centering}
\medskip{}
\par\end{centering}
\caption{\label{fig01}Geometry for the definition of Stokes parameters. The
propagation direction vectors of the incident and scattered light
are denoted by $\mathbf{k}$ and $\mathbf{k}'$, respectively. A scattering
matrix is defined in the coordinate systems where the basis vector
$\mathbf{m}$ (and $\mathbf{m}'$) is parallel to and $\mathbf{n}$
($=\mathbf{n}'$) perpendicular to the scattering plane. The definition
of the polarization angle $\chi$ is shown inside the red inset circle. }

\centering{}\medskip{}
\end{figure}

\subsection{Definition of the Stokes Vector}

We describe the polarization state of a photon packet using the Stokes
vector 
\begin{equation}
\mathbf{S}=\left(\begin{array}{c}
I\\
Q\\
U\\
V
\end{array}\right),\label{eq:10}
\end{equation}
where $I$ represents the intensity of the Ly$\alpha$ radiation field,
$Q$ and $U$ describe the linear polarization, and $V$ the circular
polarization. In the present formulation, we employ a local reference
frame of the photon packet, defined by the propagation direction $\mathbf{k}$
and two orthogonal unit vectors $\mathbf{m}$ and $\mathbf{n}$ in
the plane perpendicular to $\mathbf{k}$, as shown in Figure \ref{fig01}.
The orthogonal triad of unit vectors $(\mathbf{m},\mathbf{n},\mathbf{k})$
constitutes a right-handed coordinate system. The reference frame
is attached to the photon packet and thus comoves and corotates with
the photon as it propagates and is scattered off. The Stokes vector
is defined relative to the two vectors $\mathbf{m}$ and $\mathbf{n}$
in the photon packet's local reference frame, so $\mathbf{m}$ and
$\mathbf{n}$ are referred to as the polarization basis vectors.

The Stokes parameters are defined by two orthogonal complex electric
field components along the $\mathbf{m}$ and $\mathbf{n}$ directions,
as follows:
\begin{align}
I & =\left\langle E_{m}E_{m}^{*}+E_{n}E_{n}^{*}\right\rangle \nonumber \\
Q & =\left\langle E_{m}E_{m}^{*}-E_{n}E_{n}^{*}\right\rangle \nonumber \\
U & =\left\langle E_{m}E_{n}^{*}+E_{m}^{*}E_{n}\right\rangle \nonumber \\
V & =i\left\langle E_{m}E_{n}^{*}-E_{m}^{*}E_{n}\right\rangle ,\label{eq:11}
\end{align}
where $\left\langle \cdots\right\rangle $ denotes an ensemble average
for the quasi-monochromatic wave. The sign of Stokes $V$ should be
appropriately chosen according to the definition of the electric field
phase. In our convention, the phase is assumed to be $e^{-i\omega t}$
for the angular frequency $\omega=2\pi\nu$; $V$ should be multiplied
by $-1$ if the phase of $e^{i\omega t}$ is adopted.

The linear polarization angle $\chi$ is defined as the angle between
the direction of linear polarization and the basis vector $\mathbf{m}$.
Using $\chi$, the Stokes parameters $Q$ and $U$ can be represented
as
\begin{eqnarray}
Q & = & \sqrt{Q^{2}+U^{2}}\cos2\chi,\nonumber \\
U & = & \sqrt{Q^{2}+U^{2}}\sin2\chi.\label{eq:12}
\end{eqnarray}
The degree of linear polarization is given by 
\begin{equation}
P_{{\rm L}}=\sqrt{Q^{2}+U^{2}}/I.\label{eq:13}
\end{equation}

There have been, confusingly, many different conventions concerning
the polarization angle direction of linear polarization and the handedness
of circular polarization, i.e., the signs of the Stokes parameters
$Q$, $U$, and $V$. This paper uses the ``right-handed'' system
following the International Astronomical Union recommendation \citep[see][]{1996A&AS..117..161H}.
In the IAU standard, a polarization angle is measured counterclockwise
from the North to the East in the sky when looking at the photon source.
In LaRT, the polarization angle is defined to be an angle starting
from $\mathbf{m}$, rotated about the axis $\mathbf{k}$. The vectors
$\mathbf{m}$ and $\mathbf{n}$ do not always coincide with the North
and East in the sky. However, when recording the output signals on
a detector plane in the peeling-off stage, we rotate $\mathbf{m}$
and $\mathbf{n}$ to coincide with the North and East directions in
the sky, respectively.

\citet{2017A&A...601A..92P} compare different conventions for $U$
and $V$ adopted by various authors.\footnote{\citet{2017A&A...601A..92P} adopt the right-handed basis as in this
paper. However, their approach differs from ours in that they measure
the polarization angle starting from $\mathbf{n}$, regarding $\mathbf{n}$
and $-\mathbf{m}$ as being the North and East direction. In LaRT,
a photon packet carries all three basis vectors ($\mathbf{m},$ $\mathbf{n}$,
$\mathbf{k}$). On the other hand, \citet{2017A&A...601A..92P} use
only two of them ($\mathbf{k}$ and $\mathbf{n}$) in the hope of
reducing the number of computations. However, we found that the number
of calculations is the same if the computation sequence is well organized,
even when using all three basis vectors. It is also more convenient
to employ the three basis vectors.} For example, \citet{1960ratr.book.....C} measures the polarization
angle clockwise when looking at the source (left-handed basis; $U_{{\rm Chandra}}=-U_{{\rm IAU}}$),
while Stokes $V$ is defined in the same way as the IAU standard ($V_{{\rm Chandra}}=V_{{\rm IAU}}$).
\citet{2002ApJ...567..922A} and \citet{2018ApJ...856..156E} use
the same convention as \citet{1960ratr.book.....C} in the definition
of the density matrix. This paper and \citet{2017A&A...601A..92P}
represent the Stokes vector in the continuously moving, local reference
frame of the photon, while \citet{2002ApJ...567..922A} and \citet{2018ApJ...856..156E}
use a fixed (or laboratory) reference frame to define the density
matrix. In Appendix \ref{sec:density_matrix}, a more general density
matrix formula than that given in \citet{2002ApJ...567..922A} is
obtained from the Stokes vector scheme of the present study; the appendix
verifies the equivalence of the two approaches.

\subsection{Transform of the Stokes and Basis Vectors\label{section2.3}}

\begin{figure}[t]
\begin{centering}
\includegraphics[clip,scale=0.52]{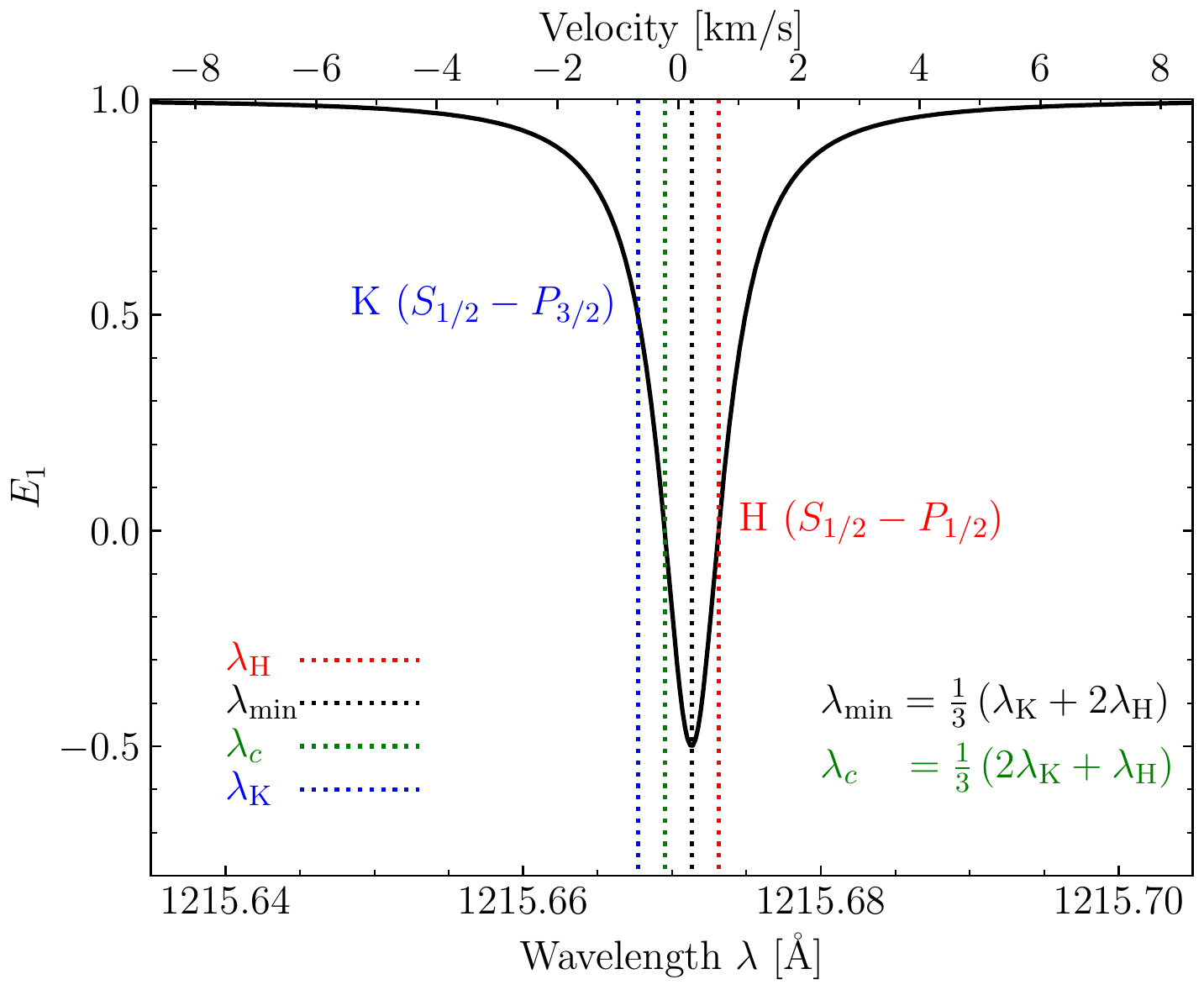}
\par\end{centering}
\begin{centering}
\medskip{}
\par\end{centering}
\caption{\label{fig02} The $E_{1}$ parameter for the scattering matrix of
Ly$\alpha$ as a function of wavelength (velocity). The wavelengths
for the K and H transitions of Ly$\alpha$ are $\lambda_{{\rm K}}$
and $\lambda_{{\rm H}}$, respectively. $\lambda_{{\rm min}}$ is
the wavelength at which $E_{1}$ has the minimum value. $E_{1}$ has
a negative value in the wavelength range of $\lambda_{c}<\lambda<\lambda_{{\rm H}}$;
a negative $E_{1}$ implies that the polarization pattern resulting
from a central source is radial rather than concentric. Interestingly,
$\lambda_{c}$ and $\lambda_{{\rm min}}$ divide the interval $\lambda_{{\rm H}}-\lambda_{{\rm K}}$
equally into three parts.}
\medskip{}
\end{figure}

After a scattering event, the polarization state is updated by multiplying
the initial Stokes vector by a scattering matrix (also known as the
M\"uller matrix), suitable for the type of scattering. The scattering
matrix elements depend on the geometrical configuration and the material
property; they also depend on the photon frequency. The scattering
matrix $\mathbf{M}(\theta)$ is usually defined so that the basis
vector $\mathbf{m}$ lies in the scattering plane, constructed by
the incoming and outgoing propagation directions $\mathbf{k}$ and
$\mathbf{k}'$ of the photon, as shown in Figure \ref{fig01}. The
scattering matrix is thus a function of only the scattering (polar)
angle $\theta$ between $\mathbf{k}$ and $\mathbf{k}'$ ($\cos\theta=\mathbf{k}\cdot\mathbf{k}'$).

Suppose a photon is scattered in a direction with a polar angle $\theta$
and azimuth angle $\phi$ in a coordinate system defined by $(\mathbf{m},\mathbf{n},\mathbf{k})$.
In that case, we first need to rotate the polarization basis vectors
$(\mathbf{m},\mathbf{n})$ by the azimuth angle $\phi$ to make $\mathbf{m}$
lie in the scattering plane before applying the scattering matrix
$\mathbf{M}$. The Stokes vector should then be expressed in the rotated,
reference frame by multiplying it by a rotation matrix $\mathbf{L}(\phi)$:
\begin{align}
\mathbf{L}(\phi) & =\left(\begin{array}{cccc}
1 & 0 & 0 & 0\\
0 & \cos2\phi & \sin2\phi & 0\\
0 & -\sin2\phi & \cos2\phi & 0\\
0 & 0 & 0 & 1
\end{array}\right).\label{eq:14}
\end{align}
The final Stokes vector after the scattering event is obtained by
mutiplying $\mathbf{M}(\theta)$ to $\mathbf{L}(\phi)\mathbf{S}$,
as follows:
\begin{equation}
\mathbf{S'}=\mathbf{M}(\theta)\mathbf{L(\phi)}\mathbf{S}.\label{eq:15}
\end{equation}

After every scattering event, we also need to update the basis vectors
($\mathbf{m}$, $\mathbf{n}$). The new basis vectors are obtained
first by rotating them by $\phi$ about $\mathbf{k}$, updating ($\mathbf{m}$,
$\mathbf{n}$), and then subsequently rotating by $\theta$ about
the updated $\mathbf{n}'$ (already rotated by $\phi$). The resulting
new basis vectors are expressed in terms of the initial basis vectors
and the scattering angles $(\theta,\,\phi)$, as follows:
\begin{align}
\mathbf{m}' & =\cos\theta\left(\cos\phi\mathbf{m}+\sin\phi\mathbf{n}\right)-\sin\theta\mathbf{k}\nonumber \\
\mathbf{n}' & =-\sin\phi\mathbf{m}+\cos\phi\mathbf{n}\nonumber \\
\mathbf{k}' & =\sin\theta\left(\cos\phi\mathbf{m}+\sin\phi\mathbf{n}\right)+\cos\theta\mathbf{\mathbf{k}}.\label{eq:16}
\end{align}
Note that the Stokes vector $\mathbf{S}'$ in Equation (\ref{eq:15})
is automatically expressed in the new local reference frame constructed
by $(\mathbf{m}',\mathbf{n}',\mathbf{k}')$.

\subsection{Initial Stokes and Basis Vectors}

We now describe how to select the initial basis vectors for a newly
emitted photon packet. When injecting a photon toward a direction
of angles $(\theta,\phi)$ in the galaxy reference frame, the initial
basis vectors can be conveniently obtained by rotating $\mathbf{m}=\left(1,0,0\right)^{T}$,
$\mathbf{n}=\left(0,1,0\right)^{T}$, and $\mathbf{k}=\left(0,0,1\right)^{T}$
by the angles $(\theta,\,\phi)$ using Equation (\ref{eq:16}). Consequently,
the initial basis vectors for the angles $(\theta,\phi)$ are chosen
to be:
\begin{eqnarray}
\mathbf{m}_{0} & = & \left(\cos\theta\cos\phi,\cos\theta\sin\phi,-\sin\theta\right)^{T}\nonumber \\
\mathbf{n}_{0} & = & \left(-\sin\phi,\cos\phi,0\right)^{T}\nonumber \\
\mathbf{k}_{0} & = & \left(\sin\theta\cos\phi,\sin\theta\sin\phi,\cos\theta\right)^{T}.\label{eq:17}
\end{eqnarray}
Note that \citet{2017A&A...601A..92P} use only one polarization basis
vector referred to as $\mathbf{n}$ and choose an initial basis vector
to be $\mathbf{n}_{0}^{{\rm Peest}}=-\mathbf{m}_{0}$ in their Equations
(38)-(40). The two polarization basis vectors of \citet{2015JKAS...48..195A}
can be expressed in terms of our definition, such that $\mathbf{e}_{1}=\mathbf{n}_{0}$
and $\mathbf{e}_{2}=\mathbf{m}_{0}$.

In this study, we assume, unless otherwise stated, that initial input
photons are unpolarized, i.e., $I=1$ and $Q=U=V=0$. Alternatively,
unpolarized light can also be modeled by adopting 100\% linearly polarized,
monochromatic photons $(P_{{\rm L}}=1)$. In this case, however, the
polarization angle $\chi$ should be chosen uniform-randomly in the
range $0\le\chi\le2\pi$. This approach of using 100\% polarized photons
is equivalent to that of \citet{1999ApJ...520L..79R} if the simulation
is limited to the Rayleigh scattering (see Appendix \ref{sec:App_100=000025-Polarized}
for further discussion).

\subsection{Scattering Matrix for the Scattering by Hydrogen}

\label{subsec:2.5}The scattering matrix for the scattering of Ly$\alpha$
photons by hydrogen atoms is expressed by
\begin{align}
\mathbf{M}(\theta) & =\left(\begin{array}{cccc}
S_{11} & S_{12} & 0 & 0\\
S_{12} & S_{22} & 0 & 0\\
0 & 0 & S_{33} & 0\\
0 & 0 & 0 & S_{44}
\end{array}\right),\label{eq:18}
\end{align}
where
\begin{eqnarray}
S_{11} & = & \frac{3}{4}E_{1}\left(\cos^{2}\theta+1\right)+E_{2}\nonumber \\
S_{12} & = & \frac{3}{4}E_{1}\left(\cos^{2}\theta-1\right)\nonumber \\
S_{22} & = & \frac{3}{4}E_{1}\left(\cos^{2}\theta+1\right)\nonumber \\
S_{33} & = & \frac{3}{2}E_{1}\cos\theta\nonumber \\
S_{44} & = & \frac{3}{2}E_{3}\cos\theta,\label{eq:19}
\end{eqnarray}
as described in Appendix \ref{sec:scattering_matrix_for_hydrogen}.
Here, $E_{1}$, $E_{2}$, and $E_{3}$ are functions of frequency
given by

\begin{eqnarray}
E_{1} & = & \frac{2\left(\nu-\nu_{{\rm K}}\right)\left(\nu-\nu_{{\rm H}}\right)+\left(\nu-\nu_{{\rm H}}\right)^{2}}{\left(\nu-\nu_{{\rm K}}\right)^{2}+2\left(\nu-\nu_{{\rm H}}\right)^{2}}=\frac{2x_{{\rm K}}x_{{\rm H}}+x_{{\rm H}}^{2}}{x_{{\rm K}}^{2}+2x_{{\rm H}}^{2}}\nonumber \\
E_{2} & = & 1-E_{1}\nonumber \\
E_{3} & = & \frac{1}{3}\left(E_{1}+2\right).\label{eq:20}
\end{eqnarray}
The second equation for $E_{2}$ is always true for the transition
between a singlet ground state and an upper doublet state with arbitrary
angular quantum numbers. But, the third equation between $E_{3}$
and $E_{1}$ is satisfied only for the transitions occurring between
the same quantum levels as for Ly$\alpha$. The matrix elements $S_{11}$,
$S_{12}$, $S_{22}$, and $S_{33}$ have been described in \citet{1980A&A....84...68S}.
The matrix element $S_{44}$ for the circular polarization is newly
derived in Appendix \ref{sec:scattering_matrix_for_hydrogen}.

\begin{figure}[t]
\begin{centering}
\includegraphics[clip,scale=0.52]{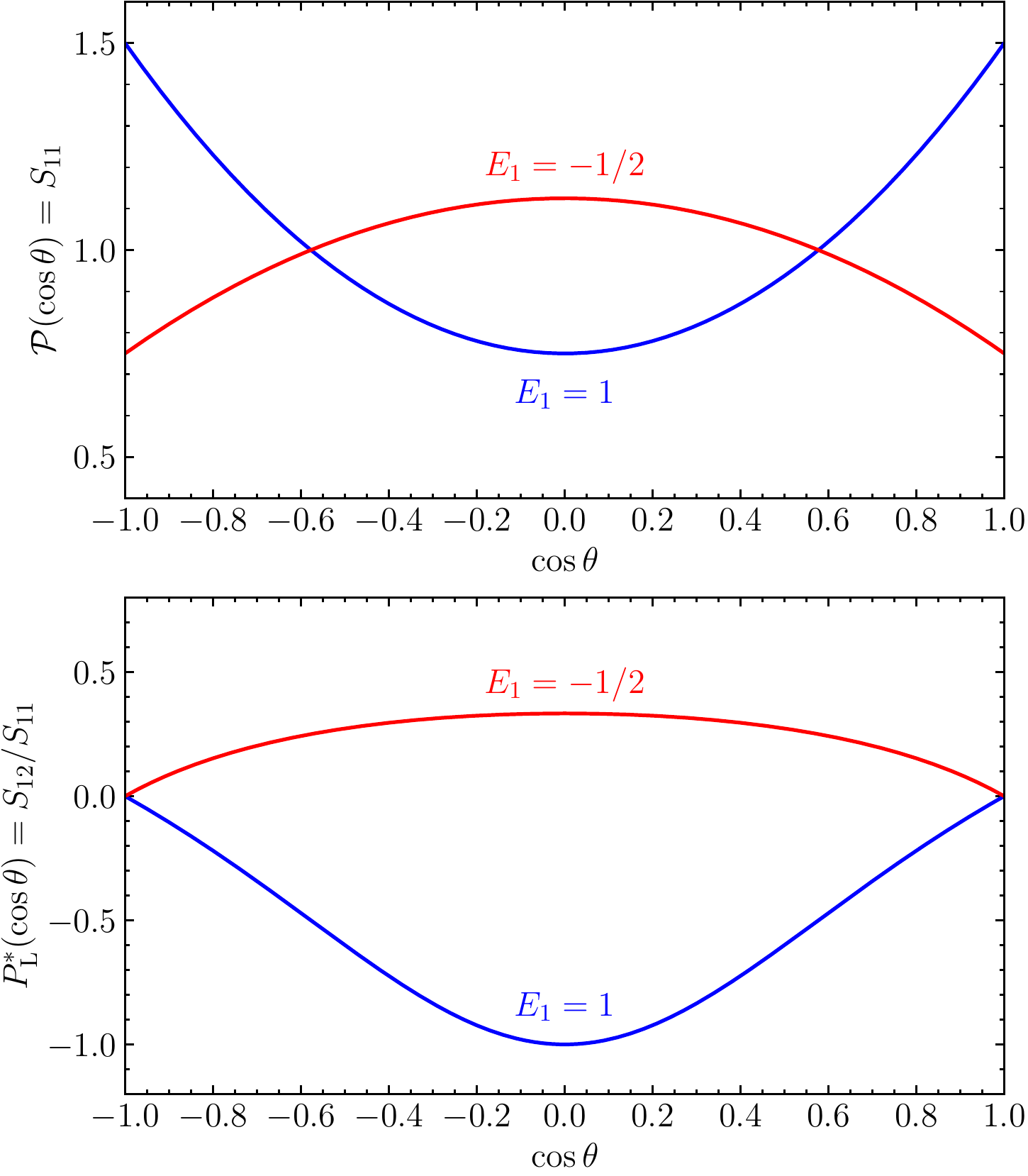}
\par\end{centering}
\begin{centering}
\medskip{}
\par\end{centering}
\caption{\label{fig03} (top) Scattering phase function for $E_{1}=-1/2$ and
$E_{1}=1$. (bottom) Degree of linear polarization for $E_{1}=-1/2$
and $E_{1}=1$. Here, $P_{{\rm L}}^{*}$ is defined in Equation (\ref{eq:23})
to allow to have a negative value to explain the direction of polarization
vector. In a spherical system where a point light source is located
at the center, a negative $P_{{\rm L}}^{*}$ indicates the tangential
polarization, and a positive $P_{{\rm L}}^{*}$ the radial polarization.}
\medskip{}
\end{figure}

The new Stokes parameters after a scattering event by a hydrogen atom
are given by
\begin{eqnarray}
I' & = & S_{11}I+S_{12}\left(Q\cos2\phi+U\sin2\phi\right)\nonumber \\
Q' & = & S_{12}I+S_{22}\left(Q\cos2\phi+U\sin2\phi\right)\nonumber \\
U' & = & S_{33}\left(-Q\sin2\phi+U\cos2\phi\right)\nonumber \\
V' & = & S_{44}V.\label{eq:21}
\end{eqnarray}
The last equation for $V$ implies that the scattering of Ly$\alpha$
by hydrogen atoms does not newly produce any circular polarization
if there was no initial circular polarization; but, it will remain
once it is created. The scattering matrix for the electron Thomson
scattering and Rayleigh scattering is obtained by setting $E_{1}=1$,
$E_{2}=0$, and $E_{3}=1$.

The parameter $E_{1}$ represents the effect of quantum interference
between the fine-structure levels in resonant Ly$\alpha$ scattering.
It determines the shape of the scattering phase function as a function
of frequency. As shown in Figure \ref{fig02}, it is intriguing that
$E_{1}<0$ for a wavelength range of $\lambda_{c}<\lambda<\lambda_{{\rm H}}$,
where $\lambda_{c}=(2\lambda_{{\rm K}}+\lambda_{{\rm H}})/3$. However,
in most wavelengths, $E_{1}$ is positive. From Equation (\ref{eq:21}),
the scattering phase function for initially unpolarized light $(Q=U=0)$
is obtained to be
\begin{equation}
\mathcal{P}(\cos\theta)=\frac{I'}{I}=\frac{3}{4}E_{1}\left(\cos^{2}\theta+1\right)+1-E_{1}.\label{eq:22}
\end{equation}
The phase functions, for instance, for $E_{1}=-1/2$ and $E_{1}=1$,
are shown in the top panel of Figure \ref{fig03}. In most cases $(E_{1}>0$),
scattering occurs dominantly in the forward and backward directions.
However, for a negative $E_{1}$, scattering occurs more frequently
in perpendicular directions to the incident direction. For a positive
$E_{1}$, the phase function is a linear superposition of Rayleigh
scattering and isotropic scattering with weights of $E_{1}$ and $1-E_{1}$.

An interesting thing to note is that when $E_{1}<0$, scattering events
give rise to a counterintuitive, radial polarization pattern in the
case of a central light source; this type of polarization is referred
to as the ``negative polarization.'' In most circumstances, the polarization
pattern made by a central source is concentric, in other words, perpendicular
to radial directions from the source. The counterintuitive, radial
polarization pattern for $E_{1}<0$ can be demonstrated by examining
the degree of linear polarization of the scattered light (for initially
unpolarized light, $Q=U=0$), which is defined to allow to have a
negative value:
\begin{equation}
P_{{\rm L}}^{*}=\frac{Q'}{I'}=\frac{S_{12}}{S_{11}}=\frac{\left(3/4\right)E_{1}\left(\cos^{2}\theta-1\right)}{\left(3/4\right)E_{1}\left(\cos^{2}\theta+1\right)+1-E_{1}}.\label{eq:23}
\end{equation}
Thus, a negative $E_{1}$ yields a positive $Q'$, meaning a radial
polarization pattern ($\left|E_{m}\right|^{2}>\left|E_{n}\right|^{2}$
in Equation (\ref{eq:11})), as shown in the lower panel of Figure
\ref{fig03}. We also note that, regardless of the sign of $E_{1}$
(or polarization direction), the maximum polarization always happens
at the right-angle scattering case.

In most astrophysical cases, a substantial diffusion of Ly$\alpha$
photons in frequency space is expected because of a large number of
resonance scatterings. Therefore, most photons will escape the system
only after scattering events occurring at the wing regime where the
Rayleigh scattering is predominant and $E_{1}$ is positive. In the
end, such a radial polarization (negative polarization) pattern that
happens when $E_{1}<0$ is unlikely to be observable in actual situations.

\subsection{Scattering Matrix for the Scattering by Dust}

For scattering by spherical dust particles, the scattering matrix
has a form of
\begin{align}
\mathbf{M}^{{\rm dust}}(\theta) & =\left(\begin{array}{cccc}
S_{11} & S_{12} & 0 & 0\\
S_{12} & S_{11} & 0 & 0\\
0 & 0 & S_{33} & S_{34}\\
0 & 0 & -S_{34} & S_{33}
\end{array}\right).\label{eq:24}
\end{align}
The matrix elements for dust scattering cannot be expressed as an
analytical function of $\theta$, unlike the scattering case by hydrogen
atoms. Instead, they should be numerically estimated using the Mie
theory for a given grain size distribution and material composition.
In general, all 16 elements of the scattering matrix for non-spherical
dust grains will not be vanishing. Non-spherical dust is beyond the
scope of this paper; however, the present work can be relatively easily
extended to non-spherical cases. The scattering matrix elements calculated
for the Milky-Way (MW), Small Magenellic Cloud (SMC), and Large MC
(LMC) dust models of \citet{2001ApJ...548..296W} are described in
Appendix \ref{sec:scattering_matrix_for_dust_scattering}. In this
paper, we adopt the MW dust model. The gas-to-dust ratio is fixed
to that of the MW ($\sim100$ by mass).

The new Stokes parameters after a dust scattering are then given by
\begin{eqnarray}
I' & = & S_{11}I+S_{12}\left(Q\cos2\phi+U\sin2\phi\right)\nonumber \\
Q' & = & S_{12}I+S_{11}\left(Q\cos2\phi+U\sin2\phi\right)\nonumber \\
U' & = & S_{33}\left(-Q\sin2\phi+U\cos2\phi\right)+S_{34}V\nonumber \\
V' & = & -S_{34}\left(-Q\sin2\phi+U\cos2\phi\right)+S_{33}V.\label{eq:25}
\end{eqnarray}
The above equation for $V'$ indicates that circular polarization
can arise naturally in dust scattering even without initial circular
polarization ($V=0$). We also note that the presence of circular
polarization can modify the Stokes $U$ parameter and thus the degree
of ``linear'' polarization. Therefore, we need to consider the possibility
of the circular polarization of Ly$\alpha$ in dusty media, at least
for theoretical completeness. The effect of circular polarization
is, however, likely to be negligible in most circumstances.

\subsection{Random Sampling of Scattering Angles\label{section2.7}}

Regardless of whether the scattering partner is a hydrogen atom or
dust grain, the scattering phase function, which gives the probability
distribution function to be scattered by angles $(\theta,\,\phi)$,
is given by
\begin{align}
\mathcal{P}(\theta,\phi) & =\frac{I'/I}{\int I'/Id\Omega}\nonumber \\
 & =\frac{1}{2\pi}\left[S_{11}(\theta)+S_{12}(\theta)\left(\frac{Q}{I}\cos2\phi+\frac{U}{I}\sin2\phi\right)\right].\label{eq:26}
\end{align}
Here, $S_{11}$ is assumed to be normalized, i.e., $\int S_{11}\sin\theta d\theta=1$;
all other scattering matrix elements are also normalized by the same
factor as that for $S_{11}$. One wants to generate two random variables
$\theta$ and $\phi$ simultaneously, according to the joint distribution
function $\mathcal{P}(\theta,\phi)$. It is, however, more convenient
to obtain $\theta$ first and then $\phi$, instead of generating
$(\theta,\phi)$ simultaneously. For this purpose, we first marginalize
$\mathcal{P}(\theta,\phi)$ over the azimuth angle $\phi$ and obtain
a random $\theta$ from the marginal distribution of $\theta$:
\begin{eqnarray}
P(\theta) & = & \int_{0}^{2\pi}\mathcal{P}(\theta,\phi)d\phi=S_{11}(\theta).\label{eq:27}
\end{eqnarray}

For a chosen $\theta$, the conditional probability distribution function
of $\phi$ is given by

\begin{align}
\overline{P}(\phi|\theta) & =\frac{\mathcal{P}(\theta,\phi)}{P(\theta)}\nonumber \\
 & =\frac{1}{2\pi}\left[1+\frac{S_{12}(\theta)}{S_{11}(\theta)}\left(\frac{Q}{I}\cos2\phi+\frac{U}{I}\sin2\phi\right)\right].\label{eq:28}
\end{align}
A random variate for $\phi$ from this distribution function can be
obtained using a rejection method. We first choose a random azimuth
angle $\phi=2\pi\xi$ from a uniform random number $\xi$ between
0 and 1. We accept it if $\xi'\le\overline{P}(\phi|\theta)/\bar{P}_{{\rm max}}$
for a new independent random variate $\xi'$ ($0\le\xi'<1$). If not,
then generate a new random $\phi$ until it is accepted. Here, the
maximum value $\bar{P}_{{\rm max}}$ of $\overline{P}(\phi|\theta)$
is given by
\begin{equation}
\bar{P}_{{\rm max}}=\frac{1}{2\pi}\left[1+\frac{S_{12}(\theta)}{S_{11}(\theta)}\sqrt{\left(\frac{Q}{I}\right)^{2}+\left(\frac{U}{I}\right)^{2}}\right].\label{eq:29}
\end{equation}
The procedure is efficient in that more than 50\% (in fact, most)
of $\phi$ is accepted.

An alternative sampling strategy is to choose a scattering angle $\theta$
according to $P(\theta)$ and draw a ``biased'' azimuth angle $\phi$
using a uniform distribution in the range of $\left[0,2\pi\right)$,
i.e., $\phi=2\pi\xi$. In this case, we need to adjust the weight
of the photon packet to $w'=w\overline{P}(\phi|\theta)/2\pi$, where
$w$ and $w'$ are weights for the incident and scattered photon packet,
respectively. The present study uses the approach of using a rejection
method for $\phi$.

For the case of scattering by hydrogen atoms, the scattering angle
$\theta$ between the incident and scattered directions is governed
by the following probability distribution function:

\begin{equation}
P(\mu)\equiv\frac{S_{11}}{\int S_{11}d\mu}=\frac{3E_{1}}{8}\mu^{2}+\frac{4-E_{1}}{8},\label{eq:30}
\end{equation}
where $\mu\equiv\cos\theta$. We use the inversion method to generate
random scattering angles from this distribution function. In other
words, a random angle $\theta$ is given by inverting the integral
\begin{equation}
\int_{0}^{\mu}P(\mu')d\mu'=\xi\label{eq:31}
\end{equation}
for a uniform random number $\xi$. The integral yields a cubic equation
of $\mu$ for given $\xi$ and $E_{1}$, which can be solved using
Cardano's method. The analytical solution of the cubic equation is
summarized, for an arbitrary $E_{1}$ $(-1/2\le E_{1}\le1)$, as follows:

\begin{align}
\mu & =\begin{cases}
\left|p\right|^{1/2}\left(W-1/W\right) & {\rm for}\;\ E_{1}>0\\
\\
2\left|p\right|^{1/2}\cos\left[\left(\cos^{-1}\mathcal{Q}+4\pi\right)/3\right] & {\rm for}\;\ E_{1}<0\\
\\
2\xi-1 & {\rm for}\;\ E_{1}=0,
\end{cases}\label{eq:32}
\end{align}
where

\begin{align}
p & \equiv\frac{4-E_{1}}{3E_{1}},\ \mathcal{Q}\equiv\frac{4\xi-2}{E_{1}\left|p\right|^{3/2}},\nonumber \\
W & \equiv\left(\mathcal{Q}+\sqrt{\mathcal{Q}^{2}+1}\right)^{1/3}.\label{eq:33}
\end{align}
The solution is detailed in Appendix \ref{sec:App_scattering_angle}.

For the scattering by dust grains, the scattering phase function for
$\theta$ is well approximated by a Henyey-Greenstein (H-G) function
or a combination of two independent H-G functions \citep{1977ApJS...35....1W}.
The integral of the H-G function is analytically invertible, as described
in \citet{1977ApJS...35....1W}; a scattering angle $\theta$ that
follows a H-G function can be obtained by
\begin{equation}
\mu=\frac{1+g^{2}}{2g}-\frac{1}{2g}\left(\frac{1-g^{2}}{1-g+2g\xi}\right)^{2}.\label{eq:34}
\end{equation}
If one adopts a sum of two H-G functions, they can use a composition
method to generate scattering angles ($\theta$). Instead of using
an analytical approximation for the phase function, the numerical
inversion technique can also be used to draw the random scattering
angle $\theta$. The scattering matrix elements, including the phase
function, for the scattering by dust are described in Appendix \ref{sec:scattering_matrix_for_dust_scattering}.
In the appendix, we also provide approximate equations for the matrix
elements of Equation (\ref{eq:25}).

\subsection{Update of the Stokes vector\label{section2.8}}

After each scattering event, we update the Stokes vector as follows:
\begin{equation}
\mathbf{S}'=\left(\begin{array}{c}
1\\
Q'/I'\\
U'/I'\\
V'/I'
\end{array}\right).\label{eq:35}
\end{equation}
The Stokes vector is updated to have a unit intensity because the
photon packet carries unit intensity in the Monte-Carlo simulation.
When calculating the new Stokes vector $\mathbf{S}'$ after a scattering
event by dust, it is more convenient to use the scattering matrix
elements divided by $S_{11}$ rather than the elements themselves,
as described in Appendix \ref{sec:scattering_matrix_for_dust_scattering}.
If we designed for photon packets to have a weight (especially for
the dust scattering case), we also need to update the photon weight
appropriately and multiply the Stokes vector by the same factor.

\subsection{Peeling-off Technique\label{section2.9}}

\begin{figure}[t]
\begin{centering}
\includegraphics[clip,scale=0.45]{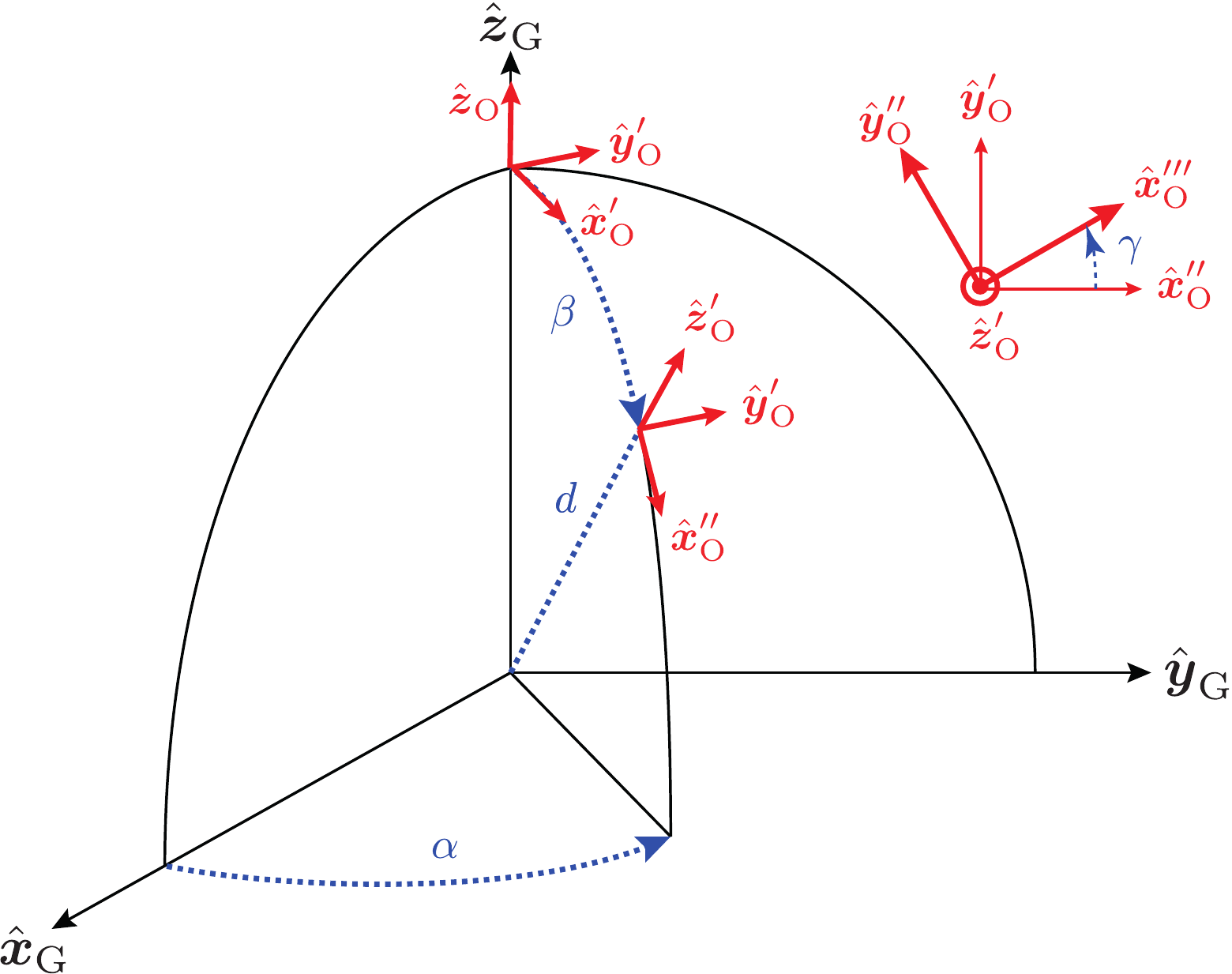}
\par\end{centering}
\begin{centering}
\medskip{}
\par\end{centering}
\caption{\label{fig04}Coordinate systems of the galaxy (or hydrogen+dust cloud,
denoted by G) and the observer (O). The coordinates of the observer
measured in the galaxy frame are given by $\mathbf{r}_{{\rm obs}}=(x_{{\rm obs}},y_{{\rm obs}},z_{{\rm obs}})$
$=$ $(d\cos\alpha\sin\beta,d\sin\alpha\sin\beta,d\cos\beta)$. The
angle $\gamma$ defines the orientation of the detector plane.}
\medskip{}
\end{figure}

A Monte Carlo RT simulation aims to obtain a spectral (or spectro-polarimetric)
image on a detector to compare with observational data. However, in
a simple Monte Carlo simulation, there will be hardly a probability
to detect a statistically significant number of photon packets on
a virtual detector located far from the system. We, therefore, use
the ``peeling-off'' (also known as ``next event estimation'' or ``shadow
rays'') technique to obtain high signal-to-noise images in a detector
plane \citep{1984ApJ...278..186Y,2007ApJ...657L..69L,2012MNRAS.424..884Y}.
When using this technique, the probability of a photon escaping in
the direction of the observer in every scattering (and emission) event
is added to the detector plane. The ``peeling-off'' technique has
been implemented in many Ly$\alpha$ RT codes, but, surprisingly,
was not well described. Here, we briefly describe a peeling-off strategy
implemented in LaRT.

\subsubsection{Detector Plane}

LaRT can place an observer in an arbitrary location and make a detector
plane to have an arbitrary orientation in the sky. As shown in Figure
\ref{fig04}, the location of the observer is defined by a luminosity
distance ($d$) and two rotation angles $(\alpha,\beta)$, or equivalently,
by the coordinates $x_{{\rm obs}}$, $y_{{\rm obs}}$, and $z_{{\rm obs}}$
of the observer in the galaxy frame; a third angle $\gamma$ then
defines the orientation of the detector plane. The observer frame,
given by $(\hat{\mathbf{x}}_{{\rm O}},\hat{\mathbf{y}}_{{\rm O}},\hat{\mathbf{z}}_{{\rm O}})$,
is assumed to initially coincide with the galaxy frame defined by
$(\hat{\mathbf{x}}_{{\rm G}},\hat{\mathbf{y}}_{{\rm G}},\hat{\mathbf{z}}_{{\rm G}})$.
Next, the observer frame is obtained by sequentially rotating the
initial observer frame by $\alpha$ about $\hat{\mathbf{z}}_{{\rm O}}$,
by $\beta$ about the rotated axis $\hat{\mathbf{y}}'_{{\rm O}}$,
and then by $\gamma$ about the rotated axis $\hat{\mathbf{z}}'_{{\rm O}}$.\footnote{This transformation is equivalent to rotating the galaxy by $-\gamma$
about $\hat{\mathbf{z}}_{{\rm G}}$, by $-\beta$ about the rotated
axis $\hat{\mathbf{y}}'_{{\rm G}}$, and by $-\alpha$ about the rotated
axis $\hat{\mathbf{z}}'_{{\rm G}}$ while fixing the observer frame.
When one considers a disk galaxy, the three angles correspond to the
position angle $(\gamma)$ of the major axis, the inclination angle
$(\beta)$ of the galactic plane, and the phase angle ($\alpha)$
of the spiral pattern of the galaxy. In most cases, it is convenient
to choose $\gamma=0$ for $\beta=0$, $\gamma=\pi/2$ for $0<\beta<\pi/2$,
and $\gamma=-\pi/2$ for $\pi/2<\beta<\pi$. \citet{2014ApJ...785L..18S}
used a similar convention to model an edge-on disk galaxy.} The detector is then located at the coordinate $(0,0,d)$ in the
observer frame. Therefore, the components of a direction vector in
the observer frame can be obtained by multiplying its components in
the galaxy frame by the following rotation matrix:
\begin{align}
\mathbf{R} & \equiv\mathbf{R}_{z}(\gamma)\mathbf{R}_{y}(\beta)\mathbf{R}_{z}(\alpha),\label{eq:36}
\end{align}
where
\begin{align}
\mathbf{R}_{z}(\alpha) & =\left(\begin{array}{ccc}
\cos\alpha & \sin\alpha & 0\\
-\sin\alpha & \cos\alpha & 0\\
0 & 0 & 1
\end{array}\right),\nonumber \\
\mathbf{R}_{y}(\beta) & =\left(\begin{array}{ccc}
\cos\beta & 0 & -\sin\beta\\
0 & 1 & 0\\
\sin\beta & 0 & \cos\beta
\end{array}\right),\nonumber \\
\mathbf{R}_{z}(\gamma) & =\left(\begin{array}{ccc}
\cos\gamma & \sin\gamma & 0\\
-\sin\gamma & \cos\gamma & 0\\
0 & 0 & 1
\end{array}\right).\label{eq:37}
\end{align}

To perform the ``peeling-off'' of a photon packet, we imagine a ``virtual''
scattering event in the direction of the observer and construct a
propagation vector $(\mathbf{k}_{{\rm peel}}=(\mathbf{r}_{{\rm obs}}-\mathbf{r}_{{\rm ph}})/\left|\mathbf{r}_{{\rm obs}}-\mathbf{\mathbf{r}}_{{\rm ph}}\right|)$
of the ``peeled-off'' photon by connecting the photon $(\mathbf{r}_{{\rm ph}})$
to the observer $(\mathbf{r}_{{\rm obs}})$. The propagation vector
is then transformed from the galaxy frame to the observer frame. The
celestial coordinates of the ``peeled-off'' photon in the detector
plane are calculated to be $\theta_{x}=\arctan(-k_{x}^{{\rm O}},k_{z}^{{\rm O}}$)
and $\theta_{y}=\arctan(-k_{y}^{{\rm O}},k_{z}^{{\rm O}})$, where
$k_{x}^{{\rm O}}$, $k_{y}^{{\rm O}}$, and $k_{z}^{{\rm O}}$ are
the coordinate components of $\mathbf{k}_{{\rm peel}}$ in the observer
frame. The celestial coordinates are finally binned into a two-dimensional
array to obtain an image of a desired observable quantity in the detector
plane.

\subsubsection{The Stokes vector of peeled-off photon}

To calculate the Stokes vector carried by a peeled-off photon packet,
let us consider a photon packet peeled off into a direction ($\mathbf{\mathbf{k}}_{{\rm peel}}$)
pointing toward the observer. We first need to find the scattering
angles $\theta_{{\rm peel}}$ and $\phi_{{\rm peel}}$ for this ``virtual''
scattering event. For this purpose, the basis vectors ($\mathbf{n}_{{\rm peel}}$
and $\mathbf{k}_{{\rm peel}}$) of the peeled-off photon are expressed
in terms of the initial basis vectors $(\mathbf{m},\mathbf{n},\mathbf{k})$
and the scattering angles $(\theta_{{\rm peel}},\phi_{{\rm peel}})$,
as follows:
\begin{align}
\mathbf{n}_{{\rm peel}} & =-\sin\phi_{{\rm peel}}\mathbf{m}+\cos\phi_{{\rm peel}}\mathbf{n}\label{eq:38}\\
\mathbf{k}_{{\rm peel}} & =\sin\theta_{{\rm peel}}(\cos\phi_{{\rm peel}}\mathbf{m}+\sin\phi_{{\rm peel}}\mathbf{n})+\cos\theta_{{\rm peel}}\mathbf{k},\label{eq:39}
\end{align}
where $\mathbf{n}_{{\rm peel}}$, $\theta_{{\rm peel}}$, and $\phi_{{\rm peel}}$
are the unknowns supposed to be found using ($\mathbf{m}$, $\mathbf{n}$,
$\mathbf{k}$) and $\mathbf{k}_{{\rm peel}}$. Here, it is not necessary
to derive $\mathbf{m}_{{\rm peel}}=\mathbf{n}_{{\rm peel}}\times\mathbf{k}_{{\rm peel}}$.

From Equation (\ref{eq:39}), the scattering polar angle $\theta_{{\rm peel}}$
is readily calculated via the scalar product of the incoming and ``virtual''
scattering direction vectors:
\begin{equation}
\cos\theta_{{\rm peel}}=\mathbf{k}\cdot\mathbf{k}_{{\rm peel}},\label{eq:40}
\end{equation}
which uniquely determines the scattering angle in the range of $0\le\theta_{{\rm peel}}\le\pi$.
To calculate the azimuth angle $\phi_{{\rm peel}}$ unambiguously,
both cosine and sine of the angle are required. They can be obtained
from Equation (\ref{eq:39}):
\begin{align}
\cos\phi_{{\rm peel}} & =\mathbf{k}_{{\rm peel}}\cdot\mathbf{m}/\sin\theta_{{\rm peel}}\nonumber \\
\sin\phi_{{\rm peel}} & =\mathbf{k}_{{\rm peel}}\cdot\mathbf{n}/\sin\theta_{{\rm peel}}.\label{eq:41}
\end{align}
 The basis vector $\mathbf{n}_{{\rm peel}}$ can then be obtained
by Equation (\ref{eq:38}). The Stokes vector of the peeled-off photon
is given by $\mathbf{S}'=\mathbf{M}(\theta_{{\rm peel}})\mathbf{L}(\phi_{{\rm peel}})\mathbf{S}$.

\begin{figure}[t]
\begin{centering}
\includegraphics[clip,scale=0.62]{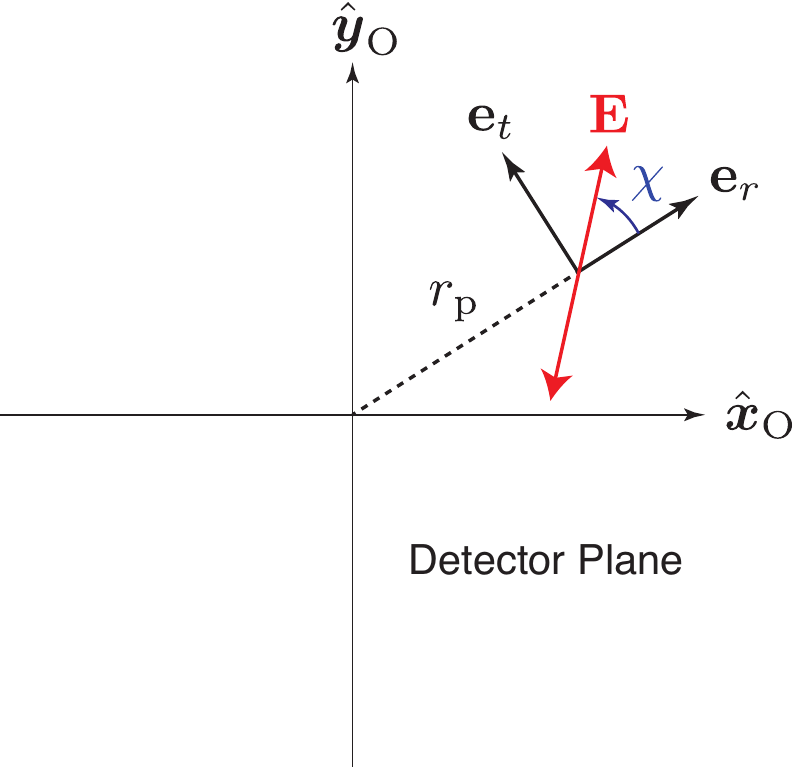}
\par\end{centering}
\begin{centering}
\medskip{}
\par\end{centering}
\caption{\label{fig05}A local coordinate system defined by the radial ($\mathbf{e}_{r}$)
and tangential ($\mathbf{e}_{t}$) vectors at a projected radius $r_{{\rm p}}$
in the detector plane. A linearly-polarized electric field vector
($\mathbf{E}$) and its polarization angle ($\chi$) in the local
coordinate system are also shown.}
\medskip{}
\end{figure}

In the above, the detector orientation was not yet taken into account.
In other words, $\mathbf{m}_{{\rm peel}}$ and $\mathbf{n}_{{\rm peel}}$
do not coincide with the axes $\hat{\mathbf{x}}_{{\rm O}}$ and $\hat{\mathbf{y}}_{{\rm O}}$
of the detector plane (see Figures \ref{fig04} and \ref{fig05}).
We, therefore, need to rotate $\mathbf{m}_{{\rm peel}}$ and $\mathbf{n}_{{\rm peel}}$
by an angle $\phi_{{\rm det}}$ between $\mathbf{m}_{{\rm peel}}$
and $\hat{\mathbf{y}}_{{\rm O}}$ (or between $\mathbf{n}_{{\rm peel}}$
and $-\hat{\mathbf{x}}_{{\rm O}}$). Here, recall that the IAU standard
measures a polarization angle counter-clockwise from the North in
the sky ($\hat{\mathbf{y}}_{{\rm O}}$ in the observer frame) when
looking at the photon source; hence, $\phi_{{\rm det}}$ is defined
as an angle between $\mathbf{m}_{{\rm peel}}$ and $\hat{\mathbf{y}}_{{\rm O}}$.
By this definition, the desired rotation angle $\phi_{{\rm det}}$
is uniquely determined by
\begin{align}
\cos\phi_{{\rm det}} & =-\mathbf{n}_{{\rm peel}}\cdot\hat{\mathbf{x}}_{{\rm O}}\nonumber \\
\sin\phi_{{\rm det}} & =\mathbf{n}_{{\rm peel}}\cdot\hat{\mathbf{y}}_{{\rm O}}.\label{eq:42}
\end{align}
It should also be noted that $\mathbf{n}_{{\rm peel}}$ in Equation
(\ref{eq:38}) was calculated in the galaxy frame. Therefore, we need
to represent the components of $\mathbf{n}_{{\rm peel}}$ in the observer
frame by multiplying the rotation matrix $\mathbf{R}$, defined in
Equation (\ref{eq:37}). The final Stokes vector, represented in the
detector plane, is then given by
\begin{equation}
\mathbf{S}''=\mathbf{L}(\phi_{{\rm det}})\mathbf{S}'=\mathbf{L}(\phi_{{\rm det}})\mathbf{M}(\theta_{{\rm peel}})\mathbf{L}(\phi_{{\rm peel}})\mathbf{S},\label{eq:43}
\end{equation}
which follows the IAU standard.

\subsubsection{Frequency of peeled-off photon}

To perform raytracing in the ``peeling-off'' procedure of scattered
light, the optical depth along the path to the observer should be
calculated at the frequency in the comoving fluid frame. On the other
hand, the spectral binning on the detector plane should be performed
using the frequency measured in the fixed (observer) frame. Therefore,
we need to consider two frequencies, the frequency $x_{{\rm tau}}$
(for optical depth) in the fluid frame and the frequency $x_{{\rm spec}}$
(for spectral binning) in the observer frame. If the photon is scattered
by a hydrogen atom with a velocity $\mathbf{u}_{{\rm atom}}$, they
are given by
\begin{align}
x_{{\rm tau}} & =x_{i}-u_{{\rm atom},\parallel}+\mathbf{k}_{{\rm peel}}\cdot\mathbf{u}_{{\rm atom}}\nonumber \\
x_{{\rm spec}} & =x_{{\rm tau}}+\mathbf{k}_{{\rm peel}}\cdot\mathbf{u}_{{\rm fluid}}.\label{eq:44}
\end{align}
Here, $x_{i}$ is the initial frequency (before the scattering event)
expressed in the fluid frame, $u_{{\rm atom},\parallel}=\mathbf{k}_{i}\cdot\mathbf{u}_{{\rm atom}}$
the velocity component parallel to the initial direction $\mathbf{k}_{i}$,
$x_{i}-u_{{\rm atom},\parallel}$ the frequency in the rest frame
of the scattering atom, $\mathbf{k}_{{\rm peel}}$ the direction vector
pointing to the observer, and $\mathbf{u}_{{\rm fluid}}$ the fluid
velocity. In the case of scattering by a dust grain, $x_{{\rm tau}}$
and $x_{{\rm spec}}$ are given by setting $\mathbf{u}_{{\rm atom}}=0$
in the above equation because dust grains are heavy and assumed to
have no thermal motion.

For the peeling-off of direct light, which escapes the system without
undergoing any scattering, we consider the two cases (injection from
the comoving fluid frame and a non-comoving fixed frame) separately.
In most cases, photons are injected in the fluid frame. LaRT also
has another option to inject photons in the fixed frame rather than
in the comoving frame. The former option is useful to simulate emissions
from the gas. The latter is suitable for the emission originating
from an object, such as stars/galaxies, that independently moves from
the ISM/IGM. For the case injected in the fluid frame, $x_{{\rm tau}}$
and $x_{{\rm spec}}$ are given by
\begin{align}
x_{{\rm tau}} & =x_{i}^{{\rm fluid}}\nonumber \\
x_{{\rm spec}} & =x_{i}^{{\rm fluid}}+\mathbf{k}_{{\rm peel}}\cdot\mathbf{u}_{{\rm fluid}},\label{eq:45}
\end{align}
where $x_{i}^{{\rm fluid}}$ is the initial photon frequency in the
fluid frame. On the other hand, when a photon is injected in the fixed
frame, they are given by
\begin{align}
x_{{\rm tau}} & =x_{i}^{{\rm fixed}}-\mathbf{k}_{{\rm peel}}\cdot\mathbf{u}_{{\rm fluid}}\nonumber \\
x_{{\rm spec}} & =x_{i}^{{\rm fixed}},\label{eq:46}
\end{align}
where $x_{i}^{{\rm fixed}}$ is the initial frequency in the fixed
frame. In this case, the comoving frequency of a photon injected in
the fixed frame is $x_{i}^{{\rm fluid}}=x_{i}^{{\rm fixed}}-\mathbf{k}_{i}\cdot\mathbf{u}_{{\rm fluid}}$
for the initial propagation vector $\mathbf{k}_{i}$.

\subsubsection{Peeled-off fraction}

The fraction to be peeled-off toward the detector is
\begin{equation}
I_{{\rm peel}}=\frac{\Phi(\theta_{{\rm peel}},\phi_{{\rm peel}})}{d^{2}}e^{-\tau}.\label{eq:47}
\end{equation}
Here, $\tau$ is the optical depth at the photon's current frequency,
measured over the path from the scattering (or injection) position
to the observer, and $d$ the luminosity distance. The phase function
$\Phi$ is normalized to one over the solid angle of $\int d\Omega=4\pi$.
This paper assumes isotropically-emitting Ly$\alpha$ sources. In
this case, the phase function for direct light, which undergoes no
interaction with the medium, is 
\begin{equation}
\Phi^{{\rm direc}}=\frac{1}{4\pi}.\label{eq:48}
\end{equation}
The phase function for scattered light is given by Equation (\ref{eq:26}),
as follows:
\begin{align}
\Phi^{{\rm scatt}}(\theta_{{\rm peel}},\phi_{{\rm peel}}) & =\mathcal{P}(\theta_{{\rm peel}},\phi_{{\rm peel}}).\label{eq:49}
\end{align}
The peeled-off Stokes parameters should also be multiplied by the
same factor as for the peeled-off intensity.

\begin{figure}[t]
\begin{centering}
\includegraphics[clip,scale=0.65]{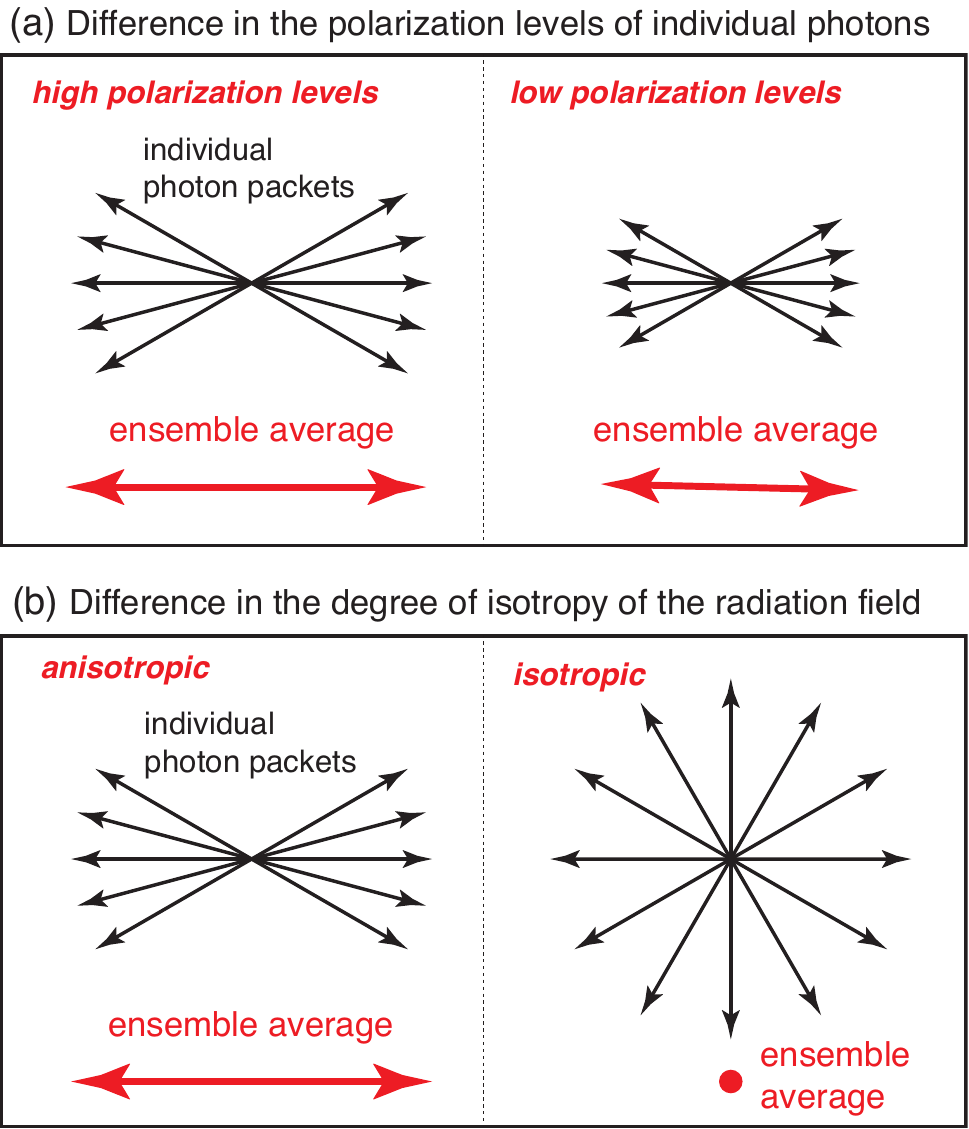}
\par\end{centering}
\begin{centering}
\medskip{}
\par\end{centering}
\caption{\label{fig06}Schematic representation of the ensemble-averaged polarization
as a superposition of the polarization vectors of individual photon
packets. (a) Effect of the difference in the polarization levels of
individual photon packets on the ``emsemble-averaged'' polarization.
The individual polarization amplitudes in the left panel are larger
than those in the right panel. However, the radiation fields in both
panels have the same degree of anisotropy. (b) Polarization signals
produced by an anisotropic or isotropic radiation field. The individual
polarization amplitudes are the same in both panels. The radiation
field is highly anisotropic in the left panel, whereas it is isotropic
in the right panel. The black arrows indicate the polarization vectors
of individual photon packets and the red arrows the ``ensemble-averaged''
polarization. The length of polarization vectors denotes the degree
of polarization, but not in exact scale. The red dot in the lower
right panel indicates a zero polarization.}
\medskip{}
\end{figure}

\begin{figure}[t]
\begin{centering}
\includegraphics[clip,scale=0.545]{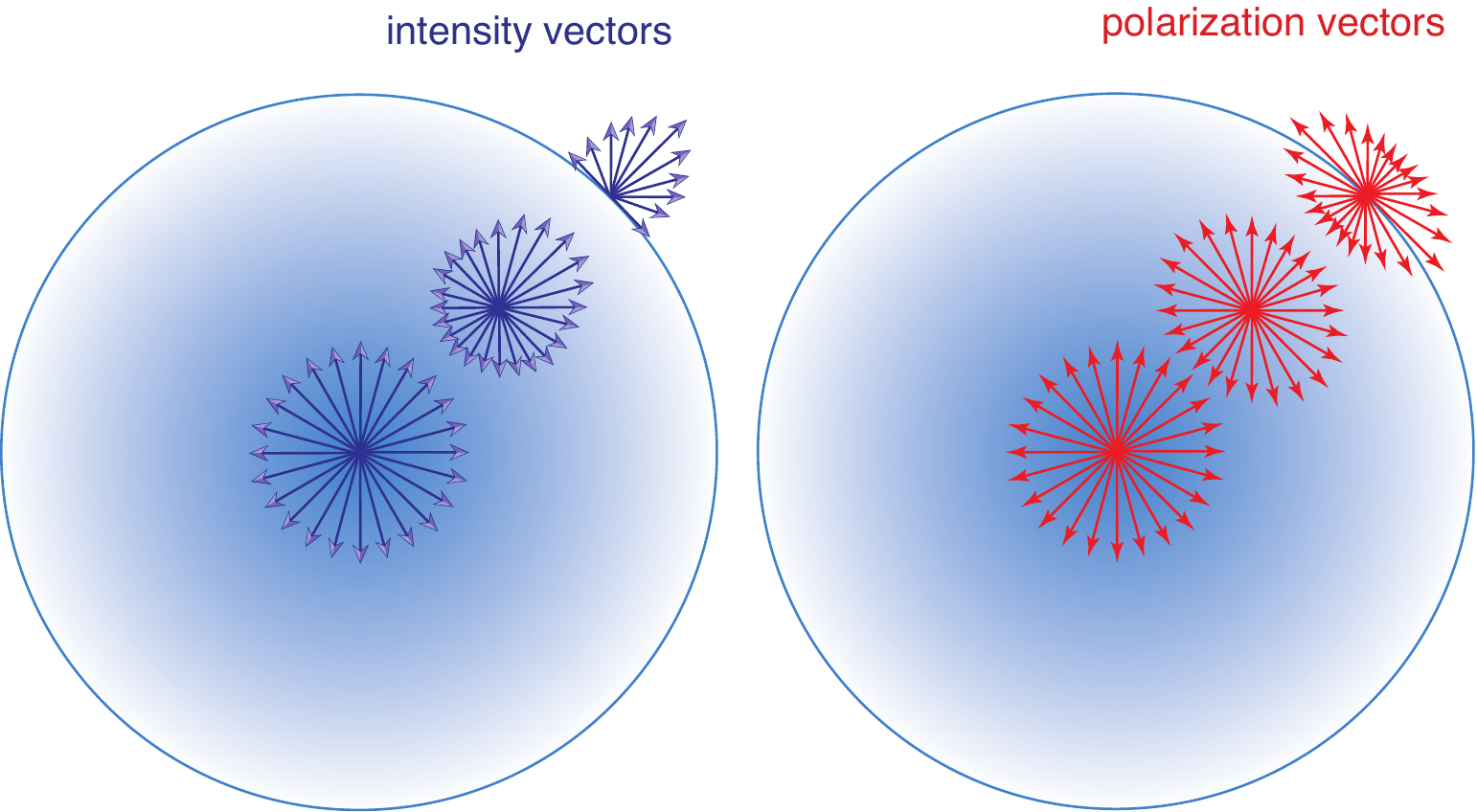}
\par\end{centering}
\begin{centering}
\medskip{}
\par\end{centering}
\caption{\label{fig07}Rise of the radiation field anisotropy and the polarization
with increasing radius. The left panel illustrates the rays passing
through three different locations in the projected plane when viewed
normal to the paper. Each blue arrow denotes a ray of photon packets
(intensity) propagating in the direction represented by the arrow.
The size of the blue arrows represents the intensity or the number
of photons passing through a unit area into a unit solid angle of
the given ray. The flux passing through the surface normal to the
radial direction increases with radius because of the increase in
anisotropy. In the right panel, the red arrows at each location denotes
the polarization vector carried by the scattered light of the rays
passing through that location toward the observer. In the projected
plane, the polarization vector of the scattered light of a ray is
perpendicular to the direction of the intensity ray, unless the scattering
occurs in the core, and its length is proportional to the polarized
intensity of the ray. The degree of polarization is given by $P_{{\rm L}}^{*}=Q=\left\langle E_{r}^{2}\right\rangle -\left\langle E_{t}^{2}\right\rangle $,
where $E_{r}$ and $E_{t}$ are the radially and tangentially oscillating
electric vectors, respectively. Therefore, the ensemble-averaged polarization
is dominated by the contribution of the outward rays and in general
increases with radius. However, note that the core scattering is cable
of producing a signifincantly different radial profile from that shown
in this figure, as is discussed in Sections \ref{section:4.2} and
\ref{section:4.4}.}
\medskip{}
\end{figure}

\section{Two Factors That Determine Polarization Profile}

\label{section:3}

Before describing the details, we here first summarize the essentials
to understand the simulation results. We explore the degree of polarization
calculated in two different ways. One is the quantities averaged over
an ``ensemble'' of photon packets, which experienced many ``different''
scattering histories, measured at a location in the detector plane.
In this case, the Stokes parameters are first averaged, as defined
in Equation (\ref{eq:11}), and then the degree of polarization is
calculated using the averaged Stokes parameters. \citet{2008MNRAS.386..492D}
refer to these quantities as the ``angle-averaged'' ones. However,
we prefer to refer to them as the ``ensemble-averaged'' ones. The
other is the polarization quantities of ``individual'' photon packets,
which represent the quasi-monochromatic waves undergone the ``same''
scattering history. When calculating the mean of the polarization
amplitudes for individual photons, we take the average ignoring the
orientation of the individual polarization vectors. The ensemble-averaged
quantities are observable through an experiment, while those of individual
photon packets are not. Dealing with individual photon packets is
theoretically advantageous to interpreting simulation results. We
distinguish the two concepts if necessary to clarify. However, the
term ``ensemble-averaged'' is often omitted when it is evident in
context. We also use ``photons'' interchangeably with ``photon packets.''

\begin{figure*}[t]
\begin{centering}
\includegraphics[clip,scale=0.66]{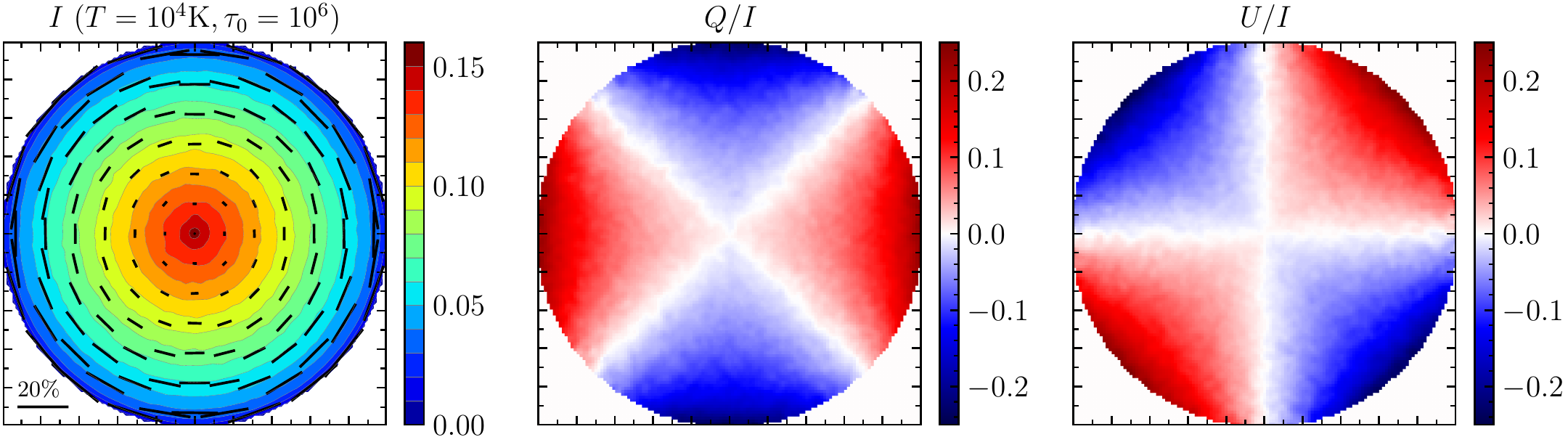}
\par\end{centering}
\begin{centering}
\medskip{}
\par\end{centering}
\caption{\label{fig08}Maps of Stokes parameters $I$, $Q$, and $U$ for a
static sphere model of $T=10^{4}$ K and $\tau_{0}=10^{6}$ ($N_{{\rm HI}}=1.7\times10^{19}$
cm$^{-2}$). The polarization vectors are also overlaid on the intensity
map in the left panel. The horizontal bar in the left panel denotes
a polarization level of 20\%. The parameters are all represented in
linear scale.}

\centering{}\medskip{}
\end{figure*}

\begin{figure*}[t]
\begin{centering}
\includegraphics[clip,scale=0.53]{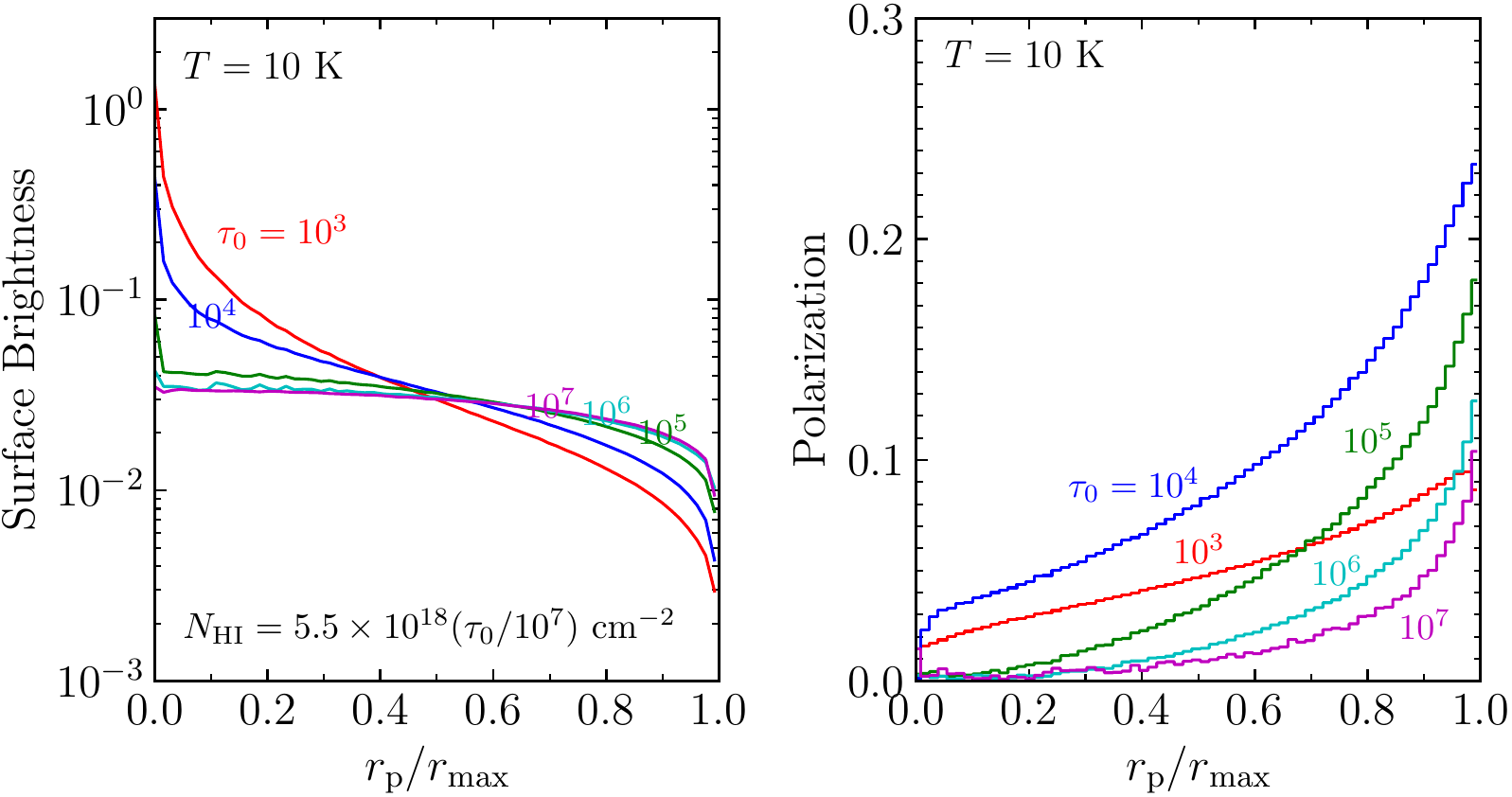}\ \ \ \includegraphics[clip,scale=0.53]{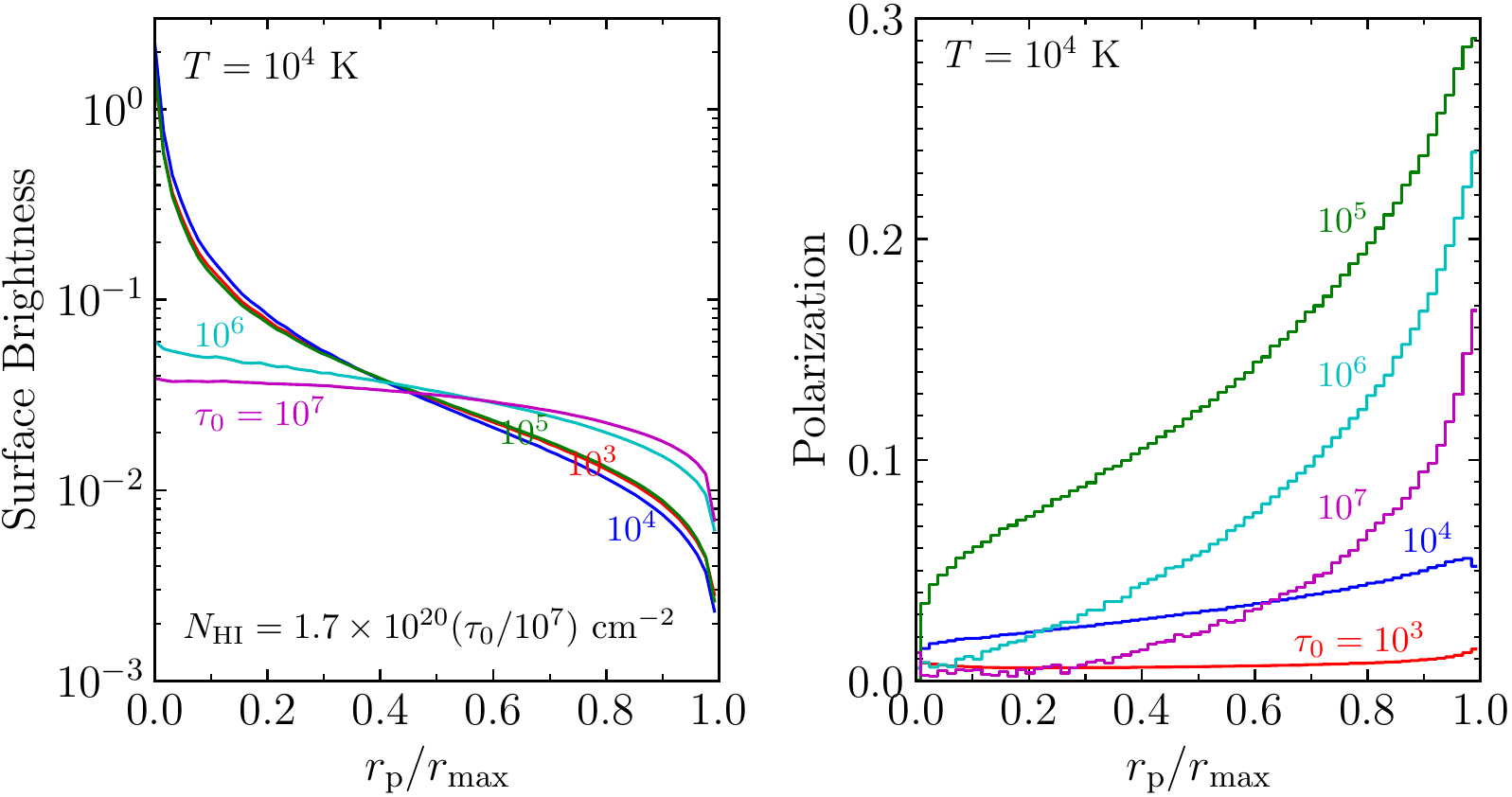}
\par\end{centering}
\begin{centering}
\medskip{}
\par\end{centering}
\caption{\label{fig09}Radial profiles of the surface brightness and polarization
for a static sphere model of $T=10$ K (the first and second panels)
and $T=10^{4}$ K (the third and fourth panels). The central optical
depth varies from $\tau_{0}=10^{3}$ to $10^{7}$. The column density
$N_{{\rm HI}}$ of each model is given by the equations shown in the
first and third panels.}
\medskip{}
\end{figure*}

The ensemble-averaged degree of polarization is primarily determined
by two factors: the polarization amplitudes of individual photon packets
and the degree of anisotropy in the Ly$\alpha$ radiation field. Figure
\ref{fig06}(a) compares two cases in which the polarization degree
of individual photons differs while the degree of anisotropy of the
radiation field is the same. The polarization amplitudes, denoted
by the length of black arrows, in the left panel are bigger than those
in the right panel. The length of the red arrows in each panel represents
the ensemble average of the polarization vectors of individual photons.
As illustrated in the figure, the stronger the polarization amplitudes
of individual photons are, the higher their ensemble average (denoted
by red arrows) will result. Figure \ref{fig06}(b) illustrates the
importance of anisotropy of the radiation field in polarization. In
the figure, the lengths of arrows are the same in both panels, while
the diversity of direction angle indicates the degree of anisotropy
(or isotropy) of the radiation field. Suppose the radiation field
is highly isotropic (as in the right panel). In that case, two perpendicular
components of the polarization vectors of individual photon packets
will significantly be canceled out, resulting in a weak polarization.
On the other hand, in a highly anisotropic radiation field (left panel),
they will be less canceled out, yielding a strong polarization.

The polarization levels of individual photons generally tend to increase
with the number of scatterings and the strength of the medium's velocity
field, as detailed in Sections \ref{section:4} and \ref{section:5.1}.
The degree of the radiation field isotropy rises with an increasing
number of scatterings but decreases in fast-moving media. In a spherical
symmetric medium with a central source, it is found that, as the optical
depth of the medium (and the number of scatterings) increases, the
polarization amplitudes of individual photon packets begin to grow
first while the radiation field is kept still anisotropic. Then, later,
the radiation field appears to become isotropic after the individual
photons gain high enough polarization amplitudes.

It is also found that the radial polarization profile increases monotonically
with increasing radius, except for some cases in Section \ref{section:4.4}.
The increase of polarization degree with increasing radius is attributable
to the rise of anisotropy in the radiation field. As illustrated in
Figure \ref{fig07}, the radiation field near the center of a sphere
will be isotropic by symmetry. On the other hand, at large radii,
the outward radiation flux will dominate over the inward one, leading
to anisotropy in the radiation field. Eventually, this anisotropy
in radiation at large radii results in an increase in polarization
with radius.

In the next section, we also show that the steepness (or slope) of
the surface brightness profile correlates well with the steepness
of the polarization profile. As optical depth increases, the number
of scatterings per photon increases, and the radiation field becomes
more isotropic due to the spatial diffusion of photons. Therefore,
the surface brightness will tend to flatten with the increase of optical
depth. In other words, a highly isotropic radiation field eventually
leads to a flatter (or shallower) surface brightness profile. On the
other hand, an anisotropic radiation field would produce a steep surface
brightness profile. Therefore, we can immediately associate the steepness
of the surface brightness and polarization profile with the degree
of anisotropy in the radiation field; a steep surface brightness profile
tends to accompany a rapid increase of polarization.

\section{Results}

\label{section:4}

This section presents the results of four types of models in total.
We first present two basic models as the first and second examples.
One is a static spherical medium with various optical depths, adopted
to examine the ``optical depth effect'' on Ly$\alpha$ polarization.
The other is a Hubble-like sphere expanding at many different velocities
to illustrate the ``kinematic or velocity effect.'' Next, we reproduce
the results for the thin shell model of \citet{2008MNRAS.386..492D}.
This model not only verifies our code but also illustrates a special
case where single-scattering is predominant. Lastly, we predict the
polarization signal from the galactic halo model of \citet{2020ApJ...901...41S},
which was utilized to fit the observed spectra and surface brightness
profiles of LAEs. This fourth example demonstrates the diversity of
the polarization pattern.

In the examples, we deal with a spherically symmetric geometry. Therefore,
it is convenient to represent the Stokes parameters in a local coordinate
system defined by the radial and tangential vectors ($\mathbf{e}_{r}$
and $\mathbf{e}_{t}$, respectively) at the photon's location in the
detector plane, as shown in Figure \ref{fig05}. We use this coordinate
system to calculate the radial profile of polarization. The polarization
angle $\chi$ is also measured in this coordinate system. For this
purpose, the basis vectors $\hat{\boldsymbol{x}}_{{\rm O}}$ and $\hat{\boldsymbol{y}}_{{\rm O}}$
of the detector plane are rotated to make them coincide with the radial
and tangential vectors, respectively. In the local coordinate system,
$Q=-1$ denotes a polarization vector perpendicular to the radial
direction, corresponding to a concentric polarization pattern. On
the other hand, $Q=+1$ represents a polarization vector parallel
to the radial direction. Here, it should be noted that the $Q=+1$
shown in this section is caused by photons traveling perpendicular
to the radial direction, not by the negative polarization ($E_{1}<0$)
discussed in Section \ref{subsec:2.5}; in other words, the $Q=+1$
case occurs due to the randomness of photon directions caused by multiple
scattering.

This study presents the results for the media of $T=10$ or $10^{4}$
K. In the first three examples, photons are injected from the center
of the sphere. In the last example, the spatial distribution of input
Ly$\alpha$ photons is assumed to be described by an exponential function
of projected radius. The input line profile is assumed to be a Voigt
function with a line width corresponding to the medium's temperature,
except for the outflowing thin shell model of Section \ref{section:4.3}
(see \citealp{2020ApJS..250....9S} for the random number generation
algorithm of the Voigt profile). The width of the input line profile
does not significantly affect the present results unless it is as
wide as those observed in active galactic nuclei or the output spectra
shown in this paper. The dust effect on the Ly$\alpha$ polarization
is discussed in the first and third examples (the static, homogeneous
sphere and the outflowing thin shell models). The fourth example (the
galactic halo model of \citealt{2020ApJ...901...41S}) also contains
dust; however, no discussion on the dust effect is given for this
example. The gas-to-dust ratio is assumed to be that of the MW. A
Cartesian grid is utilized in the present version of LaRT. The number
of cells in the simulations ranges from $200^{3}$ to 400$^{3}$.

\subsection{A Static, Homogeneous Sphere}

\label{section:4.1}

In this example, we investigate the optical depth effect on the polarization
property using a static, spherical medium. The optical depth at the
line center, measured from the center to the outer edge of the sphere,
varies from $\tau_{0}=10^{3}$ to $10^{7}$, and the gas temperature
is $T=10$ K or $10^{4}$ K. Figure \ref{fig08} shows typical images
of the Stokes parameters $I$, $Q$, and $U$, obtained for a model
of $\tau_{0}=10^{6}$ and $T=10^{4}$ K. The figure illustrates the
concentric polarization pattern that is generally expected in a configuration
with a central source.

Figure \ref{fig09} shows the surface brightness and the degree of
polarization as a function of radius. In the figure, we notice the
following. First, the surface brightness profile is relatively steep
in small optical depths ($\tau_{0}\lesssim10^{4}$ for $T=10$ K and
$\tau_{0}\lesssim10^{5}$ for $T=10^{4}$ K) but flattens quickly
as $\tau_{0}$ increases. Second, as commonly expected, the radial
profiles of polarization follow the general trend of increasing with
radius. Third, as $\tau_{0}$ increases to a certain critical value
($\tau_{0}\approx10^{4}$ for $T=10$ K and $10^{5}$ for $T=10^{4}$
K), the overall polarization level increases rapidly but decreases
as $\tau_{0}$ rises further above that value. Fourth, the radial
profile of polarization jumps at $r_{{\rm p}}\approx0$ suddenly from
zero to a non-zero value if $\tau_{0}$ is relatively small ($\tau_{0}\lesssim10^{4}$
for $T=10$ K and $\tau_{0}\lesssim10^{5}$ for $T=10^{4}$ K); but,
it gradually increases with radius if $\tau_{0}$ is large. In other
words, the polarization profile is steep near $r_{{\rm p}}\approx0$
for a small $\tau_{0}$, while it is shallow for a large $\tau_{0}$.
Lastly, both the surface brightness profile and the polarization profile
appear to begin to change in shape at the same critical optical depth
($\tau_{c}\approx10^{4}$ for $T=10$ K and $10^{5}$ for $T=10^{4}$
K), indicating that they are closely associated with each other. In
the cases where the optical depth is higher than this value, a steeper
surface brightness profile tends to be accompanied by a higher overall
level of polarization and a slightly steeper polarization profile.

\begin{figure*}[t]
\begin{centering}
\includegraphics[clip,scale=0.45]{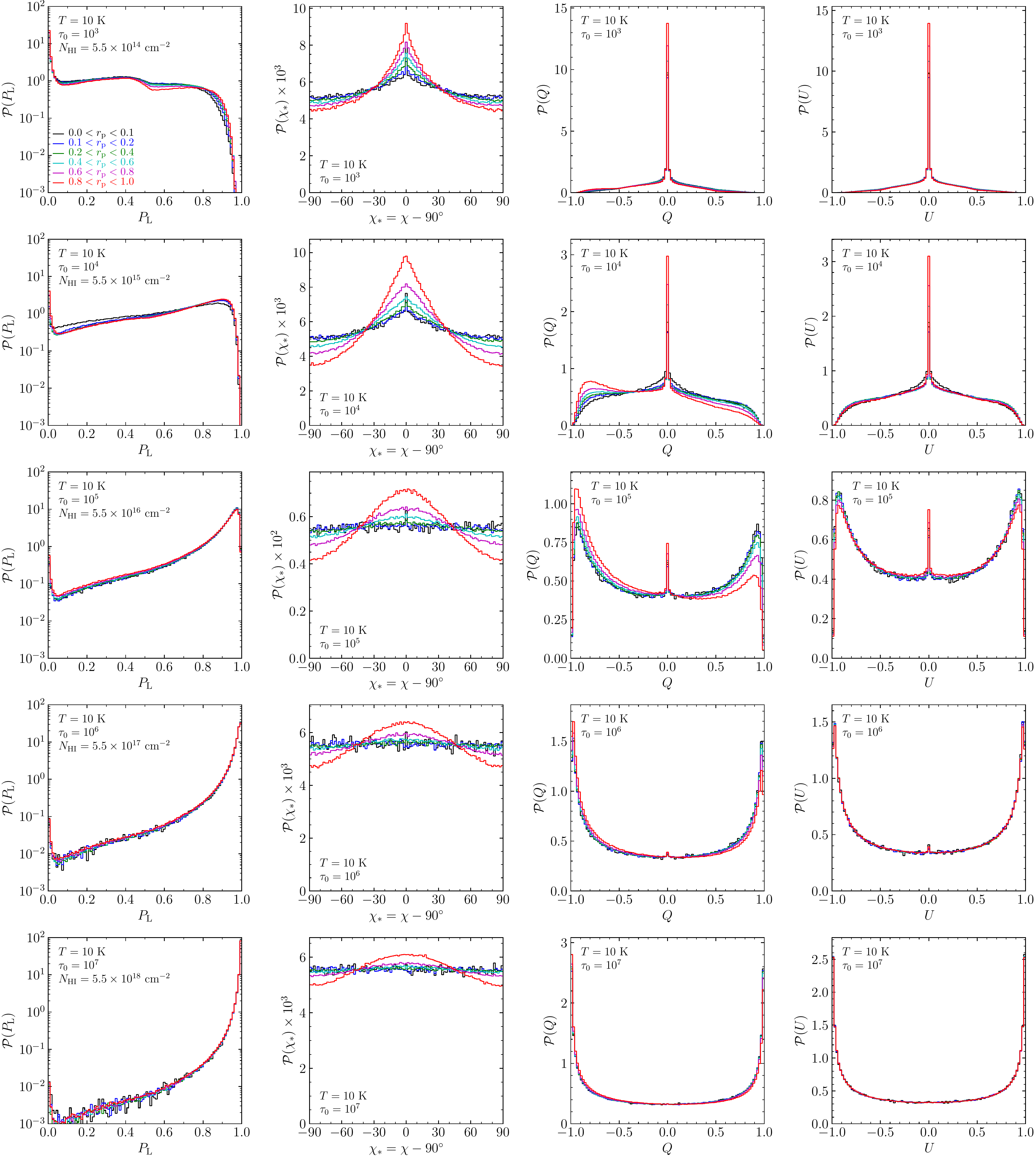}
\par\end{centering}
\begin{centering}
\medskip{}
\par\end{centering}
\caption{\label{fig10}Probability density functions of the degree of polarization
$P_{{\rm L}}$, polarization angle $\chi_{*}$, and the Stokes $Q$
and $U$ of individual photon packets for a static sphere of $T=10$
K. The model optical depth varies from $\tau_{0}=10^{4}$ (the first
row) to $10^{7}$ (the fifth row). The figure shows the distribution
functions for several radial bins ($0<r_{{\rm p}}<0.1$, $0.1<r_{{\rm p}}<0.3$,
$0.3<r_{{\rm p}}<0.5$, $0.5<r_{{\rm p}}<0.7$, $0.7<r_{{\rm p}}<0.9$,
and $0.9<r_{{\rm p}}<1$), as denoted in the upper left panel. We
use $\chi_{*}=\chi-\pi/2$, measured counterclockwise from the tangential
direction, for polarization angle, instead of $\chi$, to make its
histogram peak at $\chi_{*}=0$. In this definition, $\chi_{*}=0$
represents a tangential or concentric polarization pattern, and $\chi_{*}=\pm\pi/2$
a radial polarization pattern.}

\centering{}\medskip{}
\end{figure*}

\begin{figure*}[!t]
\begin{centering}
\includegraphics[clip,scale=0.45]{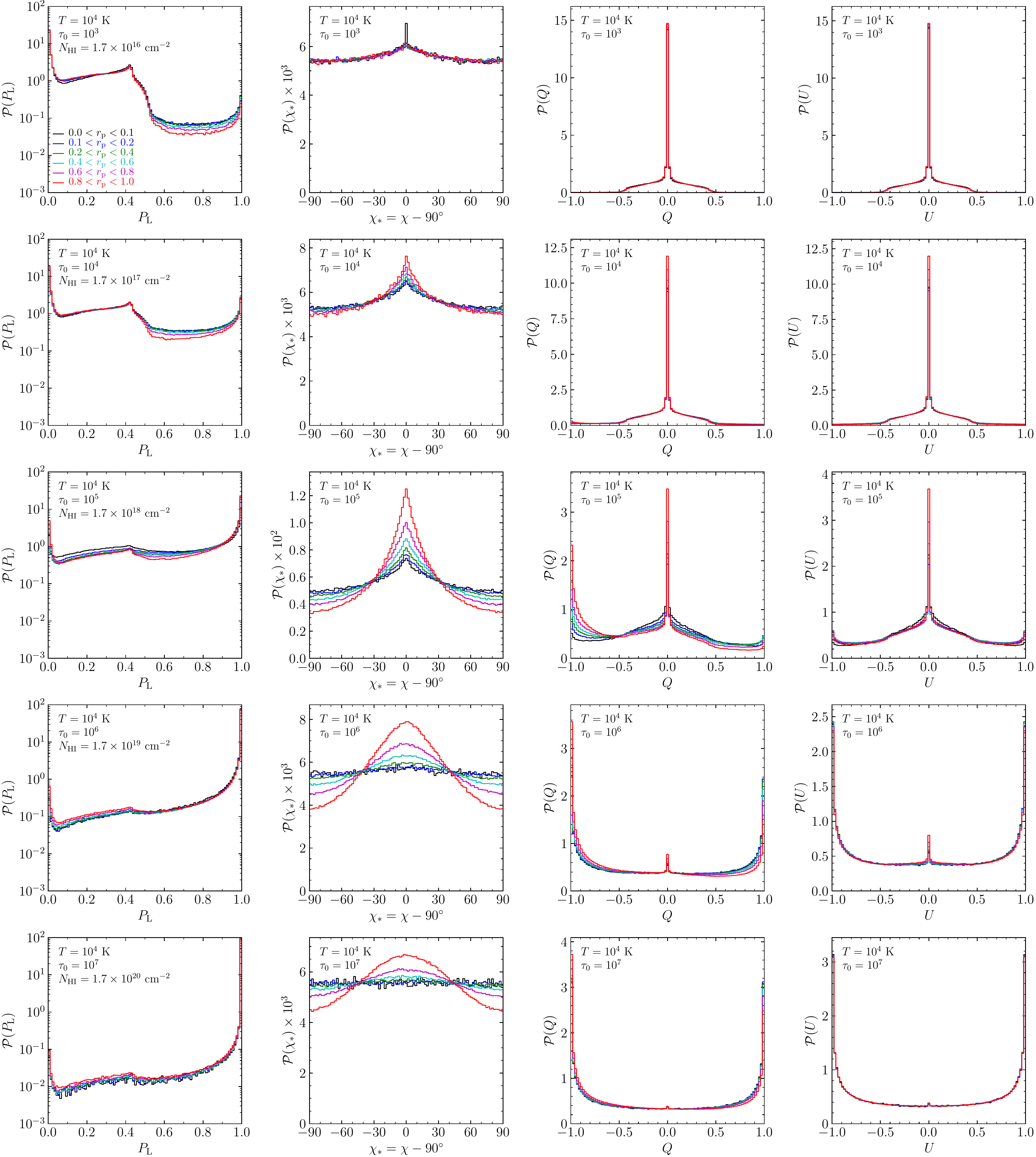}
\par\end{centering}
\begin{centering}
\medskip{}
\par\end{centering}
\caption{\label{fig11}Probability density functions of $P_{{\rm L}}$, $\chi_{*}$,
and the Stokes $Q$ and $U$ of individual photon packets for a static
sphere of $T=10^{4}$ K. The model optical depth varies from $\tau_{0}=10^{4}$
(the first row) to $10^{7}$ (the fifth row).}

\centering{}\medskip{}
\end{figure*}

\begin{figure*}[t]
\begin{centering}
\includegraphics[clip,scale=0.52]{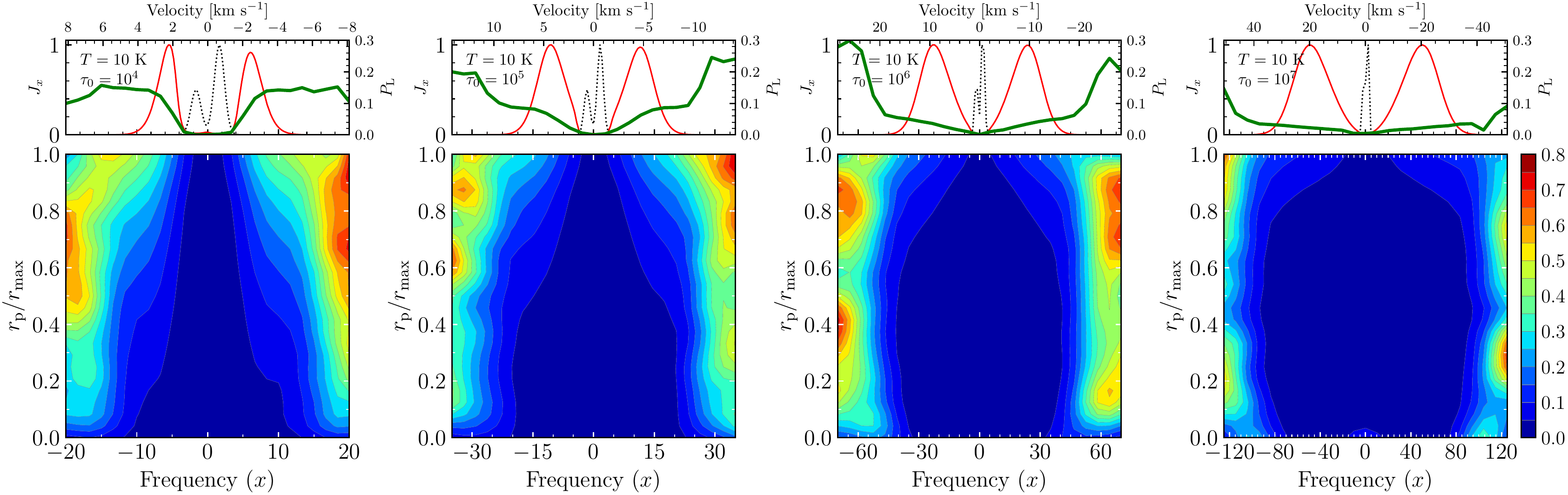}
\par\end{centering}
\begin{centering}
\medskip{}
\par\end{centering}
\begin{centering}
\includegraphics[clip,scale=0.52]{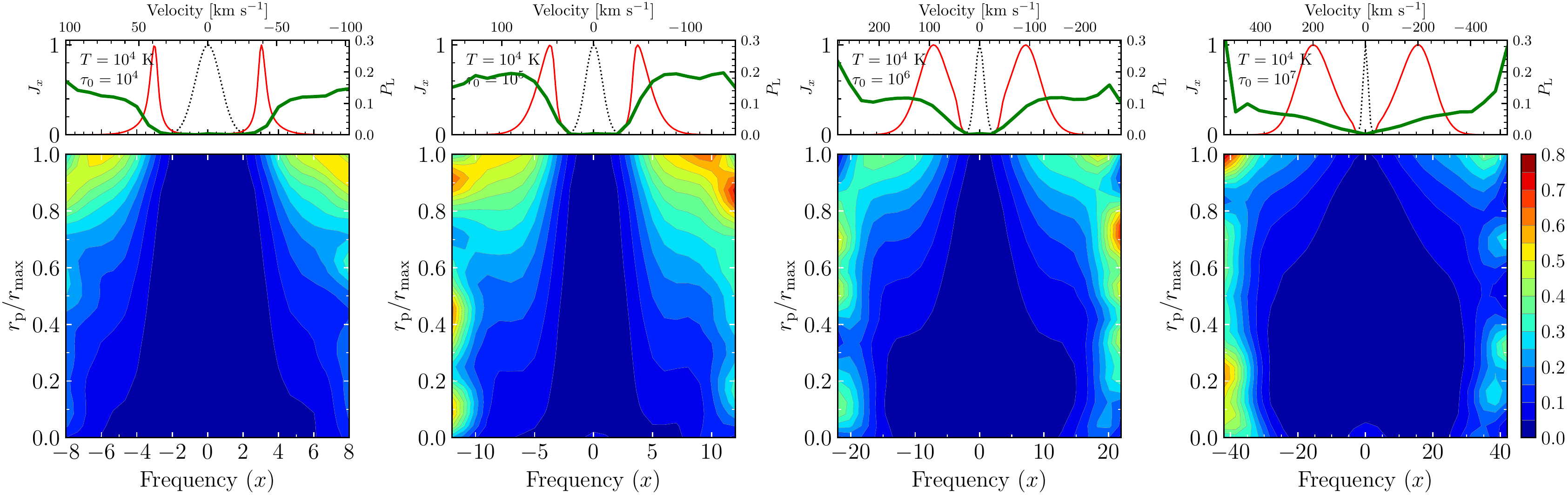}
\par\end{centering}
\begin{centering}
\medskip{}
\par\end{centering}
\caption{\label{fig12}Polarization degree for a static homogeneous spherical
medium. The first and second rows show the results for $T=10$ K and
$T=10^{4}$ K, respectively. The optical depth ranges from $\tau_{0}=10^{4}$
to $10^{7}$. In the small ancillary panels, the black dotted, red
solid curves are the initial and emergent spectra, respectively. The
green curves are polarization degrees as a function of frequency.
The main panels show the linear polarization degree in a two-dimensional
space of photon frequency and projected radius. Here, the polarization
degree is the \textquotedblleft ensemble-averaged\textquotedblright{}
one for photons detected at radius $r_{{\rm p}}$ with frequency $x$
in the detector plane. The color bar denotes the polarization levels.}

\centering{}\medskip{}
\end{figure*}

\begin{figure}[t]
\begin{centering}
\includegraphics[clip,scale=0.53]{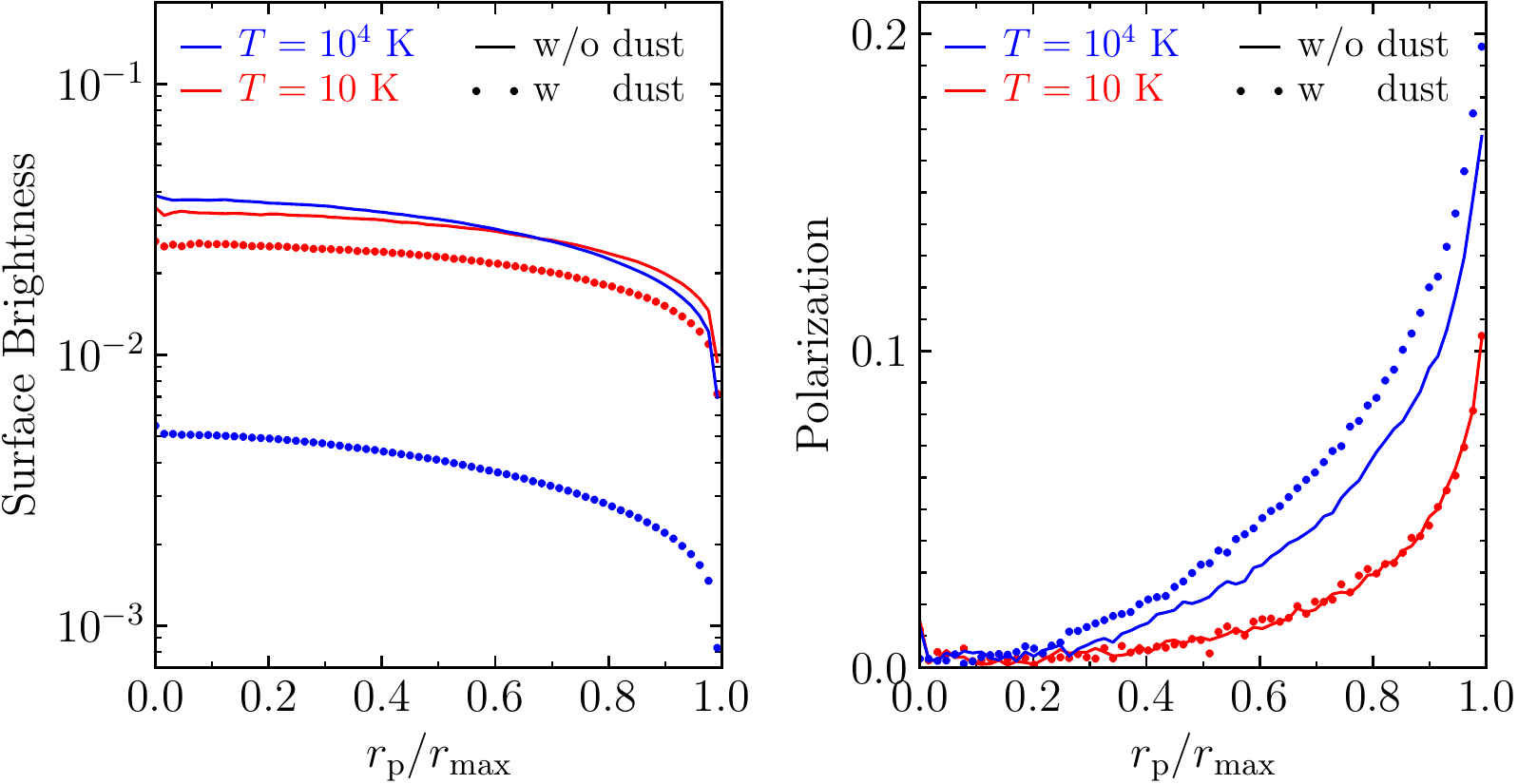}
\par\end{centering}
\begin{centering}
\medskip{}
\par\end{centering}
\caption{\label{fig13}Dust effect on the radial profiles of the surface brightness
and polarization for a static spherical medium of the \ion{H}{1}
optical depth $\tau_{0}=10^{7}$ ($N_{{\rm HI}}=5.5\times10^{18}$
and $1.7\times10^{20}$ cm$^{-2}$ for $T=10$ and $10^{4}$ K, respectively).
The optical depths due to dust are $\tau_{{\rm dust}}=8.8\times10^{-3}$
and $2.7\times10^{-1}$ for $T=10$ and $10^{4}$ K, respectively.
The blue and red colors denote the model with a gas temperature of
$T=10^{4}$ K and 10 K, respectively. The solid lines and dots represent
the model without and with dust, respectively.}

\centering{}\medskip{}
\end{figure}

The flattening of surface brightness with increasing $\tau_{0}$ (the
first result in Figure \ref{fig09}) is caused by the trend that,
as $\tau_{0}$ increases, the number of scatterings increases, and
the radiation field becomes more isotropic (see Section \ref{section:3}).
The increase of polarization degree with radius (the second result)
is due to the enhancement of anisotropy in the radiation field, as
illustrated in Figure \ref{fig07}. The third and fourth results in
Figure \ref{fig09} can be understood with the help of polarization
properties of individual photon packets. Figure \ref{fig10} shows
the probability density functions of the polarization degree $P_{{\rm L}}$,
the Stokes parameters $Q$ and $U$, and the polarization angle $\chi_{*}=\chi-\pi/2$
of individual photon packets, which were calculated at various radius
bins, for the model of $T=10$ K\footnote{Figures 10, 11, 15, 16, 19, 20, 24, and 25 describe the statistical
polarization properties of individual photon packets escaping the
system. In the figures, the Stokes parameters were calculated in the
local reference frame, defined by the last direction vector, at the
last scattering position. The peeling-off technique was not utilized
in the figures.}. To display the distribution of polarization angle conveniently,
we use $\chi_{*}$, measured counterclockwise from the tangential
vector defined in Figure \ref{fig05}; thus, $\chi_{*}=0$ denotes
the tangential direction. The results for the model of $T=10^{4}$
K are shown in Figure \ref{fig11}.

In this present study, the initial input photons are injected unpolarized.
Hence, the variation in the distribution of $P_{{\rm L}}$ with $\tau_{0}$
implies an interesting fact that individual photon packets are more
polarized as they are scattered more, which are further discussed
in Section \ref{section:5.1}. Figures \ref{fig10} and \ref{fig11}
show that, for relatively low optical depths ($\tau_{0}\lesssim10^{4}-10^{5}$),
the rise of polarization levels of individual photon packets with
increasing $\tau_{0}$ is responsible for the increase of ensemble-averaged
polarization. However, as $\tau_{0}$ increases further, the photon
packets are fully polarized, and no further increase of the individual
polarization levels occurs. Instead, the radiation field becomes more
isotropic with increasing $\tau_{0}$, leading to the decrease of
the ensemble-averaged polarization due to the cancellation of individual
polarization vectors. This polarization trend explains the third result
in Figure \ref{fig09} that, as $\tau_{0}$ increases, the overall
polarization level increases at first and then decreases later. In
the models with relatively low optical depths, a sudden breakdown
of the isotropy in the radiation field at the location right adjacent
to the center causes a sudden increase of the ensemble-averaged polarization
at that location. In contrast, in the models with relatively high
optical depths, the strong isotropy in the radiation field near the
center largely cancels the individual polarization vectors, even though
most individual photon packets bring $\sim$100\% polarization. Therefore,
the fourth result in Figure \ref{fig09} is also understood.

More detailed descriptions of the polarization properties of individual
photons are as follows. In the models with small optical depths ($\tau_{0}<10^{4}$
for $T=10$ K and $\tau_{0}<10^{5}$ for $T=10^{4}$ K), most photon
packets appear to have $P_{{\rm L}}<0.5$ (the first column in the
figures). However, as the optical depth (and the number of scatterings)
increases, most photon packets tend to be fully polarized ($P_{{\rm L}}=1$).
We also find a weak but noticeable bump shape with a peak at $P_{{\rm L}}\approx0.4$
in the distribution function of $P_{{\rm L}}$ for the model of $T=10^{4}$
K; this bump shape is associated with the pot lid shape in the distribution
function of $Q$ found in the range $\left|Q\right|\lesssim0.5$ (also
seen in $U$). It diminishes as $\tau_{0}$ increases or $T$ decreases;
it is not noticeable in the models of $T=10$ K, except for $\tau_{0}=10^{3}$,
as in Figure \ref{fig10}. In the second column of the figures, the
distribution function of polarization angle shows a peak at $\chi_{*}=0$
when $\tau_{0}$ is small, while it flattens as $\tau_{0}$ increases.
In particular, for $\tau_{0}=10^{7}$, the distribution function of
$\chi_{*}$ at $r\approx0$ is almost completely uniform. The results
imply that at small radii or as $\tau_{0}$ increases, the polarization
orientations tend to be completely randomized, and the radiation field
becomes isotropic. Therefore, even though photon packets were almost
fully polarized for a sufficiently large $\tau_{0}$, the polarization
vectors are largely canceled out due to the isotropy of the Ly$\alpha$
radiation field. The third column shows that the distribution function
of $Q$ is dominated by $Q=0$ when $\tau_{0}$ is small, but its
peak at $Q=0$ diminishes and consequently disappear, as $\tau_{0}$
increases. We also note that as $\tau_{0}$ increases, the peak at
$Q=-1$ (concentric polarization) grows first, and then the peak at
$Q=1$ (radial polarization) follows; eventually, the distribution
function of $Q$ becomes symmetric and has two sharp peaks at $Q=\pm1$.
This trend is because, as $\tau_{0}$ increases, the polarization
degree of individual photon packets increases first, and then the
radiation field becomes isotropic later. The peak at $Q=-1$ rises
as the polarization degree of individual photons increases (but while
the radiation field remains strongly anisotropic). As $\tau_{0}$
increases further, the peak at $Q=1$ also begins to grow as a result
of the enhanced isotropy of the radiation field. The distribution
of $U$ is also dominated by $U=0$ for small $\tau_{0}$ but by $\left|U\right|=1$
for large $\tau_{0}$. Unlike for $Q$, the distribution function
of $U$ is always symmetric in all models discussed in this paper
due to the spherical symmetry of the models and by definition in Equation
(\ref{eq:11}).

\begin{figure*}[t]
\begin{centering}
\includegraphics[clip,scale=0.53]{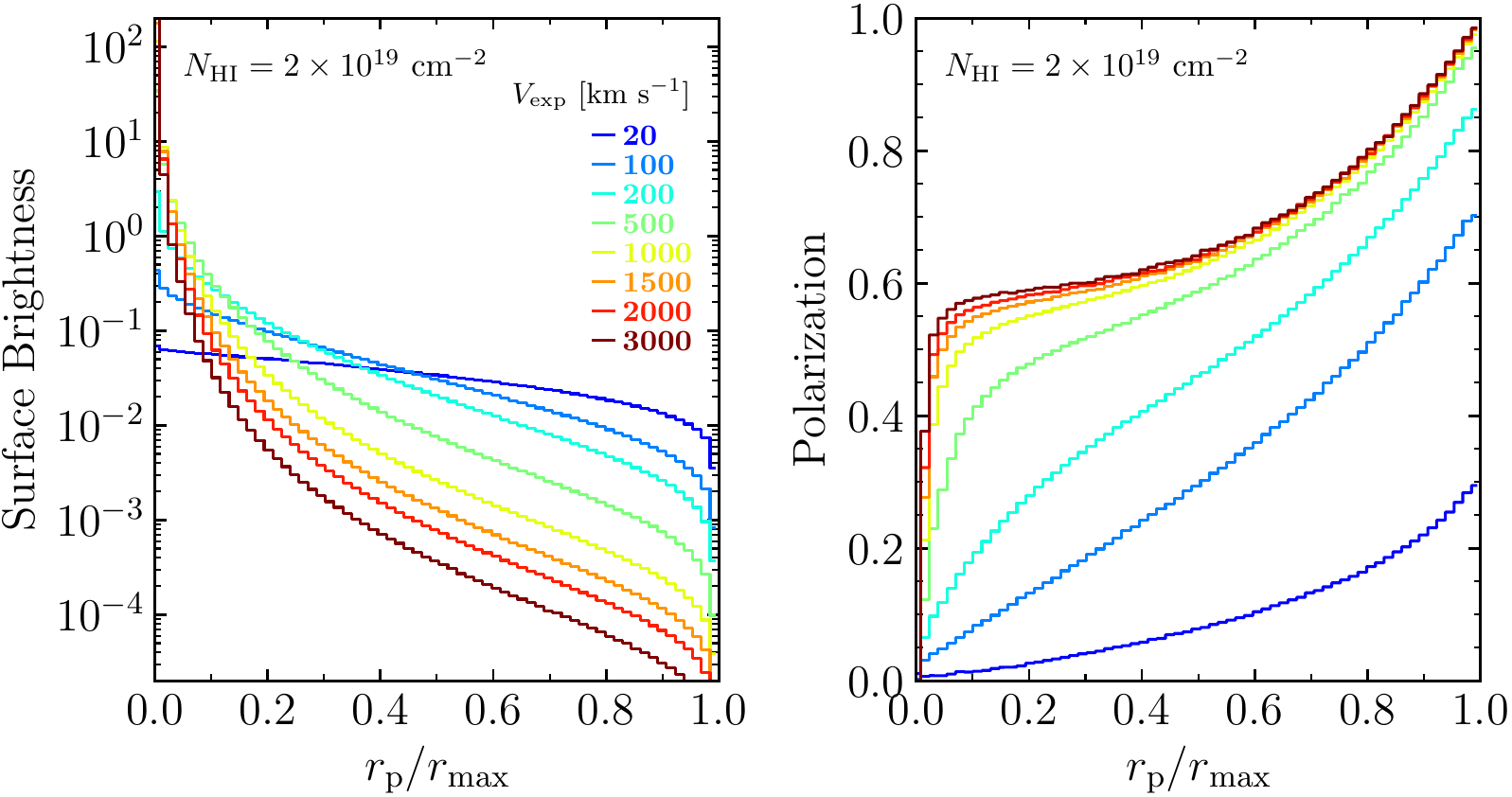}\ \ \ \includegraphics[clip,scale=0.53]{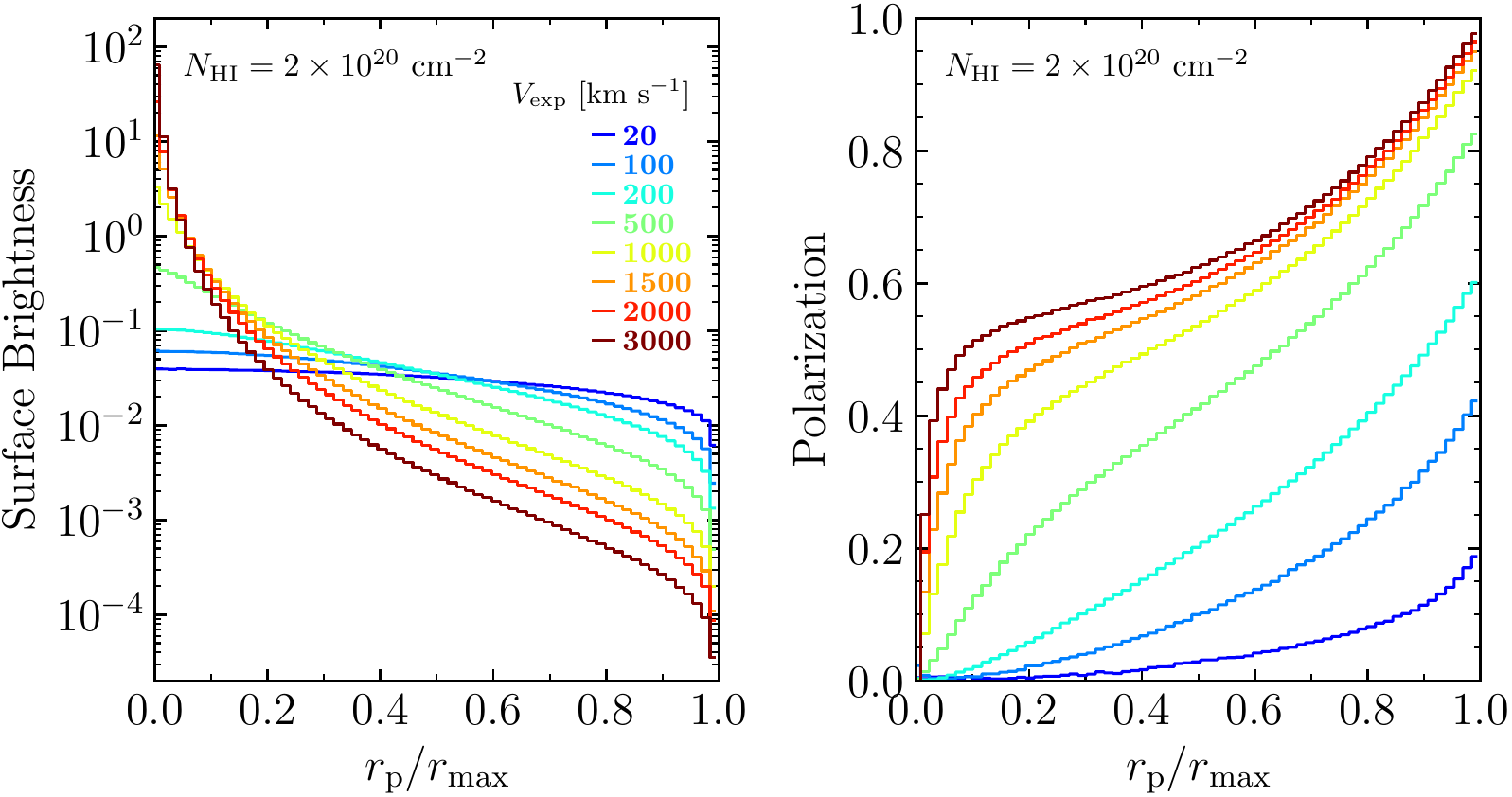}
\par\end{centering}
\begin{centering}
\medskip{}
\par\end{centering}
\caption{\label{fig14}Radial profiles of the surface brightness and polarization
for a Hubble-like expanding medium of $T=10^{4}$ K. The maximum expansion
velocity $V_{{\rm exp}}$ at the outer edge of the sphere ranges from
20 to 3000 km s$^{-1}$, as denoted in the figure. The column density
of the medium is $N_{{\rm HI}}=2\times10^{19}$ cm$^{-2}$ (the first
and second panels) or $2\times10^{20}$ cm$^{-2}$ (the third and
fourth panels).}

\centering{}\medskip{}
\end{figure*}

In summary, in the cases with a relatively low optical depth ($\tau_{0}\lesssim10^{4}-10^{5}$),
the polarization amplitudes of individual photons are the main factor
that governs the radial profile of the ensemble-averaged polarization.
On the other hand, if the optical depth is high ($\tau_{0}\gtrsim10^{4}-10^{5}$),
the degree of isotropy (or anisotropy) in the radiation field is the
prime factor that determines the polarization. It is also evident
that an isotropic radiation field will result in a shallow or flat
surface brightness profile, while an anisotropic radiation field would
produce a steep surface brightness profile. This association of the
surface brightness profile with the anisotropy of the radiation field
explains the last finding in Figure \ref{fig09}.

Figure \ref{fig12} presents the dependence of polarization on the
photon frequency and the projected radius, which may be useful to
compare with the spatially resolved spectropolarimetric observations.
The small ancillary panels show the frequency dependence of the polarization
degree averaged over the whole spatial area. The polarimetric spectra
are symmetric between the red and blue wavelengths because this model
is static. The strongest polarization is found in the line wings of
both red and blue sides. This is because in the wings, not only can
the photons easily escape, but also they tend to preserve the polarization
level of 100\% during scattering events, as discussed in Section \ref{section:5.1}.
On ther other hand, the core photons tend to be depolarized. In more
detail, the radial profile of polarization is frequency-dependent;
the farther the frequency is from the line center, the steeper the
radial polarization profile. This is because, at frequencies farther
from the line center, the Ly$\alpha$ photons tend to escape more
easily, making the radiation field more anisotropic.

\citet{2008MNRAS.386..492D} argued that the presence of dust will
boost the polarization level as a result of preferential extinction
of multiply-scattered photons. A photon scattered multiple times passes
a longer length in the medium, leading to a higher possibility of
absorption by dust. In other words, the presence of dust grains causes
less spatial diffusion of Ly$\alpha$ photons by destroying them,
particularly the multiple-scattered ones. We also note that the central
parts of the Ly$\alpha$ line are mainly absorbed by dust grains,
as shown in Figures 5 and D1 of \citet{2020ApJS..250....9S}. As a
result, the presence of dust leads to a slightly steeper surface brightness
profile and a more anisotropic Ly$\alpha$ radiation field. This effect
is expected to strengthen the polarization level eventually. The dust
effect will be more critical with higher $\tau_{0}$ or $N_{{\rm HI}}$.
We therefore examined the dust effect on the polarization and surface
brightness profiles by assuming the MW dust, as shown in Figure \ref{fig13}.
The polarization level for the model with a gas temperature of $T=10^{4}$
K ($\tau_{0}=10^{7}$) is found to increase slightly by the presence
of dust. On the other hand, no noticeable change is found in the case
of $T=10$ K. The figure also shows that the reduction of surface
brightness level by dust is more profound in the model with a high
temperature ($T=10^{4}$ K). This is because, as noted in \citet{2020ApJS..250....9S},
the ratio of dust extinction to the \ion{H}{1} scattering cross-section
is proportional to $T^{1/2}$ at the Ly\textgreek{a} line center.
Therefore, the effect is stronger when the gas temperature is higher.

\begin{figure*}[t]
\begin{centering}
\includegraphics[clip,scale=0.45]{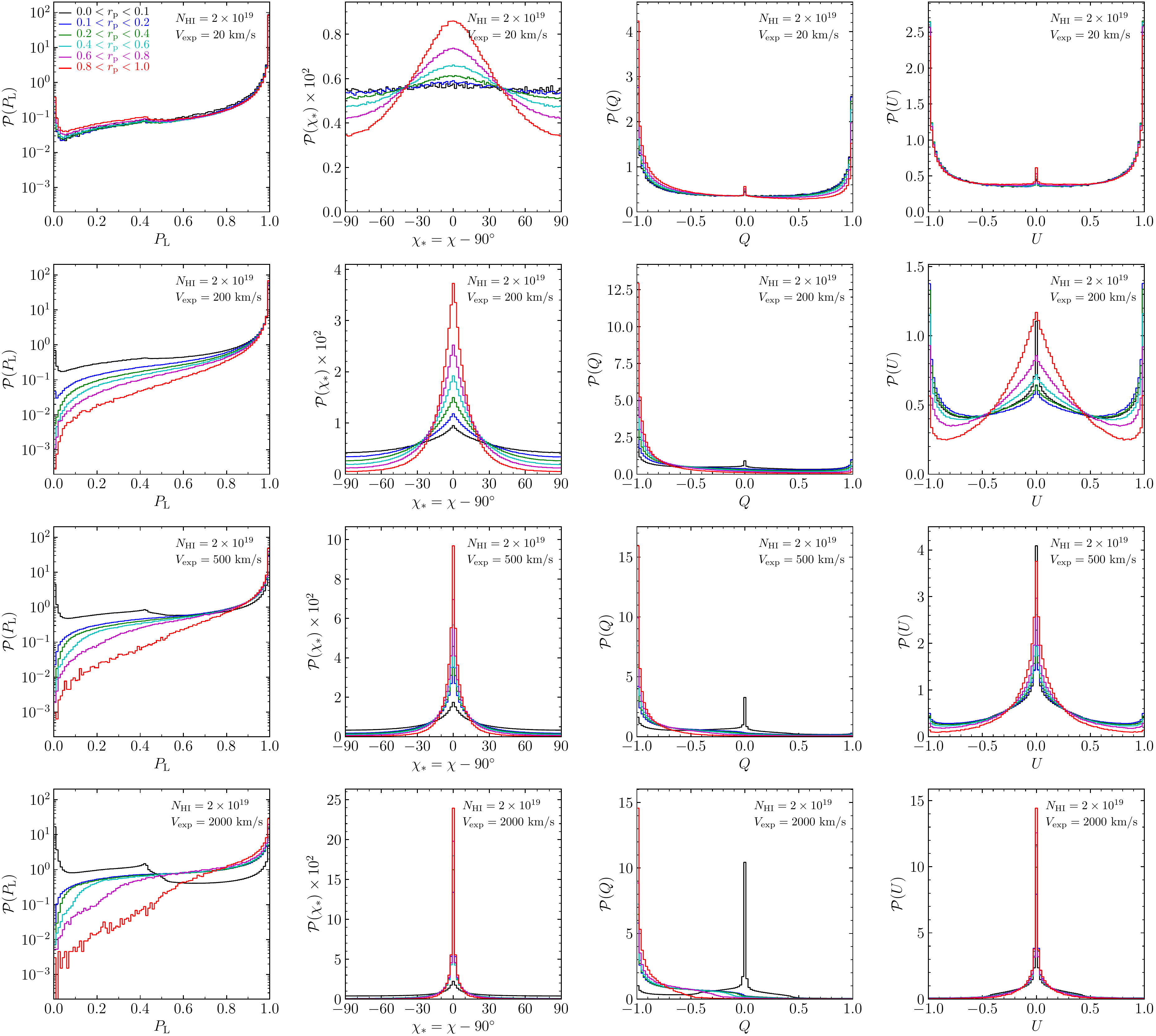}
\par\end{centering}
\begin{centering}
\medskip{}
\par\end{centering}
\caption{\label{fig15}Probability density functions of $P_{{\rm L}}$, $\chi_{*}$,
and the Stokes $Q$ and $U$ of individual photon packets for a Hubble-like
expanding medium with a column density of $N_{{\rm HI}}=2\times10^{19}$
cm$^{-2}$. The exapansion speed are 20, 200, 500, and 2000 km s$^{-1}$
from the first to last rows.}

\centering{}\medskip{}
\end{figure*}

\begin{figure*}[t]
\begin{centering}
\includegraphics[clip,scale=0.45]{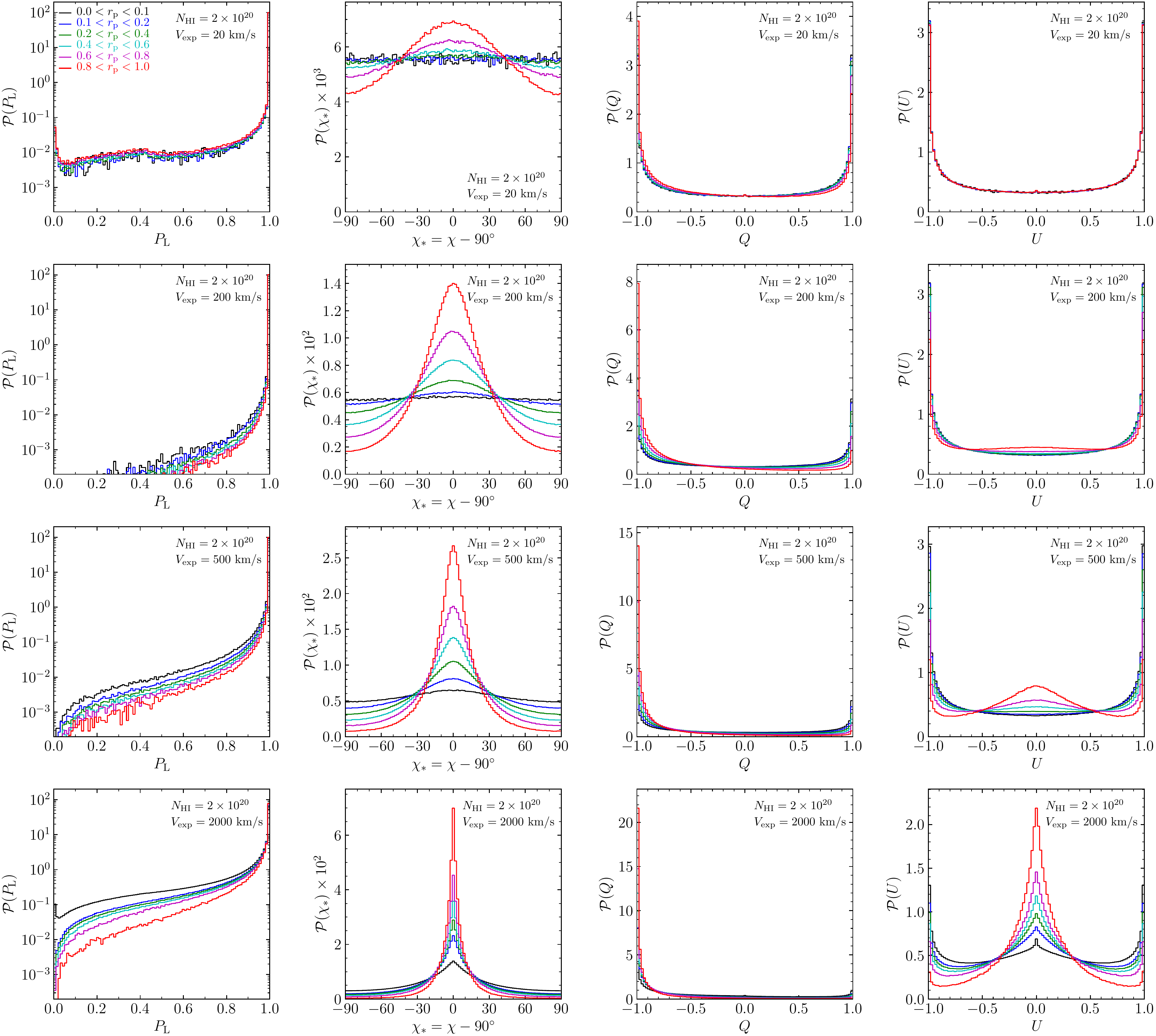}
\par\end{centering}
\begin{centering}
\medskip{}
\par\end{centering}
\caption{\label{fig16}Probability density functions of $P_{{\rm L}}$, $\chi_{*}$,
and the Stokes $Q$ and $U$ of individual photon packets for a Hubble-like
expanding medium with a column density of $N_{{\rm HI}}=2\times10^{20}$
cm$^{-2}$. The exapansion speed are 20, 200, 500, and 2000 km s$^{-1}$
from the first to last rows.}

\centering{}\medskip{}
\end{figure*}

\begin{figure*}[t]
\begin{centering}
\includegraphics[clip,scale=0.52]{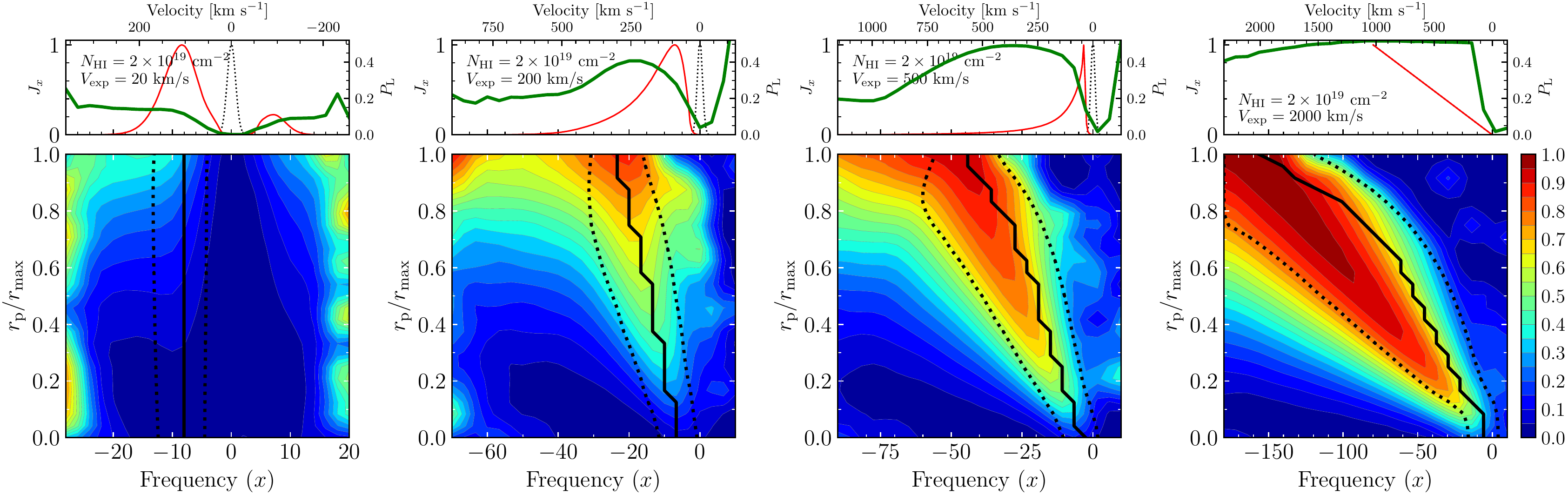}
\par\end{centering}
\begin{centering}
\medskip{}
\par\end{centering}
\begin{centering}
\includegraphics[clip,scale=0.52]{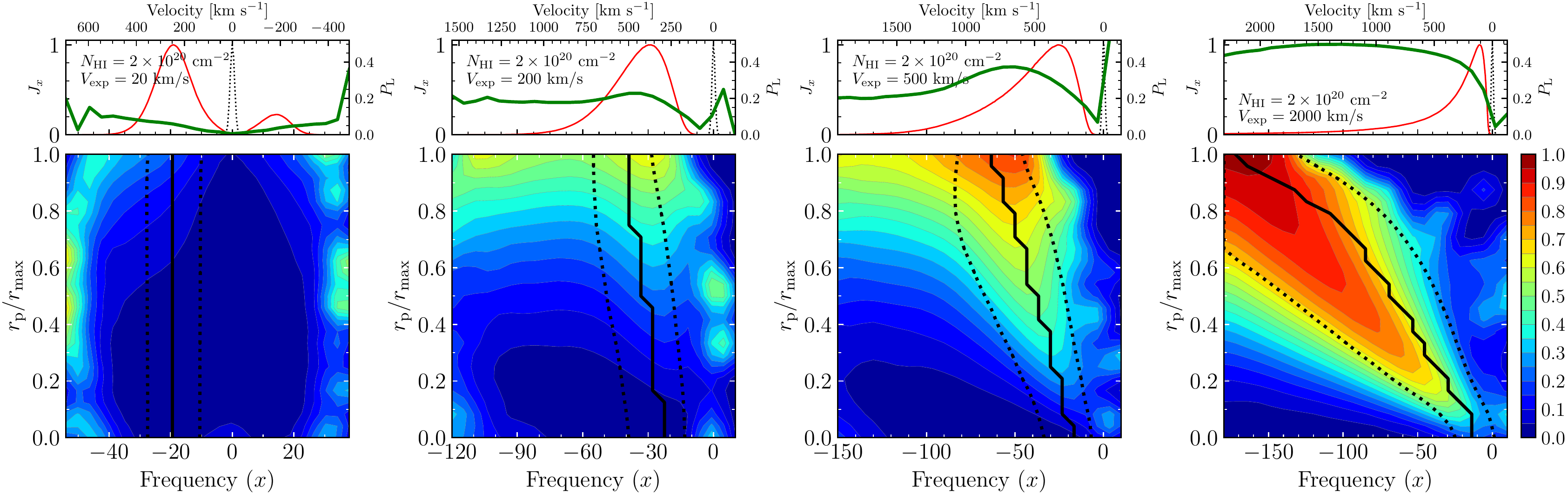}
\par\end{centering}
\begin{centering}
\medskip{}
\par\end{centering}
\caption{\label{fig17}Polarization degree for a Hubble-like expanding medium.
The first and second rows show the results for $N_{{\rm HI}}=2\times10^{19}$
cm$^{-2}$ and $2\times10^{20}$ cm$^{-2}$, respectively. In the
small ancillary panels, the black dotted and red solid curves are
the initial and emergent spectra, respectively. The green curves are
polarization degrees as a function of frequency. The main panels show
the linear polarization degree in a two-dimensional space of photon
frequency and projected radius. The color bar denotes the polarization
levels. The thick black lines denote the peak frequency where the
spectrum has maximum in each radial bin. The thick dotted lines indicate
the frequency at which the spectrum has a half maximum value.}

\centering{}\medskip{}
\end{figure*}

\subsection{A Hubble-like Flow}

\label{section:4.2}

We now investigate the velocity effect on the polarization properties
of Ly$\alpha$ using a Hubble-like expanding medium. The Hubble-like
flow model has been used as a benchmark test to predict the emergent
spectra in a moving medium \citep{2009ApJ...696..853L,2012MNRAS.424..884Y,2017MNRAS.464.2963S}.
This type of model is also proven to be useful in interpreting the
Ly$\alpha$ observations. For instance, to explain the Ly$\alpha$
halos and low-ionization metal absorption lines in Lyman break galaxies,
\citet{2010ApJ...717..289S} suggest a simple model in which Ly$\alpha$
photons are scattered in a spherical symmetric clumpy medium of outflowing
at velocity increasing with distance from the galaxy.

In this model, we consider an isothermal and homogeneous sphere, which
is isotropically expanding or contracting. The velocity of a fluid
element is assumed to be proportional to the radial distance: $V(r)=V_{{\rm exp}}\left(r/r_{{\rm max}}\right)$.
Here, $V_{{\rm exp}}$ is the maximum expansion velocity, representing
the velocity gradient in the radial direction. The neutral hydrogen
gas has a temperature of $T=10^{4}$ K and a column density of $N_{{\rm HI}}=2\times10^{19}$
cm$^{-2}$ or $2\times10^{20}$ cm$^{-2}$ (measured from $r=0$ to
$r=r_{{\rm max}}$). The sphere expands radially with a maximum velocity
ranging from $V_{{\rm max}}=20$ to 3000 km s$^{-1}$ at the outer
edge of sphere ($r=r_{{\rm max}}$). The emergent spectra for a few
cases with $N_{{\rm HI}}=2\times10^{20}$ cm$^{-2}$ were presented
in \citet{2020ApJS..250....9S}.

Figure \ref{fig14} shows the radial profiles of the surface brightness
and polarization obtained from this model. The main results are summarized
as follows. First, the polarization rises monotonically with radius.
Second, the surface brightness profile is nearly flat in the case
of the lowest velocity $V_{{\rm max}}=20$ km s$^{-1}$, as in the
static model, while it becomes steeper as $V_{{\rm max}}$ increases.
This trend is because the number of scatterings, which photons undergo
before escape the system, decreases as the medium's expansion velocity
increases. Third, the overall level of polarization is found to be
in general higher for the model with a higher expanding velocity.
Fourth, the degree of polarization in fast-moving models ($V_{{\rm max}}\gtrsim1000$
km s$^{-1}$) is found to reach almost up to 100\% at the boundary.
Fifth, in the fast-moving models, the polarization increases abruptly
at $r\approx0$, while in rather slowly moving models, it rises gradually.
Lastly, the steepening of surface brightness profile and the polarization
jumping in the highest velocity models appear to be more prominently
when the column density is lower. Similar results were obtained in
contracting media with a negative $V_{{\rm exp}}$ but not presented
here.

In a medium moving away from the source, most photons will be recognized
to have longer wavelengths in the rest frame of the gas (e.g., Figure
15 in \citealp{2020ApJS..250....9S}). Then, the photons are predominantly
scattered by the wing (Rayleigh) scattering, increasing the probability
of being 100\% polarized, as discussed in Section \ref{section:5.1}.
Moreover, photons in a fast-moving medium would experience a relatively
small number of scatterings. Hence, the resulting radiation field
will be highly anisotropic, and the surface brightness profile would
be relatively steep (the second result in Figure \ref{fig14}). If
once a photon packet acquires 100\% polarization through the Doppler
shift after only a small number of scatterings, the polarization level
of the photon packet would be kept high. Also, the ensemble-averaged
polarization level will be very high because of strong anisotropy
in the radiation field (the third and fourth results). In contrast,
in a slow-moving medium, photon packets will have yet a high probability
of being scattered in the core regime; thus, their polarization will
repeat the process of decreasing and increasing during multiple scatterings
until eventually scattered into the wing regime. This process gives
rise to a relatively lower probability of achieving 100\% polarization
than in fast-moving media. More importantly, the radiation field will
be much more isotropic than in fastly moving media, resulting in a
flatter surface brightness and a lower ensemble-averaged polarization.

In the end, the model with a faster expansion velocity (e.g., $V_{{\rm epx}}=200$
km s$^{-1}$) will achieve higher polarization than the model with
a lower velocity ($V_{{\rm exp}}=20$ km s$^{-1}$). However, if the
expansion speed is even faster ($V_{{\rm exp}}\gtrsim500$ km s$^{-1}$),
the chance of being scattered is lowered, and thus the fraction of
photons that escape the medium after being fully polarized rather
decreases, compared to that for $V_{{\rm exp}}=200$ km s$^{-1}$.
The same effect arises also in the emergent spectrum; the peak in
the spectrum for the model of $V_{{\rm exp}}\gtrsim500$ km s$^{-1}$
is less redshifted (Figure 3 of \citealp{2020ApJS..250....9S} and
the small ancillary panels of Figure \ref{fig17}), than the model
of $V_{{\rm exp}}=200$ km s$^{-1}$. Nevertheless, owing to a strongly
anisotropic radiation field, the models of $V_{{\rm exp}}\gtrsim500$
km s$^{-1}$ give rise to higher levels of ensemble-averaged polarization
than the model of $V_{{\rm exp}}=200$ km s$^{-1}$. Given that, in
the lower-velocity models ($V_{{\rm exp}}\lesssim200$ km s$^{-1}$),
more photon packets are 100\% polarized, but the ensemble-averaged
polarization is lower, it is evident that the anisotropy in the radiation
field is of primary importance to yield a polarization level as high
as those seen in the models of $V_{{\rm exp}}\gtrsim500$ km s$^{-1}$.

The abrupt jump of polarization at $r\approx0$ in the fastest models
($V_{{\rm max}}\gtrsim500$ km s$^{-1}$ for $N_{{\rm HI}}=2\times10^{19}$
cm$^{-2}$ and $V_{{\rm max}}\gtrsim1000$ km s$^{-1}$ for $N_{{\rm HI}}=2\times10^{20}$
cm$^{-2}$) results from a sudden breakdown of isotropy in the Ly$\alpha$
radiation field, as in the static models with low optical depths (the
fifth result). However, the jump in polarization occurs to a much
higher level in an expanding medium than a static medium.

Figures \ref{fig15} and \ref{fig16} show the polarization properties
of individual photons for the Hubble-like expanding media with $N_{{\rm HI}}=2\times10^{19}$
cm$^{-2}$ and $2\times10^{20}$ cm$^{-2}$, respectively. The difference
in the jump levels between expanding models and static models can
be explained by examining the figures. Figure \ref{fig15} illustrates
that the vast majority of individual photon packets in the expanding
media with $N_{{\rm HI}}=2\times10^{19}$ cm$^{-2}$ gains $\sim$100\%
polarization and the peak in the distribution function of $P_{{\rm L}}$
at $P_{{\rm L}}\approx0$ is very weak or absent, except for the inner
most region ($0\le r_{{\rm p}}\le0.1$) in fast-moving models. This
is because the photons are rapidly Doppler shifted to the wings by
the outflowing gas motion. When the column density is high ($N_{{\rm HI}}=2\times10^{20}$
cm$^{-2}$; Figure \ref{fig16}), similar results are obtained; but,
the probability of having low $P_{{\rm L}}$ is much lower. 

We now examine the difference due to the column density. In general,
individual photon packets achieve less polarization in a lower column
density model. However, for the same expansion velocity, a lower column
density model yields a higher ensemble-averaged polarization because
of stronger anisotropy in the radiation field, as shown in Figure
\ref{fig14}. The anisotropy of the radiation field plays a more critical
role than the polarization level of individual photons; the leading
role of the anisotropy explains the last finding in Figure \ref{fig14}.
The fact that fewer photons are fully polarized in lower column density
models, especially in fast-expanding cases, is illustrated in the
distribution function of $Q$ of Figures \ref{fig15} and \ref{fig16}.
In Figure \ref{fig15} ($N_{{\rm HI}}=2\times10^{19}$ cm$^{-2}$),
the peak at $P_{{\rm L}}=1$ in the distribution function of $P_{{\rm L}}$
drops rapidly with increasing $V_{{\rm exp}}$, together with the
increase in the probability of being $0.5<P_{{\rm L}}<1$; on the
other hand, in the model of $N_{{\rm HI}}=2\times10^{20}$ cm$^{-2}$,
the peak at $P_{{\rm L}}=1$ drops only mildly. However, the peaks
in the distribution functions of $\chi_{*}$ and $U$ in Figure \ref{fig15}
(the lower column density model) are much sharper than those in Figure
\ref{fig16} (the higher column density model), indicating that the
radiation field in the lower column density model is more anisotropic.

More detailed descriptions of the polarization properties of individual
photons are as follows. Figures \ref{fig15} and \ref{fig16} show
that the overall shapes of the distribution functions of $P_{{\rm L}}$,
$Q$, and $U$ change gradually as the expansion velocity increases.
For instance, the shape of the distribution function of $P_{{\rm L}}$
for the model of $V_{{\rm max}}=2000$ km s$^{-1}$ varies significantly
with the projected radius $r_{{\rm p}}$, compared to that of $V_{{\rm max}}=20$
km s$^{-1}$, which shows no significant variation. The models expanding
faster than $V_{{\rm max}}=500$ km s$^{-1}$ show a sharp peak at
$\chi_{*}=0$ in the distribution function of $\chi_{*}$. Moreover,
their distribution function of $U$ appears to significantly differ
in shape from that of $Q$, unlike the slowly-expanding or static
models. The peak at $U=0$ in the distribution function of $U$ tends
to grow as the expansion speed of the medium increases. A similar
trend is also found in the case of static media, but in this case,
it occurs as the optical depth decreases. However, the peak in the
expanding models is fairly broad, compared to that found in the static
models. Also, the peak in $U$ is not accompanied by the peak in $Q$,
except for the case of $0\le r_{{\rm p}}\le0.1$ in the models of
$N_{{\rm HI}}=2\times10^{19}$ cm$^{-2}$, unlike the static models
in which the peaks in $Q$ and $U$ seem to coexist. These differences
are mainly because the peak of $P(U)$ at $U=0$ in the static models
with a low optical depth is caused by the dominance of $P_{{\rm L}}=0$
(weak polarization of individual photons), while that in the expanding
models is due to the distribution of $\chi$, i.e., anisotropy of
the radiation field. In Figure \ref{fig15}, the peak of the distribution
function of $Q$ at $Q=0$ in the radial bins of $r_{{\rm p}}\lesssim0.1$
becomes bigger with increasing $V_{{\rm exp}}$, because of the lowering
of the effective optical depth. This trend in the small radii can
also be found by noticing that the probability of being $P_{{\rm L}}\lesssim0.2$
rises with $V_{{\rm exp}}$. On the other hand, for larger radii ($r_{{\rm p}}\gtrsim0.6$),
the probability of being $P_{{\rm L}}\lesssim0.2$ decreases gradually
as $V_{{\rm exp}}$ increases. In Figure \ref{fig16}, we find no
apparent peak of the distribution function of $Q$ at $Q=0$, unlike
the case of Figure \ref{fig15}. This difference is attributed to
that photons in higher column density models gain higher polarization
amplitudes.

Figure \ref{fig17} shows the change of the polarization degree when
the expansion velocity varies, as a two-dimensional function of radius
and frequency. The figure also shows the frequencies where the spectrum
has maximum (thick solid lines) and half maximum (thick dotted lines)
values in each radius bin. The model with the lowest expansion velocity
shows no significant difference in polarization from the static media,
except for a weak tendency of being more polarized in the red wavelengths.
For the intermediate velocity ($V_{{\rm max}}\approx200$ km s$^{-1}$),
the polarization degree shows a relatively weak dependence on frequency
in a frequency range of $x\lesssim-10$ at a fixed radius ($r\gtrsim0.4$);
however, it strongly depends on the radius. The degree of polarization
is in general weak at $r\approx0$. As the expansion velocity increases,
the polarization begins to develop near the central frequencies and
$r\approx0$. For instance, for the highest velocity ($V_{{\rm max}}=2000$
km s$^{-1}$) model, the frequency at which the maximum $P_{{\rm L}}$
is found to change linearly with the radius $r$; the peak frequency
shifts to red as $r$ increases. This is because photons are more
easily Doppler-shifted at smaller radii when the velocity gradient
is larger. It is also noticeable that the peak polarization, denoted
by the reddest color in the figure, occurs at a slightly lower frequency
than the peak frequency, indicated by the thick black lines, especially
in the highest velocity models. The variation in the polarization
properties with the expansion velocity illustrates that the spatially-resolved
spectropolarimetric observations would be of great importance to understand
the kinematics of hydrogen gas in the galactic halos.

\begin{figure*}[t]
\begin{centering}
\includegraphics[clip,scale=0.65]{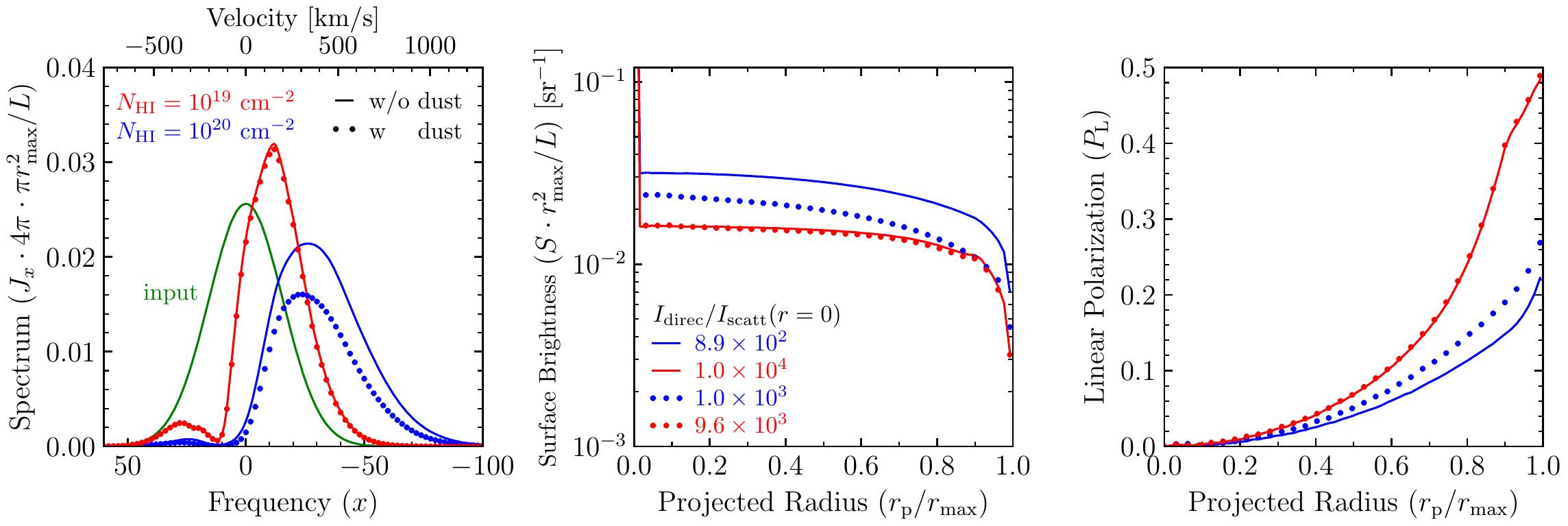}
\par\end{centering}
\begin{centering}
\medskip{}
\par\end{centering}
\caption{\label{fig18} Observable properties for an expanding thin shell model
of \citet{2008MNRAS.386..492D}. The emergent spectrum ($J_{x}$),
surface brightness ($S$), and degree of linear polarization ($P_{{\rm L}}$)
are shown in the left, middle, and right panels, respectively. In
the middle panel, the intensity ratio of the direct ($I_{{\rm direc}}$)
to the scattered ($I_{{\rm scatt}}$) intensity at the center $(r=0)$
is also denoted. The blue and red represent the models with column
densities of $N_{{\rm HI}}=10^{19}$ and $10^{20}$ cm$^{-2}$, respectively.
Models without and with dust are denoted by lines and circles, respectively.
Here, $r_{{\rm max}}$ and $L$ are the radial size and Ly$\alpha$
luminosity of the system, respectively. The models
containing dust grains have the dust optical depths of $\tau_{{\rm dust}}=1.6\times10^{-2}$
($N_{{\rm HI}}=10^{19}$ cm$^{-2}$) and $1.6\times10^{-1}$ ($N_{{\rm HI}}=10^{20}$
cm$^{-2}$).}

\centering{}\medskip{}
\end{figure*}

\begin{figure*}[t]
\begin{centering}
\includegraphics[clip,scale=0.65]{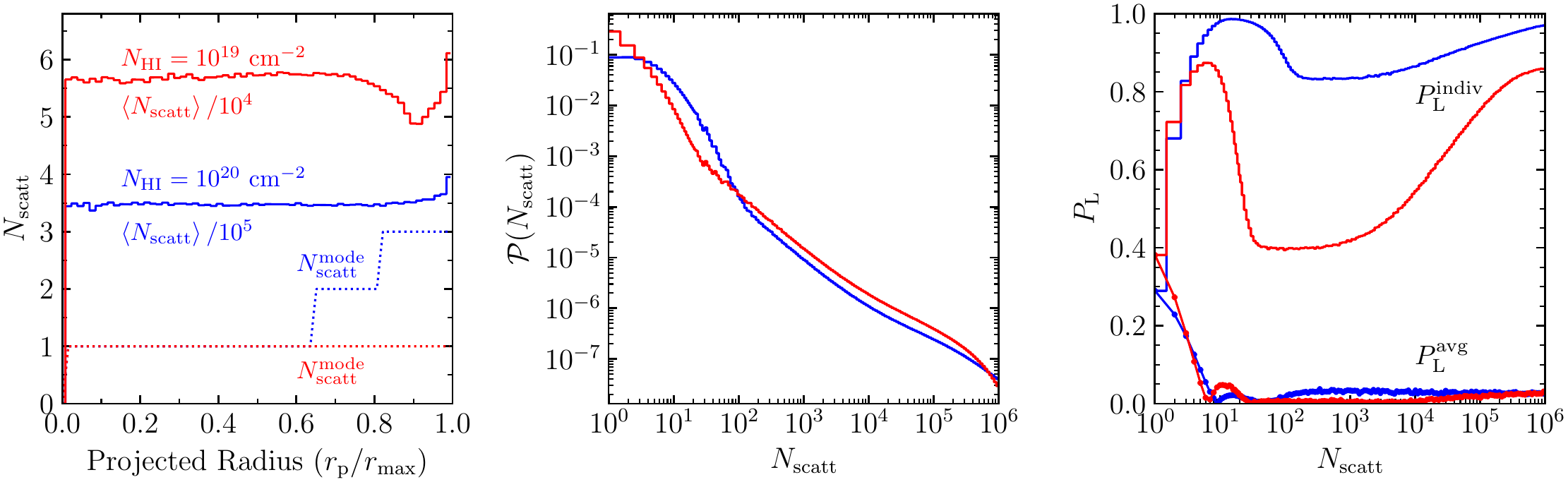}
\par\end{centering}
\begin{centering}
\medskip{}
\par\end{centering}
\caption{\label{fig19} Results of an expanding thin shell model of \citet{2008MNRAS.386..492D}.
The left panel shows the mean ($\left\langle N_{{\rm scatt}}\right\rangle $)
and mode ($N_{{\rm scatt}}^{\text{mode}}$) of the number of scatterings,
which a photon experiences before escaping the system, as a function
of projected radius (or impact parameter). The middle panel represents
the probability distribution function of the number of scatterings.
The right panel shows the degree of linear polarization as a function
of the number of scatterings that photons experience. Here, $P_{{\rm L}}^{\text{indiv}}$
denotes the mean value of the ``individual'' linear polarization amplitudes
of photons that are scattered $N_{{\rm scatt}}$ times, while $P_{{\rm L}}^{\text{avg}}$
is the ``ensemble-averaged'' polarization level. The blue and red
represent the models with column densities of $N_{{\rm HI}}=10^{19}$
and $10^{20}$ cm$^{-2}$, respectively.}

\centering{}\medskip{}
\end{figure*}

\begin{figure*}[t]
\begin{centering}
\includegraphics[clip,scale=0.45]{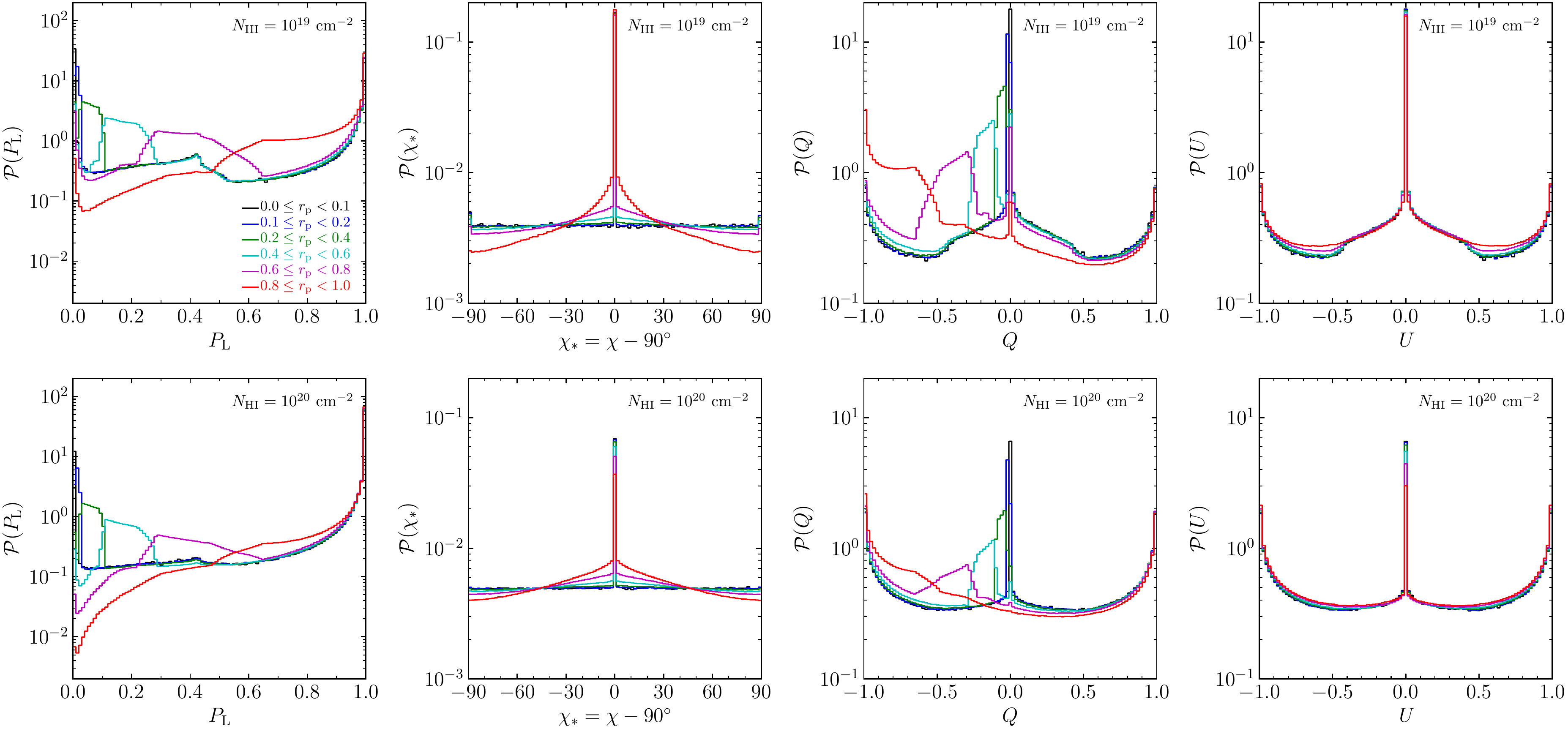}
\par\end{centering}
\begin{centering}
\medskip{}
\par\end{centering}
\caption{\label{fig20}Probability density functions of $P_{{\rm L}}$, $\chi_{*}$,
and the Stokes $Q$ and $U$ of ``individual'' photons for the expanding
thin shell model of \citet{2008MNRAS.386..492D}. They are all measured
in a local coordinate system defined by radial and tangential vectors
in the detector plane. Note that the ordinates were plotted in logarithmic
scale to highlight weak changes in their values. Only
the photon packets that have experienced scatterings at least once
($N_{{\rm scatt}}\ge1$) are included in calculating the probability
distribution functions.}

\centering{}\medskip{}
\end{figure*}

\begin{figure*}[t]
\begin{centering}
\includegraphics[clip,scale=0.52]{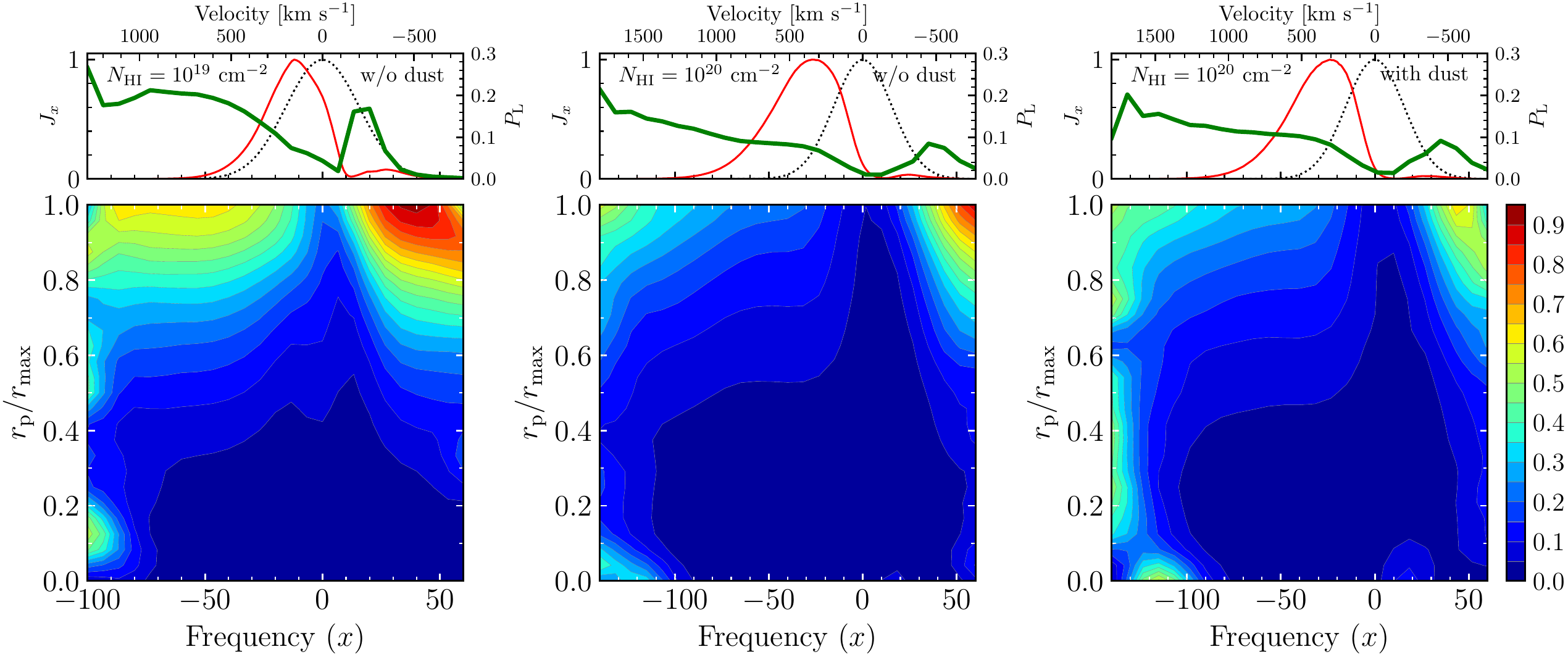}
\par\end{centering}
\begin{centering}
\medskip{}
\par\end{centering}
\caption{\label{fig21}Polarization degree for the expanding thin shell model
of \citet{2008MNRAS.386..492D}. The first and second columns show
the results for the models without dust, while the third column shows
the results for the model with dust. In the small ancillary panels,
the black dotted and red solid curves are the initial and emergent
spectra, respectively. The green curves represent polarization degrees
as a function of frequency. The main panels show the linear polarization
degree in a two-dimensional space of photon frequency and projected
radius. The color bar denotes the polarization levels.}

\centering{}\medskip{}
\end{figure*}

\subsection{An Outflowing Thin Shell}

\label{section:4.3}

The galactic superwind model of \citet{2008MNRAS.386..492D} was developed
to explain the Ly$\alpha$ spectra redshifted relative to the systemic
velocity, measured using non-resonant lines, of Lyman break galaxies
(LBGs) and LAEs. \citep{2003MNRAS.340..863A,2006A&A...460..397V,2008A&A...491...89V}.
In the model, the redshifted spectra are attributed to the Doppler
boost that Ly$\alpha$ photons undergo when they are scattered back
toward the observer by an outflowing shell on the far side of the
galaxy. The outflow is modeled by a radially expanding thin shell.
The shell expands at a constant speed of $V_{{\rm exp}}=200$ km s$^{-1}$
and has a column density of $N_{{\rm HI}}=10^{19}$ or $10^{20}$
cm$^{-2}$. The initial Ly$\alpha$ spectrum originating from a central
source is assumed to be a Gaussian with a velocity width $\sigma=200$
km s$^{-1}$, which corresponds to a typical circular velocity of
a dark matter halo with a total mass of $M_{{\rm tot}}=3\times10^{11}$
M$_{\odot}$. This example was adopted to verify our code and investigate
a case in which single scattering is predominant.

Figure \ref{fig18} shows the resulting emergent spectrum, surface
brightness, and degree of linear polarization, consistent with those
of \citet{2008MNRAS.386..492D}.\footnote{We found that a variation of the shell thickness gives rise to small
yet recognizable changes. As the thickness decreases, the emergent
spectrum, particularly for $N_{{\rm HI}}=10^{20}$ cm$^{-2}$, was
found to be less redshifted, the surface brightnesses became flatter,
and the maximum degree of polarization decreased slightly. A shell
thickness of $\sim0.1r_{{\rm max}}$ was found to yield surface brightness
profiles that best match \citet{2008MNRAS.386..492D}; hence, a thickness
of $0.1r_{{\rm max}}$ was adopted in this paper.} The left panel of the figure shows that the spectrum becomes more
redshifted and broader as $N_{{\rm HI}}$ increases. The green line
in the same panel denotes the initial input spectrum. The middle panel
shows the profile of surface brightness as a function of projected
radial distance $r_{{\rm p}}$, which is relatively flat up to $r_{{\rm p}}\sim0.7r_{{\rm max}}$.
We note that a considerable fraction of Ly$\alpha$ flux (49\% for
$N_{{\rm HI}}=10^{19}$ cm$^{-2}$ and 8\% for $N_{{\rm HI}}=10^{20}$
cm$^{-2}$) escapes out of the system without experiencing any scattering.
As a result, a sharp central peak is apparent in the surface brightness
profile, but it is, unfortunately, not consistent with the shapes
observed in LAEs. Note that, in this model, the ISM can be regarded
to be concentrated in a small central region together with the Ly\textgreek{a}
emission source, and the galactic halo is assumed to be geometrically
thin and far from the ISM. The presence of a peak in the surface brightness
indicates that the galactic wind should be dealt with a spatially
extended medium rather than a geometrically thin shell so that the
ISM and galactic halo are seamlessly connected without an extreme
gap. The right panel shows that the degree of linear polarization
tends to increase with radius, as expected. The degree of polarization
at the center is zero due to symmetry. In the outermost part, the
polarization reaches as high as $\sim50$\% for $N_{{\rm HI}}=10^{19}$
cm$^{-2}$ and $\sim20$\% for $N_{{\rm HI}}=10^{20}$ cm$^{-2}$.

\citet{2008MNRAS.386..492D} found that the probability distribution
of the number of scattering events ($N_{{\rm scatt}}$) peaks at $N_{{\rm scatt}}=1$
and rapidly decreases with increasing $N_{{\rm scatt}}$. This is
because, in this model, most photons reside on the wing of the line
profile when they reach the expanding shell; thus, they escape the
system without any scattering or after only one scattering. The left
panel of Figure \ref{fig19} shows the average ($\left\langle N_{{\rm scatt}}\right\rangle $)
and most frequent value ($N_{{\rm scatt}}^{{\rm mode}}$) of the number
of scatterings as a function of projected radius in the detector plane.
In the middle panel, we also show the probability distribution function
of the number of scatterings, $\mathcal{P}(N_{{\rm scatt}})$. The
distribution function peaks at $N_{{\rm scatt}}=1$ for $N_{{\rm HI}}=10^{19}$
cm$^{-2}$ and $N_{{\rm scatt}}=1-3$ for $N_{{\rm HI}}=10^{20}$
cm$^{-2}$, and decreases sharply with increasing $N_{{\rm scatt}}$.
The median number of scatterings is found to be $N_{{\rm scatt}}^{\text{median}}=3$
for $N_{{\rm HI}}=10^{19}$ cm$^{-2}$ and $N_{{\rm scatt}}^{\text{median}}=9$
for $N_{{\rm HI}}=10^{20}$ cm$^{-2}$, considering only the photons
that have experienced scatterings at least once (but not shown in
the figure). Even though most photons are scattered only a few times
at most, the distribution function of $N_{{\rm scatt}}$ shows a very
long tail. Thus, $\left\langle N_{{\rm scatt}}\right\rangle $ is
found to be very high. Interestingly, $\left\langle N_{{\rm scatt}}\right\rangle $
remains nearly constant as $r_{{\rm p}}$ increases, except for a
slight rise near $r_{{\rm p}}=r_{{\rm max}}$ due to the rim effect.
The highest optical thickness arises toward the line of sight of $r_{{\rm p}}\approx r_{{\rm max}}$.
The dip at $r_{{\rm p}}\approx0.9r_{{\rm max}}$ in the model of $N_{{\rm HI}}=10^{19}$
cm$^{-2}$ is also a geometrical effect due to a finite shell thickness
of $0.1r_{{\rm max}}$, adopted in our calculation. In the Rayleigh
regime, scatterings toward the observer at the right angle at $r_{{\rm p}}=0.9r_{{\rm max}}$
are least probable, causing the dip. The dip disappears for a higher
$N_{{\rm HI}}$ because of an increase in the number of scatterings.

\citet{2008MNRAS.386..492D} also noticed that the ``ensemble-averaged''
polarization level of photons that experienced more than 10 scatterings
($N_{{\rm scatt}}>10$) nearly vanishes. They also found that a group
of photons scattered only once gives rise to the highest (ensemble-averaged)
polarization, while photons scattered multiple times yield much lower
polarization. To confirm the results, we computed the ``ensemble-averaged''
degree of polarization for photon packets escaping after a specific
number of scatterings, as shown in the right panel of Figure \ref{fig19}.
For comparison, we also present the polarization degree of ``individual''
photons. In the figure, $P_{{\rm L}}^{\text{indiv}}$ is the mean
polarization amplitude of individual photon packets, measured regardless
of the polarization direction, and $P_{{\rm L}}^{\text{avg}}$ the
ensemble-averaged polarization, i.e., the polarization calculated
after averaging the Stokes $Q$ and $U$ of photon packets that were
scattered $N_{{\rm scatt}}$ times. As in \citet{2008MNRAS.386..492D},
$P_{{\rm L}}^{\text{avg}}$ peaks at $N_{{\rm scatt}}=1$ and decreases
as $N_{{\rm scatt}}$ increases. On the other hand, $P_{{\rm L}}^{\text{indiv}}$
increases with $N_{{\rm scatt}}$ in a range of $N_{{\rm scatt}}\lesssim10$,
and then decreases; but, as $N_{{\rm scatt}}$ further increases ($N_{{\rm scatt}}\gtrsim10^{2}$),
it increases again until it finally saturates. The saturated level
of the polarization of individual photons reaches on average as high
as $\sim85$\% for $N_{{\rm HI}}=10^{19}$ cm$^{-2}$ and $\sim100$\%
for $N_{{\rm HI}}=10^{20}$ cm$^{-2}$.

Figure \ref{fig20} shows the distribution functions of the polarization
angle $\chi_{*}$, the Stokes $Q$ and $U$, and the linear polarization
degree $P_{{\rm L}}$ of individual photons at different radii in
the detector plane. In this figure, only photon packets that experienced
scatterings at least once or more were considered. The polarization
angle $\chi_{*}$ is uniformly distributed at the central part ($r_{{\rm p}}\sim0$),
except for a very sharp peak at $\chi_{*}=0$. The peaks at $\chi_{*}=0$
found in all radius bins are mainly due to photons escaping after
only a single or couple of scatterings. The peaks of the distribution
functions of $\chi_{*}$, $Q$, and $U$ found in this model are much
stronger and sharper than those shown in other models. This is because
the thin shell model is predominated by single-scattering, while the
other models are by multiple scattering. The polarization degree of
individual photons tends to have $P_{{\rm L}}<0.5$ for $r_{{\rm p}}\lesssim0.7r_{{\rm max}}$
but increases with increasing $r_{{\rm p}}$. At $r_{{\rm p}}\sim r_{{\rm max}}$,
most photons have a polarization of $P_{{\rm L}}\sim1$. It also appears
that, as $N_{{\rm HI}}$ increases, more photons tend to have the
polarization vectors aligned to the tangential direction ($\chi_{*}=0$)
and be 100\% polarized.

The distribution functions of $Q$ are similar to those in the static
sphere models with $\tau_{0}\lesssim10^{5}$, except for two aspects:
the presence of a rectangle shape bump and the weakness (or absence)
of the peak near $Q=0$ in radii $r_{{\rm p}}\gtrsim0.7r_{{\rm max}}$.
The rectangle bump, which is caused by photons escaping after a single
scattering ($N_{{\rm scatt}}=1$), is also found in the distribution
function of $P_{{\rm L}}$. When we ignored the single-scattered photons,
the bumps disappeared, leading to smoothly varying distribution functions.
The bump tends to move toward higher $P_{{\rm L}}$ and $\left|Q\right|$,
as $r_{{\rm p}}$ increases. This trend can be understood with the
help of the Rayleigh scattering phase function in Equation (\ref{eq:22}).
Suppose photons that escape after a single Rayleigh scattering. As
the projected radius at which a photon is scattered increases from
$r_{{\rm p}}=0$ to $r_{{\rm max}}$, the scattering angle toward
the observer also increases from $\theta=0$ to $\pi/2$. The increase
in $\theta$ leads to higher $P_{{\rm L}}$ and $\left|Q\right|$
because the right angle scattering gives rise to the highest polarization
level. The extent of $Q$ in the rectangular bump corresponds to the
range of radii from which the photons escape. In Equation (\ref{eq:21}),
we also note that $Q'=S_{12}I$ (for $Q=U=0$) is negative. This is
why the rectangular shapes occur only at negative $Q$ values. The
shapes of the distribution functions of $Q$ at large radii resemble
those for the Hubble-like expanding media that show no peak at $Q=0$,
while the sharp peaks at small radii are more like those seen in the
static models.

We also note that the polarization degree of individual photon packets
shows a rather complex variation of increasing and decreasing, as
shown in the right panel of Figure \ref{fig19}. However, eventually,
photons that experienced a sufficiently large number of scatterings
end up having a very high polarization level of up to $\sim$100\%.
As noted in Section \ref{section:4.1}, if once photon packets were
diffused out into the wings, further scatterings lead to an increase
in the degree of polarization of individual photon packets. However,
photons that have traversed through many different routes will have
randomly oriented polarization directions. The randomness of the polarization
direction will largely cancel out the Stokes $Q$ and $U$ parameters.
In the end, the final ``ensemble-averaged'' polarization level will
be much lowered than those of individual photons. The model with $N_{{\rm HI}}=10^{20}$
cm$^{-2}$ produces higher polarizations for individual photon packets
than the model with $N_{{\rm HI}}=10^{19}$ cm$^{-2}$ (Figures \ref{fig18}
and \ref{fig20}) but gives a lower ``ensemble-averaged'' polarization
because of higher isotropy ($P_{{\rm L}}^{\text{avg}}$ in Figure
\ref{fig18}).

\begin{figure*}[t]
\begin{centering}
\includegraphics[clip,scale=0.53]{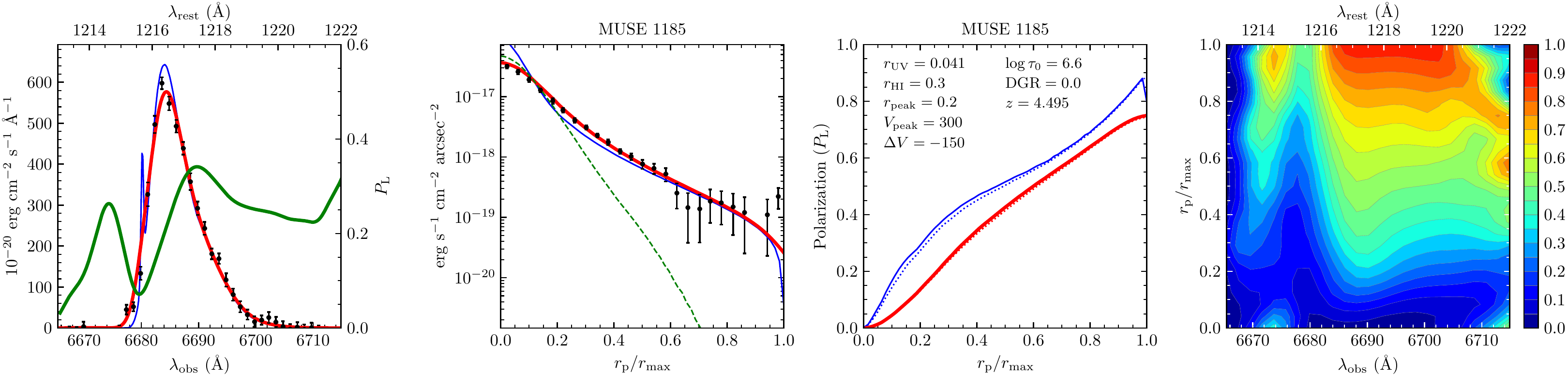}
\par\end{centering}
\begin{centering}
\medskip{}
\par\end{centering}
\begin{centering}
\includegraphics[clip,scale=0.53]{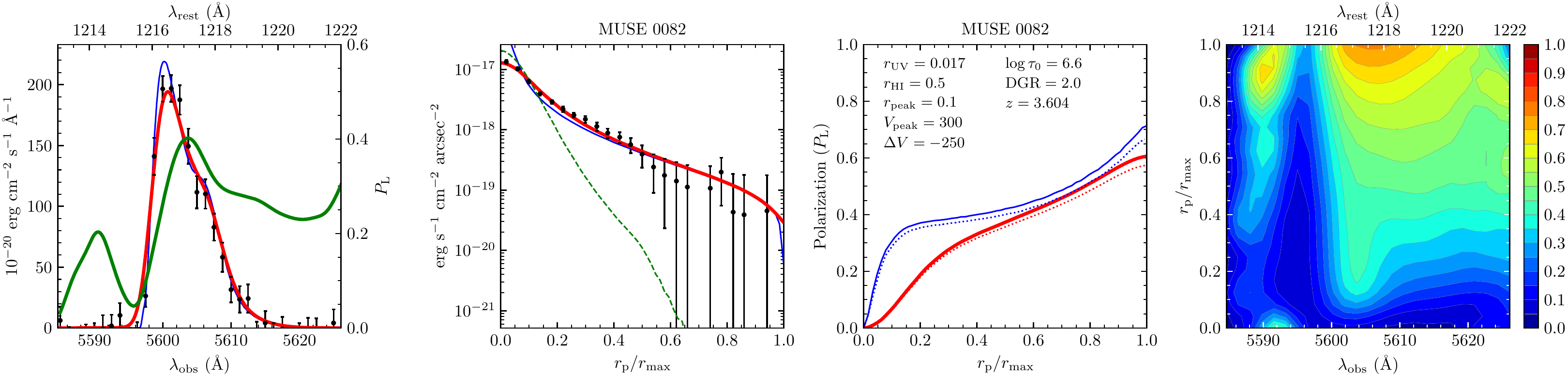}
\par\end{centering}
\begin{centering}
\medskip{}
\par\end{centering}
\begin{centering}
\includegraphics[clip,scale=0.53]{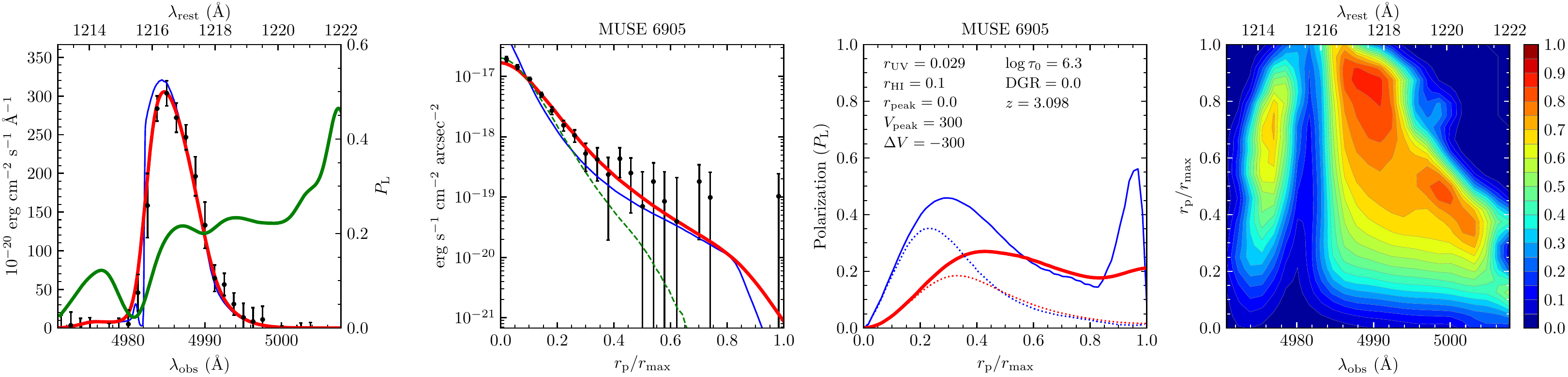}
\par\end{centering}
\begin{centering}
\medskip{}
\par\end{centering}
\begin{centering}
\includegraphics[clip,scale=0.53]{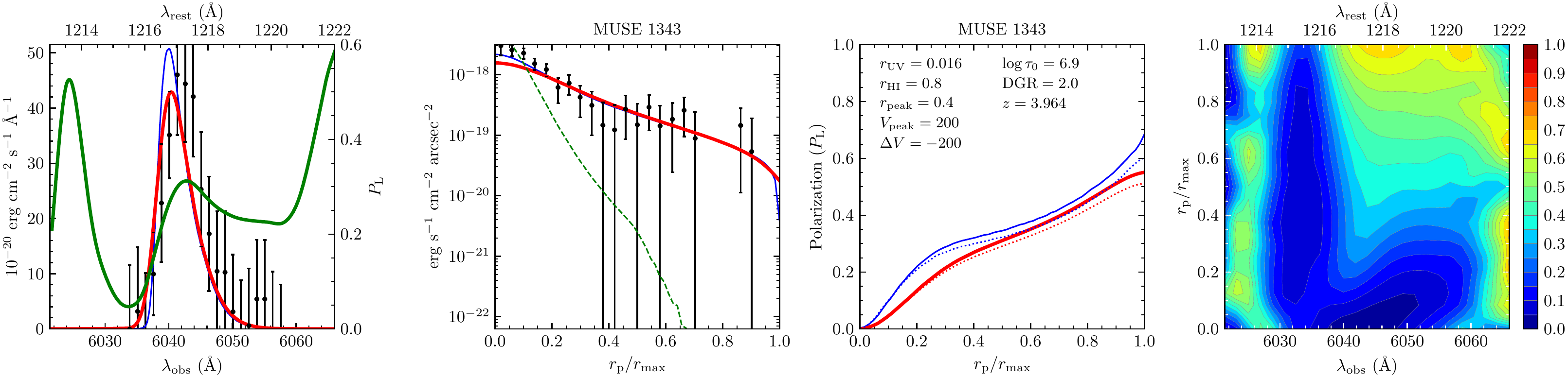}
\par\end{centering}
\begin{centering}
\medskip{}
\par\end{centering}
\caption{\label{fig22}Spectrum and polarization degree as a function of wavelength
(first column), surface brightness profile (second column), polarization
vs. projected radius (third column), and polarization degree as a
function of wavelength and projected radius (fourth column) of four
LAEs (MUSE 1185, 0082, 6905, and 1343), which are obtained using a
momentum-driven galactic wind model of \citet{2020ApJ...901...41S}.
In the first and second columns, the filled circles with error bars
are the observational data. The blue/red lines in the first, second,
and third panels are the best-fit models before/after convolution
with the instrumental line or point spead function. The green lines
in the first column represent the polarization degree vs. wavelength,
convolved with the instrumental line spread function. The green dashed
lines in the second column is the point spread function of MUSE. The
spectra are shown as a function of the observational wavelength as
well as the rest frame wavelength. The third column show the degree
of linear polarization, predicted using the best-fit model. The theoretical
spectra and surface brightness/polarization profiles were obtained
after applying the aperture sizes and wavelength ranges, respectively,
described in \citet{2017A&A...608A...8L} and \citet{2020ApJ...901...41S}.
In the third column, the polarization profiles constructed using all
photons are also shown in dotted lines; the blue and red lines denote
the profiles before and after the convolution with the PSF, respectively.
The best-fit model parameters are also shown in the panels of the
third column. Here, $r_{{\rm UV}}$ and $r_{{\rm HI}}$ are the scale
parameters of the distribution functions of Ly$\alpha$ source and
gas density. DGR is the dust-to-gas ratio relative to that of MW,
$\tau_{0}$ the optical depth, and $z$ the redshift of galaxy. $r_{{\rm peak}}$,
$V_{{\rm peak}}$, and $\Delta V$ define the velocity profile of
the galactic wind, as in Equation (\ref{eq:50}); $V_{{\rm peak}}$
is the peak velocity, $r_{{\rm peak}}$ is the radius at which the
fluid velocity peaks, and $\Delta V$ is the difference between $V_{{\rm peak}}$
and the velocity at $r=r_{{\rm max}}$. All size variables ($r_{{\rm UV}},$$r_{{\rm HI}}$,
and $r_{{\rm peak}}$) are normalized to the size of halo ($r_{{\rm max}}$).
The velocity is in units of km s$^{-1}$. The rightmost panels are
the color bars to represent the polarization level.}

\centering{}\medskip{}
\end{figure*}

\begin{figure*}[t]
\begin{centering}
\includegraphics[clip,scale=0.53]{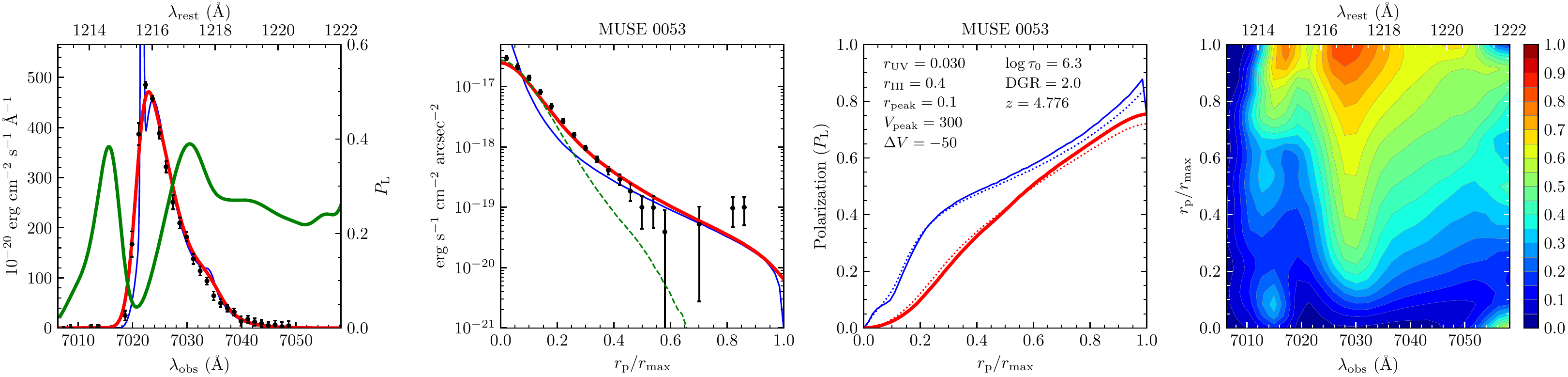}
\par\end{centering}
\begin{centering}
\medskip{}
\par\end{centering}
\begin{centering}
\includegraphics[clip,scale=0.53]{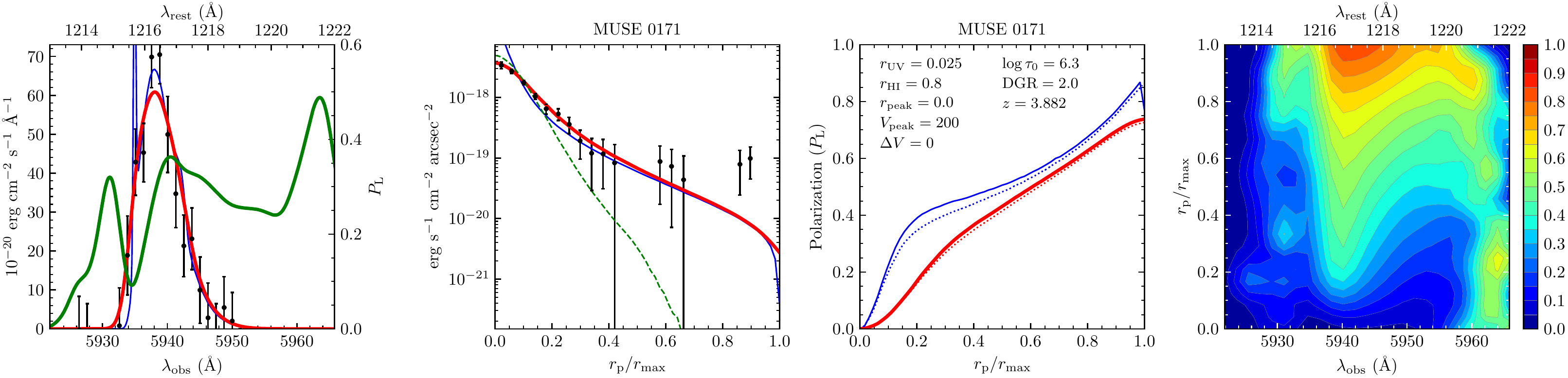}
\par\end{centering}
\begin{centering}
\medskip{}
\par\end{centering}
\begin{centering}
\includegraphics[clip,scale=0.53]{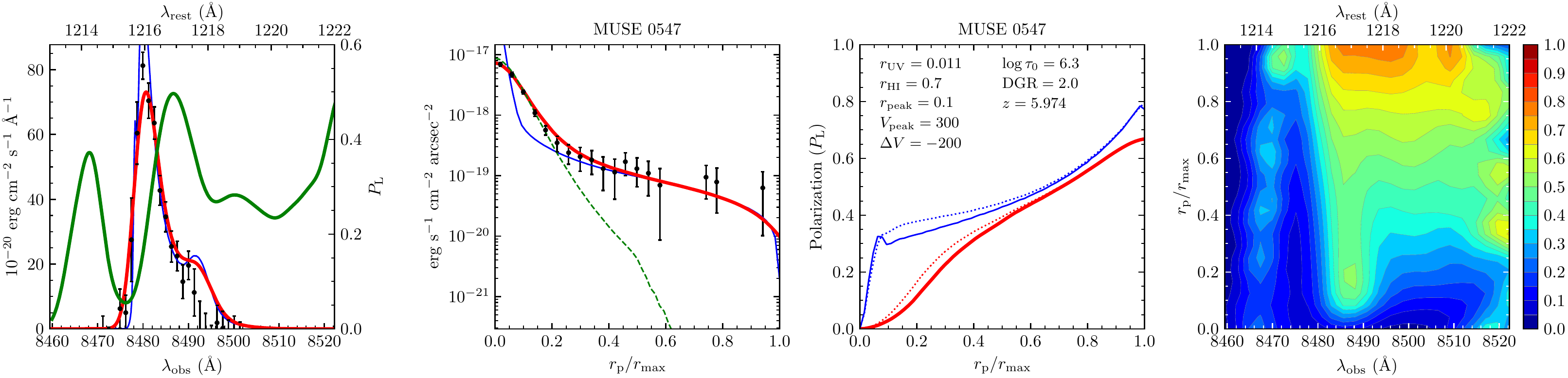}
\par\end{centering}
\begin{centering}
\medskip{}
\par\end{centering}
\begin{centering}
\includegraphics[clip,scale=0.53]{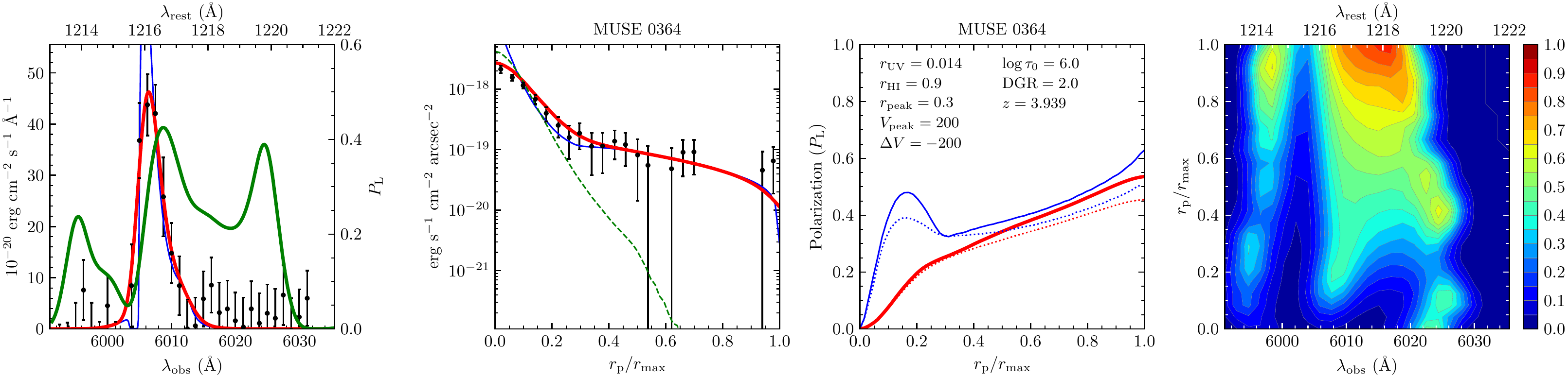}
\par\end{centering}
\begin{centering}
\medskip{}
\par\end{centering}
\caption{\label{fig23}Spectrum and polarization degree as a function of wavelength
(first column), surface brightness profile (second column), polarization
vs. projected radius (third column), and polarization degree as a
function of wavelength and projected radius (fourth column) of four
LAEs (MUSE 0053, 0171, 0547, and 0364). See the caption of Figure
\ref{fig22} for the definitions of symbols, lines, and variables.}

\centering{}\medskip{}
\end{figure*}

We additionally computed the thin shell models that include dust grains;
the filled circles in Figure \ref{fig18} show the results. The dust-to-gas
ratio was set to that of the MW. In the figure, especially for the
higher $N_{{\rm HI}}$, the presence of dust appears to make the surface
brightness profile slightly steeper and raise the polarization level;
no significant enhancement in the polarization is found for $N_{{\rm HI}}=10^{19}$
cm$^{-2}$, as in the static model of $\tau_{0}=10^{6}$ because of
relative weakness of the dust extinction effect. We also found that
ignoring single-scattered photons significantly reduced the ensemble-averaged
degree of polarization (but not shown in figures); this is because
the radiation field's anisotropy in this model is mainly attributed
to single scattered photons, and ignoring them decreases the anisotropy
of the radiation field.

In Figure \ref{fig21}, we show the dependence of polarization degree
on radius and frequency. As in the Hubble-like outflow model, the
polarization level is found to increase toward the red wavelength.
This is because the redder photons are farther from resonance in the
frame of the expanding gas; therefore, they are scattered less, yield
a relatively steeper radial profile compared to the bluer photons,
and eventually achieve a higher polarization. However, the result
appears more complicated if we examine the variation of the frequency
dependence of polarization with radius. At small radii, red photons
tend to give a higher polarization. In contrast, near $r_{{\rm p}}=r_{{\rm max}}$,
higher polarization is found at blue wavelengths. This is because
blue photons are scattered more (10 times more than red photons in
the radius range $r_{{\rm p}}>0.9r_{{\rm max}}$). The radiation field
at $r\approx r_{{\rm max}}$ is highly anisotropic, independently
of the wavelength, because of the geometrical effect. In the boundary
region, the ``ensemble-averaged'' polarization level appears to be
primarily determined by the polarization of ``individual'' photons
rather than by the anisotropy of the radiation field. However, despite
this opposite trend at large radii, the spatially averaged polarization
is higher in red wavelengths because fluxes at smaller radii dominate
the total flux. In the right panel, we note that the model with dust
reveals a lower maximum polarization than the case without dust (middle
panel). This is because of the preferential extinction of blue photons
scattered more. However, at small radii, the model with dust gives
a higher polarization than the model without dust; therefore, dust
grains yield, in the end, a higher polarization in the spatially averaged
spectrum than the dustless case.

\subsection{A Momentum-Driven Galactic Wind Model}

\label{section:4.4}

\citet{2020ApJ...901...41S} recently simultaneously modeled the emergent
spectra and surface brightness profiles of eight LAEs at $z=3-6$
observed with the Multi-Unit Spectroscopic Explorer \citep{2015A&A...575A..75B,2020A&A...635A..82L}.
In the study, we considered a galaxy as an outflowing halo in which
the distributions of Ly$\alpha$ source and gas are described by exponential
functions. The outflow of the medium is described by a simplified
momentum-driven galactic wind; rather than adopting the functional
form in \citet{2012MNRAS.424.1672D}\footnote{Adopting the analytical solution of \citet{2012MNRAS.424.1672D} yielded
unrealistic spectra with a strong peak at the Ly$\alpha$ line center
($x=0$) due to an extremely large velocity gradient at $r=0$. This
was why we adopted a linearly increasing and then decreasing velocity
profile instead of their solution.}, obtained by solving a momentum equation in a gravitational potential
well, we instead considered a linearly increasing and then decreasing
velocity profile, as follows:
\begin{equation}
V(r)=\begin{cases}
V_{{\rm peak}}r/r_{{\rm peak}} & {\rm if}\ \ r\le r_{{\rm peak}}\\
V_{{\rm peak}}+\Delta V(r-r_{{\rm peak}})/(r_{{\rm max}}-r_{{\rm peak}}) & \text{otherwise},
\end{cases}\label{eq:50}
\end{equation}
where $V_{{\rm peak}}$ is the peak velocity, $r_{{\rm peak}}$ the
radius where the velocity peaks and $\Delta V$ the difference between
velocities at $r_{{\rm peak}}$ and $r_{{\rm max}}$. In the model,
we assumed $\Delta V\le0$ to consider only the case in which the
gas accelerates and then decelerates.

In the model calculation, \citet{2020ApJ...901...41S} adopted uniform
distribution functions for the initial input spectrum and spatial
distribution of the Ly$\alpha$ source. Then, to reduce the number
of simulations, they obtained the final results by multiplying the
intermediate results by the desired input profiles (i.e., Voigt and
Bessel profiles for the spectrum and UV surface brightness profile,
respectively). The resulting spectrum and surface brightness profile
were convolved by the spectral and spatial kernels appropriate to
the instrument, and were compared with the observational data to find
the best-fit parameters. For this paper, however, we directly generated
the desired input spectrum and spatial distribution of injected photons
using the best-fit parameters instead of adopting a uniform distribution
and then adjusting the inputs by multiplying appropriate distribution
functions. The present study also took into account the polarization
state of Ly$\alpha$ in the model calculation. This model is presented
to demonstrate the diversity of the polarization pattern.

\begin{figure}[t]
\begin{centering}
\includegraphics[clip,scale=0.435]{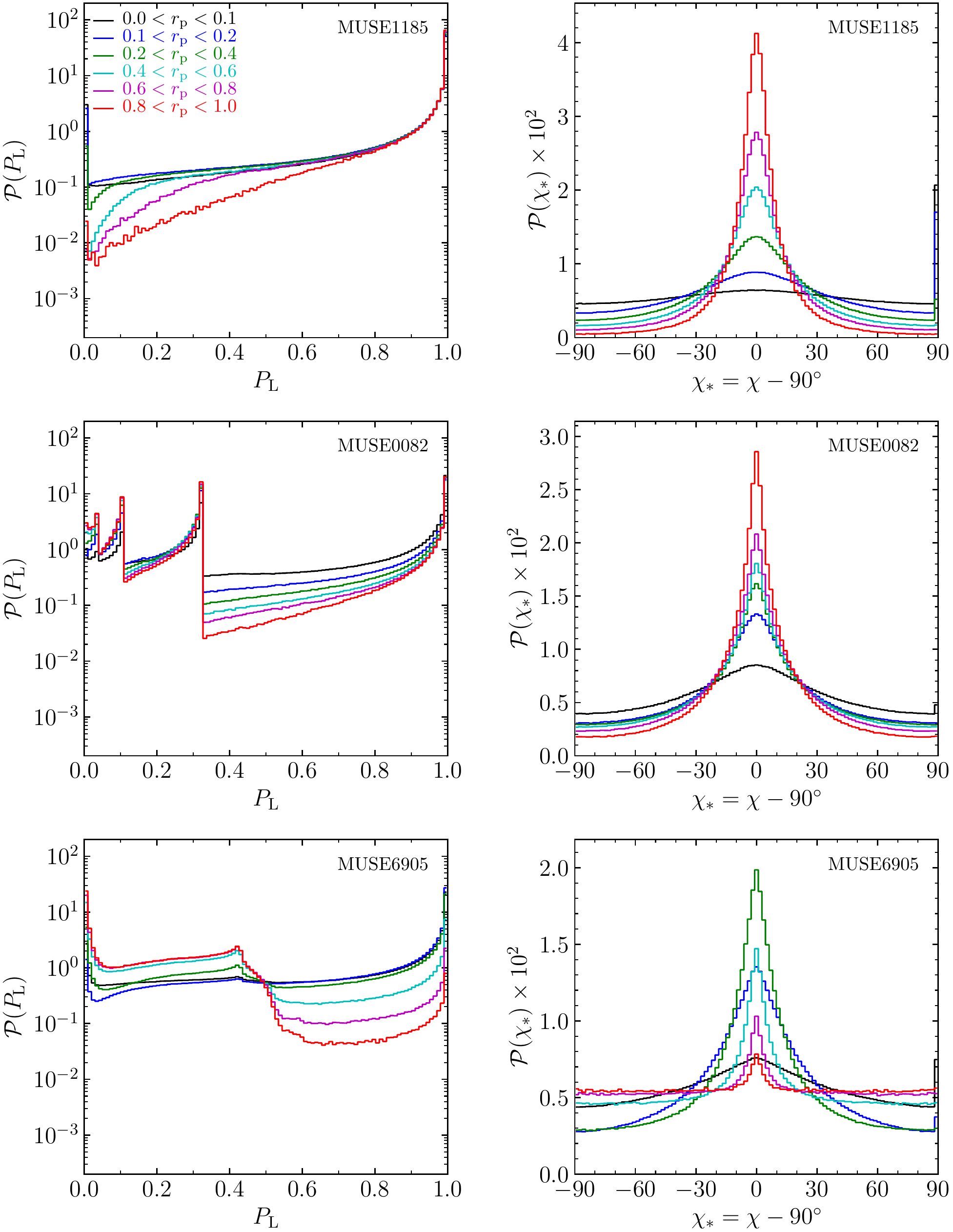}
\par\end{centering}
\begin{centering}
\medskip{}
\par\end{centering}
\caption{\label{fig24}Probability density functions of $P_{{\rm L}}$ and
$\chi_{*}$ of individual photons for three LAEs (MUSE 1185, 0082,
and 6905), obtained using a momentum-driven galactic wind model.}

\centering{}\medskip{}
\end{figure}

The first and second columns in Figures \ref{fig22} and \ref{fig23}
compare the observational spectra and surface brightness profiles
with those obtained by simulations for the LAEs. The output spectra
and surface brightness profiles calculated in this study are confirmed
to reproduce those of \citet{2020ApJ...901...41S} well. We also present
the polarization profiles predicted from the best-fit models for each
LAE in the third and fourth columns of the figures; the third column
shows the radial profile of polarization degree, and the fourth column
the degree of polarization as a function of wavelength and projected
radius. To compare with the observational data, the calculated spectra
and surface brightness/polarization profiles were obtained after applying
the aperture sizes and wavelength ranges described in \citet{2017A&A...608A...8L}
and \citet{2020ApJ...901...41S}. The radial profiles were also convolved
with the MUSE point spread function (PSF). In the third column, the
polarization profiles obtained using all photons, including those
having wavelengths outside the wavelength boundaries, are also shown
in dotted lines.

In the figures, it is noteworthy that the polarization profile is
fairly diverse in shape. In previous models, the polarization level
monotonically increased with radial distance, even though the rate
of increase (or slope) was dependent on model parameters. However,
in the models of MUSE 6905, 0547, and 0364, the radial profile of
polarization is found not to be a simple monotonic function. In particular,
the model of MUSE 6905 shows an interesting feature, even after the
convolution with the instrumental PSF. The polarization profiles for
MUSE 0082, 0171, 0547, and 0364 show a rapid jump near $r\approx0$,
although the convolution by the instrumental PSF smoothes out the
profiles, erasing the jump shapes. The third column also demonstrates
that the polarization profile can be significantly altered, at least
in some cases (MUSE 6905 and 0364), depending on the wavelength range
adopted to measure the polarization signal.

The polarization property of individual photons is too complex to
interpret, unlike the other models. Detailed analysis on the variation
of polarization pattern depending on the model parameters is beyond
of the scope of this paper. However, it is clear that the polarization
profile depends on the kinematic properties of the galactic wind,
as shown in Figure \ref{eq:24}. The distribution functions of $P_{{\rm L}}$
and $\chi_{*}$ for MUSE 1185 are somewhat similar to those obtained
for the Hubble-like flow model. MUSE 0082 shows several spiky peaks,
whose cause is not clear, in the distribution function of $P_{{\rm L}}$.
The distribution functions of $\chi_{*}$ for MUSE 1185 and 0082 indicate
that the radiation field becomes more anisotropic as the radius increases,
yieding the rise of polarization.

The distribution function of $P_{{\rm L}}$ for MUSE 6905 reveals
a more or less similar trend to those for the static models of $T=10^{4}$
K and $\tau_{0}\lesssim10^{5}$ (Figure \ref{fig11}), except that
the static models barely show the dependence on radius. However, unlike
the static model, the degree of isotropy of the radiation field shows
a non-monotonic trend, decreasing and then later increasing with increasing
radius, as shown in the distribution of $\chi_{*}$. This trend in
the model of MUSE 6905 is primarily due to the velocity profile of
outflow; the outflow monotonically decelerates, starting with the
maximum velocity of $V_{{\rm exp}}=300$ km s$^{-1}$ at the center,
as the radius increases. However, except for MUSE 0171, the other
outflow models accelerate, starting at rest, until a radius ($r_{{\rm peak}}>0$)
and then decelerate. The difference between the models for MUSE 0171
and 6905 is that the outflow of MUSE 0171 expands at a constant speed,
while that of MUSE 6905 monotonically decelerates with increasing
radius. The peculiarity in the polarization pattern of MUSE 6905 is
attributable to this monotonically decelerating velocity profile.
Ly$\alpha$ photons are recognized as wing photons at small radii
because of the high velocity there. However, the wing photons that
gained high polarization at small radii will be identified to be in
the core regime when they reach large radii. Then, their polarization
levels will be reduced by the core scattering at large radii, as can
be seen in the distribution function of $P_{{\rm L}}$. Furthermore,
the radiation field becomes highly isotropic due to many scatterings,
as shown in the distribution function of $\chi_{*}$. Eventually,
the ensemble-averaged polarization level decreases at large radii.
Note that the model of \citet{2020ApJ...901...41S} is a simplified
version to avoid an exceedingly abrupt jump in outflow speed. Therefore,
the model of MUSE 6905 may be applicable to the systems where a rapid
outflow is launched near the galactic center and then decelerate monotonically.

\section{DISCUSSION}

\label{sec:5}

The energy levels of hydrogen are splitted not only by the coupling
between the electon wavefunctions but also by the interference between
the electron and proton wavefunctions. In the present paper, we considered
only the fine structure of hydrogen atom. The hyperfine structure
was taken into account in \citet{1990Icar...84..106C}, \citet{1998A&A...332..732B},
and \citet{2006MNRAS.367..259H}. It is relatively straightforward
to extend the present RT algorithm to include the hyperfine structure.
However, the present scope is sufficient for the polarization RT in
most cases interested in astrophysics.

In our study, we used the quasi-monochromatic photon packets that
represent superpositions of mutually independent, incoherent waves.
We found that Ly$\alpha$ photon packets tend to be fully polarized,
unlike the dust scattering case, when repeatedly scattered. This result
indicates that the quasi-monochromatic Ly$\alpha$ light that has
undergone the same scattering history will be fully polarized as if
they are purely monochromatic waves. In the following, we further
discuss the tendency of photon packets to get 100\% polarized, together
with other topics relevant to Ly$\alpha$ but not addressed or fully
discussed in the previous sections. The topics include the distribution
function of the Stokes parameter $Q$, an implication of the correlation
between the surface brightness profile and the polarization profile
to the observations of the polarization of LABs, the negative polarization,
and a sudden jump of polarization. We also discuss possibilities to
break the degeneracies between different models of LAEs using the
polarization data, and then a misleading concept relevant to the peeling-off
technique.

\subsection{Arise of 100\% Polarization of Ly$\alpha$ by Scatterings}

\label{section:5.1}

In the above sections, we noticed that ``individual'' Ly$\alpha$
photon packets gradually approach an almost 100\% polarized state
when they are repeatedly scattered. Here, we discuss in more detail
how this process can occur. We first need to note that the degree
of polarization is preserved under a rotation of the axes about the
photon\textquoteright s direction vector. Therefore, it is convenient
to define the Stokes vector in a reference frame where the polarization
basis vector $\mathbf{m}$ lies in the scattering plane (i.e., $\phi=0$).
In that coordinate system, the Stokes parameters of scattered light
are given by:
\begin{align}
I' & =S_{11}I+S_{12}Q,\nonumber \\
Q' & =S_{12}I+S_{22}Q,\label{eq:51}\\
U' & =S_{33}U.\nonumber 
\end{align}
Now, we can readily show that
\begin{align}
I'^{2}-Q'^{2}-U'^{2} & =\left(\frac{3}{2}E_{1}\cos\theta\right)^{2}\left(I^{2}-Q^{2}-U^{2}\right)\nonumber \\
 & \ \ \ \ +\left[E_{2}^{2}+\frac{3}{2}E_{1}E_{2}\left(\cos^{2}\theta+1\right)\right]I^{2}\nonumber \\
 & \ \ \ \ +\frac{3}{2}E_{1}E_{2}\left(\cos^{2}\theta-1\right)IQ.\label{eq:52}
\end{align}
Here, $I^{2}-Q^{2}-U^{2}$ denotes the intensity of unpolarized light.
The above equation indicates that isotropic scattering ($E_{1}=0$)
always completely depolarizes photons ($Q'=U'=0$) regardless of their
initial polarization states.\footnote{This conclusion can more easily be obtained using Equations (\ref{eq:19})
and (\ref{eq:21}) by setting $E_{1}=0$.} It can also be proven that if the incident light was 100\% polarized,
the scattered light is 100\% polarized if and only if $E_{1}=1$ (Rayleigh
scattering).

We note that Ly$\alpha$ photons diffuse out in frequency space while
being scattered, approaching the wing part of the line profile. If
once the photon reaches the wing, where $E_{1}=1$, the above equation
is reduced to the following:

\begin{align}
I'^{2}-Q'^{2}-U'^{2} & =\left(\frac{3}{2}\cos\theta\right)^{2}\left(I^{2}-Q^{2}-U^{2}\right).\label{eq:53}
\end{align}
Note that the factor ($9/4\cos^{2}\theta$) in the right-hand side
results in 0.9 when multiplied by the scattering phase function for
Rayleigh scattering and averaged over the scattering angle. This indicates
that the intensity of unpolarized light gradually decreases; in other
words, the degree of polarization will gradually rise while photon
packets are being Rayleigh scattered. 

Moreover, we can immediately recognize that if a photon packet in
the wing regime is scattered at the ``right'' angle ($\cos\theta=0$),
the photon will be 100\% polarized regardless of its initial polarization
state. After then, the photon will keep the polarization state of
$\sim$100\%, independently of the scattering angle, unless it happens
to be scattered back into the core regime. In this stage, the photon
is highly likely to escape the system in a single long-excursion while
trapped in the wing \citep{1972ApJ...174..439A}; consequently, the
individual photon packets escaping in the wings will have a very high
polarization degree of \ensuremath{\sim}100\%. In a static medium,
photons need a lot of scatterings to diffuse into the wings. On the
other hand, in a fast-moving medium, photons will be scattered into
the wings relatively easily with only a small number of scatterings.
Therefore, Ly$\alpha$ photon packets will approach a 100\%-polarized
state in an outflowing or infalling medium quickly, unless the gas
motion is too fast.

It is also expected that individual photon packets of broad Ly$\alpha$
lines, as observed in AGNs, could readily be fully polarized because
most of the line photons will be wing-scattered even in static media.
At the same time, the depolarization effect due to the core scattering
may play essential roles in the broad lines if the velocity gradient
is considerable in the outflow or inflowing medium.

\begin{figure}[t]
\begin{centering}
\includegraphics[clip,scale=0.53]{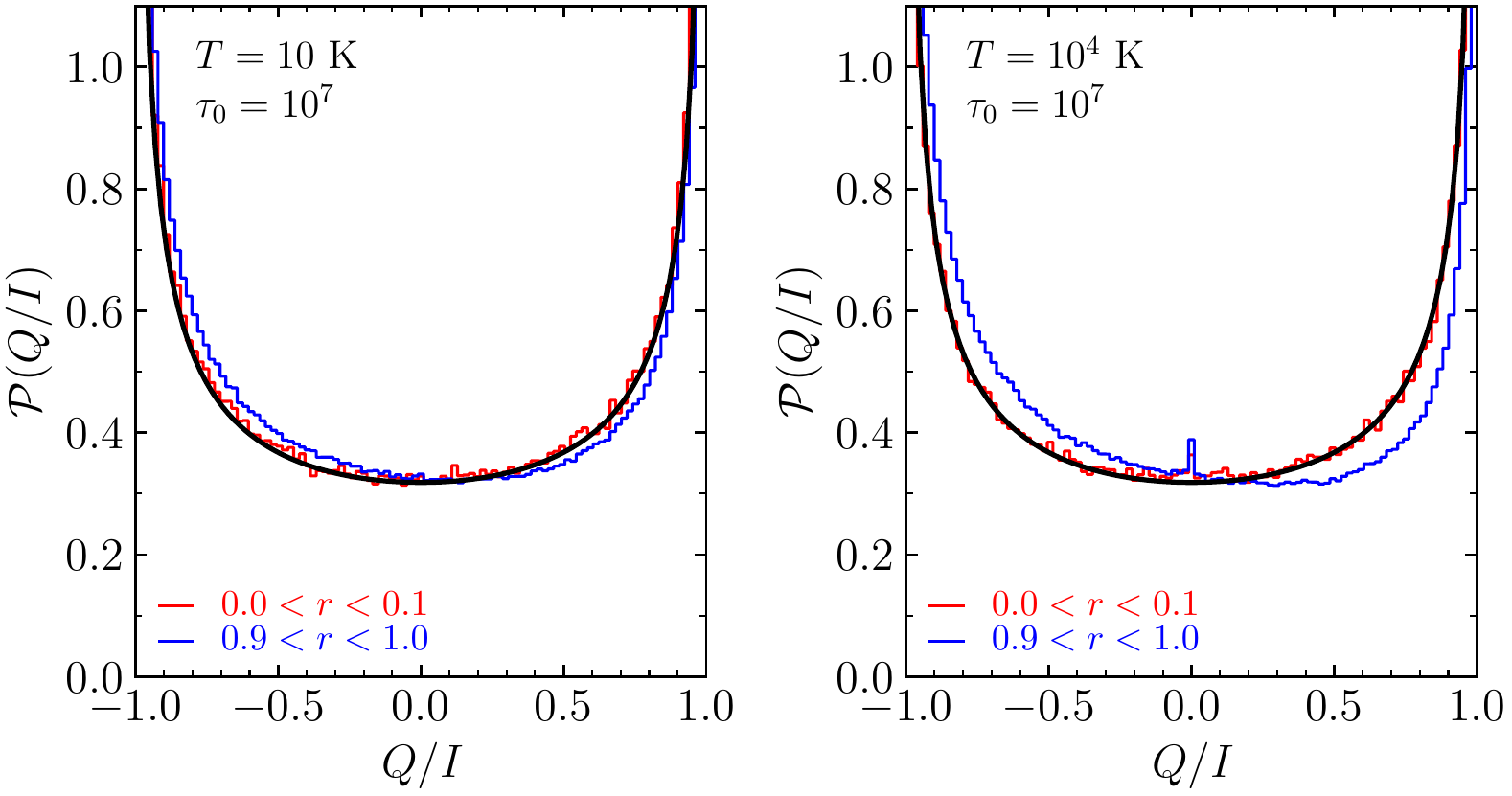}
\par\end{centering}
\begin{centering}
\medskip{}
\par\end{centering}
\caption{\label{fig25}Probability distribution function of $Q/I$ for the
static, homogeneous sphere model. The gas temperature is $T=10$ K
in the left panel and $T=10^{4}$ K in the right panel. The optical
depth for both panels is $\tau_{0}=10^{7}$. The red and blue curves
are the distribution functions obtained in the radial bins $0.1<r<0.1$
and $0.9<r<1.0$, respectively. The black curve denotes the theoretical
distribution function of Equation (\ref{eq:55}).}

\centering{}\medskip{}
\end{figure}

\subsection{Distribution Function of the Stokes Parameters $Q$}

It is evident that, at the center ($r=0$) of a sphere, the polarization
angle of individual photons will be uniformly distributed because
of the isotropy of the radiation field. Our results also indicate
that the distribution function of the Stokes parameter $Q$, measured
at the central region, is symmetric. As the optical depth and number
of scatterings increase, the polarization of individual photon packets
approaches 100\%. At the same time, the radiation field tends to be
isotropic, particularly even up to relatively large radii in static
media. In this case, it can be found that a simple analytic function
describes the distribution function of $Q$.

To derive the probability distribution function for $Q$, we assume
an isotropic radiation field and that the polarization angle $\chi$
is uniformly distributed in a range of $0\le\chi\le2\pi$, as given
by:
\begin{equation}
\mathcal{P}(\chi)=\frac{1}{2\pi}\ \ \ \left(0\le\chi\le2\pi\right).\label{eq:54}
\end{equation}
From Equation (\ref{eq:12}), we obtain the distribution function
of the normalized Stokes parameter $Q/I$, as follows:
\begin{align}
\mathcal{P}\left(Q/I\right) & =4\mathcal{P}\left(\chi\right)\left|\frac{d\chi}{d\left(Q/I\right)}\right|=\frac{2\mathcal{P}\left(\chi\right)}{\sqrt{1-\left(Q/I\right)^{2}}}\nonumber \\
 & =\frac{1}{\pi\sqrt{1-\left(Q/I\right)^{2}}}.\label{eq:55}
\end{align}
Here, the coefficient ``4'' is adopted to take into account the fact
that $Q/I$ has the same value four times in the range $0\le\chi\le2\pi$.
The obtained distribution function of $Q$ has two peaks at $Q=-1$
(concentric polarization) and $Q=+1$ (radial polarization). It is
evidently far from a Gaussian or Poisson function that is generally
adopted to analyze the intensity $I$. We also note that the Stokes
parameter $U/I$ follows the same distribution function. As we already
noted, the individual photon packets tend to achieve $\sim100$\%
polarization and thus the distribution function of $Q$ (and $U$)
is well approximated by the distribution function of $Q/I$ (and $U/I$).
In other words, $\mathcal{P}(Q)\approx\mathcal{P}(Q/I)$. Figure \ref{fig25}
shows that the theoretical distribution function derived above indeed
well reproduces those obtained in the static sphere model, especially
in the central region. We note that the distribution function of $Q$
appears to be reasonably well described, even near the boundary (r
\ensuremath{\approx} 1), by the same function. This probability density
function can be a starting point to understand the statistical properties
of the Stokes parameters $Q$ and $U$ in actual observations.

We note that similar shapes with two peaks at $Q=\pm1$ are also commonly
found in the distribution functions of $Q$ of Figures \ref{fig15}
and \ref{fig16}. The relative strengths of the two peaks are closely
associated with the degree of the isotropy of the radiation field.
The peak at $Q=1$ weakens while that of $Q=-1$ becomes stronger
as the gas motion boosts up the anisotropy of the radiation field.
Related to this point, it is noteworthy that the polarization profiles
observed in LABs rise somewhat slowly with radial distance from the
center \citep{2011Natur.476..304H,2017ApJ...834..182Y,2020ApJ...894...33K}.
This trend suggests that the Ly$\alpha$ radiation field in the LABs
might be more or less isotropic, and the distribution function of
$Q$ may have a relatively strong peak at $Q=1$. Such a peak in $\mathcal{P}(Q)$
at $Q=1$ would make the observed polarization signals noisy.

\subsection{An Implication to the Polarization Pattern in LAB1}

In our results, we noticed that there exists a correlation between
the surface brightness profile and the polarization profile. In a
static spherical model, as the optical depth increases, the overall
shape of the surface brightness profile flattens out and, at the same
time, the polarization profile becomes shallower near $r=0$. This
result is associated with the tendency that as the optical depth increases,
the Ly$\alpha$ radiation field becomes more isotropic and the polarization
vectors of near-100\%-polarized individual photons largely cancels
out. A similar correlation was also found in a Hubble-like expanding
modium. In this case, however, the polarization levels of individual
photons are primarily boosted up by the Doppler shift and a faster
medium gives rise to a steeply declining surface brightness profile
and a rapidly rising polarization pattern.

In the well-studied Ly$\alpha$ blob LAB1, the degree of polarization
is observed to increase relatively slowly with radius. \citet{2016A&A...593A.122T}
utilized a hydrodynamic simulation in the context of the gravitational
cooling and presented radial polarization patterns predicted by assuming
central sources and diffuse ionized gas surrounding the sources. The
polarization pattern for the central sources was found to be relatively
steep, which is more or less similar to those obtained in fastly expanding
media discussed in this paper. On the other hand, the polarization
obtained using the in-situ diffuse emission gas appeared to show a
slowly varying pattern. Based on these results, they showed that the
observed polarization in LAB1 is consistent with the pattern predicted
by the circumgalactic diffuse gas or by the combination of two source
types.

However, it is worth noting that the observed surface brightness profile
in \citet{2011Natur.476..304H} is shallower than that predicted by
\citet{2016A&A...593A.122T}. Shallow surface brightness profiles
are expected for fast-moving gases or media with small optical depths.
In that sense, their hydrodynamic simulation seems to predict a somewhat
too fast gas motion or a too-small optical depth. Recalling that a
slow rise in polarization usually accompanies a shallower surface
brightness profile, the observed polarization and surface brightness
patterns accord well, at least qualitatively, with the central engine
scenario in which Ly\textgreek{a} photons originating from central
sources are scattered by the hydrogen gas surrounding the sources
yielding the Ly$\alpha$ blob. Quantitative comparisons of the observational
data with more extensive theoretical calculations, taking into account
inhomogeneous and clumpy media, would be required to understand the
origin of these polarization signals.

\subsection{The Negative Polarization and Polarization Jump}

LaRT is superb, compared to the preexisting codes, in that it uses
a smoothly and seamlessly varying phase function as frequency changes.
Our approach provides a theoretically complete framework in dealing
with the polarization signals of resonance doublet lines with the
same fine structure as Ly$\alpha$. The other approaches that use
a discrete phase function or 100\%-polarized photons cannot reproduce
the ``negative'' polarization signals perpendicular to the radial
direction. Regarding the Ly$\alpha$ RT, however, our approach appears
to yield no significant difference from the preexisting approaches;
this is because the number of scatterings of Ly$\alpha$ is enormous,
and the frequency gap between the fine structure is very narrow so
that the chance of being negatively polarized is rare, as described
in Section \ref{subsec:2.5}. The present method would be advantageous
in studying the resonance lines that have the same fine structure
as Ly$\alpha$ but relatively low optical depths, such as \ion{C}{4}
$\lambda\lambda1548,1550$, \ion{Mg}{2} $\lambda\lambda2795,2803$,
and \ion{Ca}{2} $\lambda\lambda3934,3968$ doublets. The negative
polarization was indeed observed in the \ion{Ca}{2} doublet line
in the Sun \citep[e.g.,][]{1980A&A....84...68S}.

In Sections \ref{section:4.1} and \ref{section:4.2}, we discovered
an abrupt jump in the polarization level at the center of the spherical
models. However, it should be noted that the polarization jump may
not be readily observable in actual observations. No galaxies are
perfectly symmetric, and currently available telescopes do not seem
to have sufficient spatial resolutions to resolve the abrupt change
in the polarization pattern. The effect of the telescope PSF is demonstrated
for the models of LAEs in Section \ref{section:4.4}. Figures \ref{fig22}
and \ref{fig23} show a rapid jump near $r\approx0$ in the polarization
profiles of MUSE 0082, 0171, 0547, and 0364. However, these jump features
are found to disappear after the convolution with the telescope PSF
mostly. Asymmetries in the Ly$\alpha$ source and medium may also
blur the central polarization profile and thus significantly reduce
the jump level. Clumpiness in the medium might also tend to smooth
out the abrupt change in the polarization pattern.

\subsection{Breaking the Degeneracies between Models of LAEs}

\citet{2020ApJ...901...41S} demonstrated the existence of degeneracies
between different models of LAEs when analyzing only either the spectrum
or the surface brightness profile, and at least some of them could
be broken by utilizing both the spectrum and surface brightness profile
simultaneously. For instance, analyzing only the surface brightness
profile was found to show degeneracies of $\Delta V-rs_{{\rm HI}}$
and $\tau_{0}-V_{{\rm peak}}$ for most galaxies. The output spectrum
was revealed to be relatively independent of the scale length of the
gas density to a large extent. Analysis of the polarization signal,
together with the spectrum and surface brightness profile, may provide
additional clues, which cannot be revealed by analyzing only spectra
and surface brightness profiles, and help us gain a deeper understanding
of the Ly\textgreek{a} galaxies and blobs \citep{2018ApJ...856..156E}.
As discussed in Section \ref{section:4.2}, the variation in the polarization
properties with the outflow speed demonstrates the importance of the
spatially-resolved spectropolarimetric observations in understanding
the kinematics of the galactic halos.

The galactic halo model in Section \ref{section:4.4} illustrates
the diversity of the polarization patterns. Therefore, a more detailed
understanding of the polarization patterns may help us better constrain
the galactic halos' kinematic properties. For this purpose, it would
be worth investigating how the polarization pattern changes by varying
each model parameter while keeping other parameters fixed, as done
for the spectrum and surface brightness profile in \citet{2020ApJ...901...41S}.
However, we would like to defer such a detailed analysis to a future
study.

\subsection{A Note on the Peeling-Off Technique}

We here would like to clarify a misconception with regards to the
peeling-off technique. \citet{2016MNRAS.459.1710M} and \citet{2019SAAS...46....1D}
claim that one has to subtract the peeled-off fraction from the photon's
weight whenever a peeling-off procedure is performed. \citet{2019SAAS...46....1D}
also states that the portion to be subtracted is too small to affect
the simulation result, and thus one can, in practice, safely ignore
the subtraction. However, in principle, the peeled-off fraction must
not be subtracted from the photon's weight.

The basic idea behind this technique is to calculate the number (or
probability) of experiments (realizations) in which the photon is
scattered towards the observer if a sufficiently large number of identical
experiments have been performed. Suppose that we trace the trajectory
of a single photon and perform a countless number of Monte Carlo experiments.
We construct an ensemble made up of an almost infinite number of Monte
Carlo experiments performed under identical conditions and record
the whole history of every (emission and scattering) event in each
experiment. Next, we choose a subset of the ensemble, which is composed
of experiments where the photon's history is exactly the same until
a specific time $t_{0}$. Performing a scattering of the photon toward
a randomly chosen direction in the next simulation step ($t_{0}+\delta t$)
is equivalent to picking a particular experiment (a ``realization'')
among the subset by drawing a random number. Similarly, performing
a peeling-off procedure is simply to count the number of experiments
in which the photon is scattered toward the detector. Therefore, the
peeling-off procedure is nothing to do with performing the next scattering
of the photon by drawing a random scattering direction for a specific
simulation realization. Thus the photon weight should not be altered
after the peeling-off procedure.

In this sense, the term ``peeling-off'' may be misleading. The \textquotedblleft next
event estimation\textquotedblright{} would be a better term, although
it is unfamiliar in the astrophysics community. We also note that
if one performs the peeling-off technique toward a detector plane,
they should not count the photons that happen to be actually detected
in the detector plane by chance. However, fortunately, even if one
adopts the wrong method, it is unlikely to affect the result in most
cases significantly because the peeled-off fraction would be tiny.

\section{SUMMARY}

\label{sec:6}

This paper describes an RT method using the Monte Carlo technique
to investigate the Ly$\alpha$ polarization. Our results demonstrate
that the spectropolarimetry data can be used to supplement the spectrum
and surface brightness profile of the LAEs and LABs to tightly constrain
the physical origin of Ly$\alpha$ halos. Furthermore, spatially-resolved
spectropolarimetry will provide a deep understanding of the detailed
nature of Ly$\alpha$ emitting sources. For this purpose, we need
to not only take observations for many different objects but also
develop theoretical models for various possible situations. This paper
provides mainly a primary tool and fundamental results from the theoretical
point of view. We expect more detailed models will be developed in
future works.

The principal conclusions of this paper are summarized as follows:
\begin{itemize}
\item The degrees of polarization of individual photon packets increase
while they are scattered repeatedly, approaching the wing regime.
If once the photon packets diffuse out to red or blue wavelengths
in the Rayleigh scattering regime, they can easily achieve 100\% polarization
and their polarization level is persistent until they escape, unless
they are scattered into the core of the line profile.
\item In outflowing or contracting media, the polarization degree of individual
photon packets rises fairly quickly by the Doppler shift effect.
\item The ``ensemble-averaged'' degree of polarization, the observable quantity,
is primarily determined by (1) the polarization levels of individual
photon packets and (2) the degree of the isotropy (or anisotropy)
of the Ly$\alpha$ radiation field. They are both eventually controlled
by the number of scatterings and the Doppler boosting due to the gas
motion. In most cases, the polarization degree of individual photons
approaches 100\%, unless the medium's optical depth is too low.
\item There exists a correlation between the surface brightness profile
and the polarization profile. This is because a shallow surface brightness
profile is associated with a relatively isotropic radiation field,
while a steeply declining profile is due to an anisotropic radiation
field. At the center of spherical media, the polarization vectors
are canceled out due to the symmetry. If the surface brightness profile
is steep, the polarization immediately survives as soon as the isotropy
of the radiation field is broken at a location in the vicinity of
the source, yielding a rapidly rising polarization profile. On the
other hand, for a shallow surface brightness profile, the radiation
field is somewhat isotropic even at a location far from the source,
so the polarization vector is significantly canceled there, leading
to a slowly varying polarization pattern.
\item The correlation between the radial profiles of the surface brightness
and polarization can help interpret the observed data of LABs. The
relatively shallow profiles of the surface brightness and polarization
observed in the Ly$\alpha$ blob LAB1 seem to support the central
engine scenario of the Ly$\alpha$ halo pretty well. However, further
quantitative studies would be required to understand better the origin
of the observed polarization patterns of LABs.
\item The polarization angle of individual photon packets can be regarded
as a proxy to measure the degree of isotropy of the radiation field.
Assuming a uniform distribution for the polarization angle, we can
derive a distribution function for the Stokes parameter $Q$, which
may be useful in understanding the statistics of observational data.
\item Dust grains slightly steepen the surface brightness and increase the
degree of polarization by selectively destroying the core photons
that hold relatively low amplitudes of polarization. The effect appears
more prominent in a medium of a higher temperature.
\item We also found that the polarization profile can exhibit a non-monotonically
increasing pattern with radius in some models developed for LAEs in
\citet{2020ApJ...901...41S}. The unusual profiles are primarily due
to the depolarization by the core scattering in a decelerating medium.
Hence, the spatially-resolved polarization observations may play a
critical role in understanding the galactic outflow or inflow by providing
complementary information on the kinematics of the galactic halo gas,
in addition to the spectrum and surface brightness profile.
\end{itemize}

\appendix{}

\section{Scattering Matrix for Ly$\alpha$ Line Scattering}

\begin{figure}[t]
\begin{centering}
\medskip{}
\includegraphics[clip,scale=0.45]{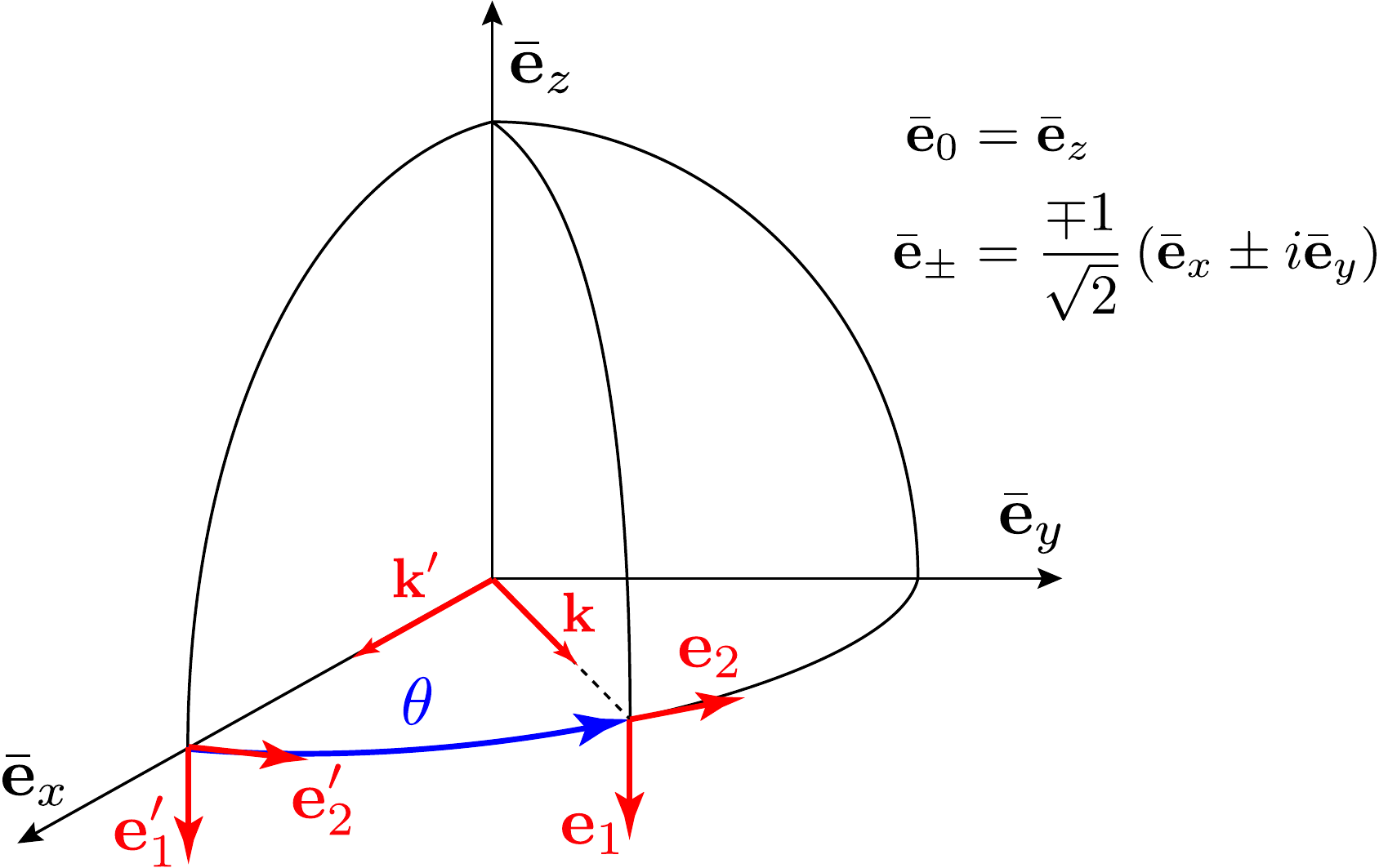}
\par\end{centering}
\begin{centering}
\medskip{}
\par\end{centering}
\caption{\label{fig_appA1}Coordinate system and polarization basis vector
to calculate the scattering matrix of Ly$\alpha$. The spherical unit
vectors ($\bar{\mathbf{e}}_{0}$, $\bar{\mathbf{e}}_{\pm}$) are defined
in a fixed reference frame (determined by the scattering plane and
the initial propagation vector) and the polarization basis vectors
($\mathbf{e}_{1}$, $\mathbf{e}_{2}$) in the photon's local frame.
The propagation vectors of the incident and scattered photon are denoted
by $\mathbf{k}'$ and $\mathbf{k}$, respectively. The primed variable
refers to the incident photon but the non-primed the scattered photon.
See Figure 3.3 of \citet{1994ASSL..189.....S} for a more general
definition. For simplicity, we choose the angles defined in \citet{1994ASSL..189.....S}
to be $\theta_{{\rm S94}}=\theta_{{\rm S94}}'=90^{\circ}$, $\phi_{{\rm S94}}=0$,
and $\phi_{{\rm S94}}'=\theta$.}
\medskip{}
\end{figure}

\label{sec:scattering_matrix_for_hydrogen}

\citet{1980A&A....84...68S} and \citet{1994ASSL..189.....S} provide
the formulae for the scattering matrix for the Ly$\alpha$ scattering
by hydrogen atoms, but without the term for circular polarization.
We, therefore, rederived the formulae, including the circular polarization
term. The time reversal term of the Quantum Field theory, which is
negligible, will be ignored in the Kramers-Heisenberg dispersion formula.
\citet{1994ASSL..189.....S} adopts the convention of $\mathcal{E}_{1}=E_{n}$
and $\mathcal{E}_{2}=-E_{m}$ and defines the Stokes parameters as
in Equation (\ref{eq:11}), but replacing ($\mathcal{E}_{1}$, $\mathcal{E}_{2}$)
by ($E_{m}$, $E_{n}$). They define the scattering plane as the $y-z$
plane (in the photon's local frame), as shown in Figure \ref{fig_appA1},
while we define it as the $x-z$ plane. They define the handedness
of circular polarization in the opposite direction. Therefore, their
Stokes parameters are expressed as $(I_{{\rm S94}},\ Q_{{\rm S94}},\ U_{{\rm S94}},\ V_{{\rm S94}})=(I_{{\rm IAU}},\ -Q_{{\rm IAU}},\ -U_{{\rm IAU}},\ -V_{{\rm IAU}})$.
The scattering matrix elements are then the same as those obtained
for the IAU standard except for only one; $M_{12}^{{\rm IAU}}=-M_{12}^{{\rm S94}}$
for $(i,j)=(1,2)$ and $M_{ij}^{{\rm IAU}}=M_{ij}^{{\rm S94}}$ otherwise.
We follow the definition of \citet{1994ASSL..189.....S} to avoid
confusion when deriving the scattering matrix elements and then later
transform the obtained results to the IAU standard.

The 4-dimensional coherency vector is defined by
\begin{equation}
\mathbf{D}^{(4)}=\left(\begin{array}{c}
D_{11}\\
D_{12}\\
D_{21}\\
D_{22}
\end{array}\right)=\left(\begin{array}{c}
\mathcal{E}_{1}\mathcal{E}_{1}^{*}\\
\mathcal{E}_{1}\mathcal{E}_{2}^{*}\\
\mathcal{E}_{2}\mathcal{E}_{1}^{*}\\
\mathcal{E}_{2}\mathcal{E}_{2}^{*}
\end{array}\right)=\frac{1}{2}\left(\begin{array}{c}
I+Q\\
U+iV\\
U-iV\\
I-Q
\end{array}\right)^{{\rm S94}},\label{eq:appA1}
\end{equation}
where the superscript ``S94'' indicates the definition of \citet{1994ASSL..189.....S}.
The Stokes vector is related to the coherency vector by
\begin{align}
\mathbf{S} & =\mathbf{T}\mathbf{D}^{(4)},\nonumber \\
\mathbf{T} & \equiv\left(\begin{array}{cccc}
1 & 0 & 0 & 1\\
1 & 0 & 0 & -1\\
0 & 1 & 1 & 0\\
0 & -i & i & 0
\end{array}\right),\label{eq:appA2}
\end{align}
where $\mathbf{T}$ is the matrix to transform the 4-dimensional coherency
vector to the Stokes vector. The scattering matrix for the 4-dimensional
coherency vector $\mathbf{D}^{(4)}$ is given by Equation (2.39) and
(9.23) in \citet{1994ASSL..189.....S}, as follows:
\begin{align}
\mathbf{W} & =\left(\begin{array}{cccc}
W_{11} & 0 & 0 & W_{14}\\
0 & W_{22} & W_{23} & 0\\
0 & W_{23} & W_{22} & 0\\
W_{14} & 0 & 0 & W_{44}
\end{array}\right)\nonumber \\
 & =\left(\begin{array}{cccc}
w_{11}w_{11}^{*} & 0 & 0 & w_{12}w_{12}^{*}\\
0 & w_{11}w_{22}^{*} & w_{12}w_{21}^{*} & 0\\
0 & w_{21}w_{12}^{*} & w_{22}w_{11}^{*} & 0\\
w_{21}w_{21}^{*} & 0 & 0 & w_{22}w_{22}^{*}
\end{array}\right),\label{eq:appA3}
\end{align}
where other matrix elements vanish because $\mu_{i}=\mu_{f}$ and
$\mu_{i}=\mu_{f}\pm1$ cannot be simultaneously satisfied. The scattering
amplitude is given by Equation (8.115) of \citet{1994ASSL..189.....S}
(see also \citet{1998A&A...338..301S}):
\begin{align}
w_{\alpha\beta} & =\sum_{J_{m},\ \mu_{m}}\left(-1\right)^{r_{im}+r_{fm}}\left(2J_{i}+1\right)^{1/2}\left(2J_{f}+1\right)^{1/2}\left(f_{J_{i}J_{m}}f_{J_{f}J_{m}}\right)^{1/2}\nonumber \\
 & \ \ \ \ \ \ \ \ \ \ \times\left(\begin{array}{ccc}
J_{m} & J_{i} & 1\\
-\mu_{m} & \mu_{i} & \mu_{m}-\mu_{i}
\end{array}\right)\left(\begin{array}{ccc}
J_{m} & J_{f} & 1\\
-\mu_{m} & \mu_{f} & \mu_{m}-\mu_{f}
\end{array}\right)\nonumber \\
 & \ \ \ \ \ \ \ \ \ \ \times\left(-1\right)^{\mu_{m}-\mu_{i}}\varepsilon_{\mu_{i}-\mu_{m}}^{\beta'}\varepsilon_{\mu_{m}-\mu_{f}}^{\alpha}\Phi_{mf}.\label{eq:appA4}
\end{align}
Here, the $2\times3$ matrices are the Wigner 3-$j$ symbols, $J$
and $\mu$ indicate the total angular momentum quantum number and
magnetic quantum number, $i$ and $f$ denote the initial and final
levels (both are lower levels), and $m$ represents the intermediate
excited levels (upper levels). $f_{J_{i}J_{m}}$ is the oscillator
strength between the lower and upper (intermediate) levels. The exponents
$r_{im}$ and $r_{fm}$ determine the sign of each term in the summation
and are given in \citet{1994ASSL..189.....S,1997A&A...324..344S}.
For Ly$\alpha$ scattering, $J_{i}=J_{f}$ and $\left(-1\right)^{r_{im}+r_{fm}}=1$.

The components of the polarization basis vectors ($\mathbf{e}_{1}$
and $\mathbf{e}_{2}$) represented in terms of the ``complex'' spherical
unit vectors, $\bar{\mathbf{e}}_{0}=\mathbf{e}_{z}$ and $\bar{\mathbf{e}}_{\pm1}=\mp\left(\bar{\mathbf{e}}_{x}\pm i\bar{\mathbf{e}}_{y}\right)/\sqrt{2}$,
are defined by

\begin{align}
\varepsilon_{q}^{\alpha} & =\mathbf{e}_{\alpha}\cdot\bar{\mathbf{e}}_{q},\nonumber \\
\varepsilon_{q}^{\alpha'} & =\mathbf{e}_{\alpha}'\cdot\bar{\mathbf{e}}_{q}\label{eq:appA5}
\end{align}
for $\alpha=1$, 2 and $q=0,$ $\pm1$. Here, the primed quantities,
as in $\varepsilon_{q}^{\alpha'}$, refer to the incident photon and
the unprimed quantities the scattered photon. To derive explicit expressions
for $w_{\alpha\beta}$, we adopt a coordinate system and basis vectors
similar to those of \citet{1994ASSL..189.....S}, which are shown
in Figure \ref{fig_appA1}. Assuming that the photon is incident in
the $x$-direction and that the scattering takes place in the $x-y$
plane into a scattering angle of $\theta$, the spherical vector components
of the polarization vectors of the incident ($\varepsilon'$) and
scattered ($\varepsilon$) photons are given by
\begin{align}
\varepsilon_{0}^{1} & =\varepsilon_{0}^{1'}=-1\nonumber \\
\varepsilon_{0}^{2} & =\varepsilon_{0}^{2'}=\varepsilon_{\pm1}^{1}=\varepsilon_{\pm1}^{1'}=0\nonumber \\
\varepsilon_{\pm1}^{2} & =\frac{-ie^{\pm i\theta}}{\sqrt{2}}\nonumber \\
\varepsilon_{\pm1}^{2'} & =\frac{-i}{\sqrt{2}}.\label{eq:appA6}
\end{align}

The line profile, in terms of frequency $\nu$, of Ly$\alpha$ is
given by
\begin{equation}
\Phi_{mf}\propto\frac{1}{\nu-\nu_{mf}+i\left(\Gamma/4\pi\right)}.\label{eq:appA7}
\end{equation}
Here, we ignored a proportional constant.

For Ly$\alpha$, we use $J_{i}=J_{f}=1/2$, $J_{m}=1/2$, 3/2, $\mu_{i}=\pm1/2$,
and $\mu_{m}=\pm1/2$ (for $J_{m}=1/2$) or $\pm1/2,\ \pm3/2$ (for
$J_{m}=3/2$). Then, after a straightforward mathematical manipulation
of the 3-$j$ symbols, we obtain the following scattering matrix elements
for $\mathbf{D}^{(4)}$:
\begin{align}
W_{11} & =\frac{1}{18}\left|f_{{\rm H}}\Phi_{{\rm H}}+f_{{\rm K}}\Phi_{{\rm K}}\right|^{2}\nonumber \\
W_{14} & =\frac{1}{18}\left|f_{{\rm H}}\Phi_{{\rm H}}-\frac{1}{2}f_{{\rm K}}\Phi_{{\rm K}}\right|^{2}\nonumber \\
W_{22} & =W_{11}\cos\theta\nonumber \\
W_{23} & =-W_{14}\cos\theta\nonumber \\
W_{44} & =W_{11}\cos^{2}\theta+W_{14}\sin^{2}\theta.\label{eq:appA8}
\end{align}
The scattering matrix for the Stokes vector $\mathbf{S}^{{\rm S94}}$
is obtained to be
\begin{align}
\mathbf{M} & =\mathbf{T}\mathbf{W}\mathbf{T}^{-1}\nonumber \\
 & =\left(W_{11}-W_{14}\right)\mathbf{P}_{1}+W_{14}\mathbf{P}_{2}+\left(W_{22}'-W_{23}'\right)\mathbf{P}_{3},\label{eq:appA9}
\end{align}
where
\begin{align}
\mathbf{P}_{1} & \equiv\left(\begin{array}{cccc}
1+\cos^{2}\theta & 1-\cos^{2}\theta & 0 & 0\\
1-\cos^{2}\theta & 1+\cos^{2}\theta & 0 & 0\\
0 & 0 & 2\cos\theta & 0\\
0 & 0 & 0 & 0
\end{array}\right),\nonumber \\
\mathbf{P}_{2} & \equiv\left(\begin{array}{cccc}
1 & 0 & 0 & 0\\
0 & 0 & 0 & 0\\
0 & 0 & 0 & 0\\
0 & 0 & 0 & 0
\end{array}\right),\nonumber \\
\mathbf{P}_{3} & \equiv\left(\begin{array}{cccc}
0 & 0 & 0 & 0\\
0 & 0 & 0 & 0\\
0 & 0 & 0 & 0\\
0 & 0 & 0 & 2\cos\theta
\end{array}\right),\label{eq:appA10}
\end{align}
and 
\begin{align}
W_{22}' & =W_{22}/\cos\theta,\nonumber \\
W_{23}' & =W_{23}/\cos\theta.\label{eq:appA11}
\end{align}

Note that the above scattering matrix is not normalized over the whole
solid angle. The normalization factor can be obtained by integrating
the scattered intensity, given by
\begin{align}
I' & =\left[\left(W_{11}-W_{14}\right)\left(1+\cos^{2}\theta\right)+4W_{14}\right]I\nonumber \\
 & \ \ \ +\left(W_{11}-W_{14}\right)\left(1-\cos^{2}\theta\right)\left(Q\cos2\phi+U\sin2\phi\right),\label{eq:appA12}
\end{align}
over the solid angle. The resulting normalization factor is
\begin{align}
\mathcal{N} & =\frac{\int I'd\Omega}{\int Id\Omega}=\frac{4}{3}\left(W_{11}+2W_{14}\right).\label{eq:appA13}
\end{align}
Finally, we obtain the ``normalized'' scattering matrix for the Stokes
vector:
\begin{align}
\mathbf{M} & =\frac{3}{4}E_{1}\mathbf{P}_{1}+E_{2}\mathbf{P}_{2}+\frac{3}{4}E_{3}\mathbf{P}_{3},\label{eq:appA14}
\end{align}
where
\begin{align}
E_{1} & =\frac{W_{11}-W_{14}}{W_{11}+2W_{14}},\nonumber \\
E_{2} & =\frac{3W_{14}}{W_{11}+2W_{14}}=1-E_{2},\nonumber \\
E_{3} & =\frac{W_{22}'-W_{23}'}{W_{11}+2W_{14}}=\frac{1}{3}\left(E_{1}+2\right).\label{eq:appA15}
\end{align}

Note that the oscillator strength (for a electric dipole transition
$j\rightarrow i$) is related to the line strength $S$:
\begin{equation}
f_{ij}\propto\left(2J_{j}+1\right)\nu_{ij}S.\label{eq:appA16}
\end{equation}
Hence, the relative ratio between the oscillator strengths of the
H and K transitions is
\begin{equation}
\frac{f_{{\rm K}}}{f_{{\rm H}}}=\frac{2J_{{\rm K}}+1}{2J_{{\rm H}}+1}\frac{\nu_{{\rm K}}}{\nu_{{\rm H}}}=2\frac{\nu_{{\rm K}}}{\nu_{{\rm H}}}\simeq2.\label{eq:appA17}
\end{equation}
Using the oscillator strength ratio, we obtain $E_{1}$:
\begin{align}
E_{1} & =\frac{\left(\nu_{{\rm K}}/\nu_{{\rm H}}\right)\left(q_{{\rm H}}^{*}q_{{\rm K}}+q_{{\rm H}}q_{{\rm K}}^{*}\right)+\left(\nu_{{\rm K}}/\nu_{{\rm H}}\right)\left|q_{{\rm H}}\right|^{2}}{\left|q_{{\rm K}}\right|^{2}+2\left(\nu_{{\rm K}}/\nu_{{\rm H}}\right)^{2}\left|q_{{\rm H}}\right|^{2}},\label{eq:appA18}
\end{align}
where
\begin{align}
q_{{\rm K}} & =\left(\nu-\nu_{{\rm K}}\right)+i\left(\Gamma/4\pi\right),\nonumber \\
q_{{\rm H}} & =\left(\nu-\nu_{{\rm H}}\right)+i\left(\Gamma/4\pi\right).\label{eq:appA19}
\end{align}
Therefore, we obtain $E_{1}$ as a function of frequency:
\begin{eqnarray}
E_{1} & \simeq & \frac{2(\nu-\nu_{{\rm K}})(\nu-\nu_{{\rm H}})\left(\nu_{{\rm K}}/\nu_{{\rm H}}\right)+(\nu-\nu_{{\rm H}})^{2}\left(\nu_{{\rm K}}/\nu_{{\rm H}}\right)^{2}}{(\nu-\nu_{{\rm K}})^{2}+2(\nu-\nu_{{\rm H}})^{2}\left(\nu_{{\rm K}}/\nu_{{\rm H}}\right)^{2}}\nonumber \\
 & \simeq & \frac{2\left(\nu-\nu_{{\rm K}}\right)\left(\nu-\nu_{{\rm H}}\right)+\left(\nu-\nu_{{\rm H}}\right)^{2}}{\left(\nu-\nu_{{\rm K}}\right)^{2}+2\left(\nu-\nu_{{\rm H}}\right)^{2}}.\label{eq:appA20}
\end{eqnarray}
We stress that the scattering matrix element $M_{12}^{{\rm S94}}$
in Equation (\ref{eq:appA14}) has a different sign from that for
the IAU standard: $M_{12}^{{\rm IAU}}=-M_{12}^{{\rm S94}}$. The above
scattering matrix can also be derived using the elegant formulae of
\citet{1983SoPh...85....3L} and \citet{1984SoPh...91....1L}.

We now obtain the scattering cross-section, as a function of frequency,
by integrating over the scattering angle:
\begin{align}
\sigma_{\nu} & \propto W_{11}+2W_{14}\nonumber \\
 & \propto\left|\Phi_{{\rm H}}\right|^{2}+\frac{1}{2}\left(\frac{f_{{\rm K}}}{f_{{\rm H}}}\right)^{2}\left|\Phi_{{\rm K}}\right|^{2}\label{eq:appA21}
\end{align}
Because $\nu_{{\rm H}}\approx\nu_{{\rm K}}$, the cross-section is
approximated by a linear combination of two Lorentzian functions with
weights of 1:2, as follows:
\begin{equation}
\sigma_{\nu}\propto\frac{1}{\left(\nu-\nu_{{\rm H}}\right)^{2}+\left(\Gamma/4\pi\right)^{2}}+2\frac{1}{\left(\nu-\nu_{{\rm K}}\right)^{2}+\left(\Gamma/4\pi\right)^{2}}.\label{eq:appA22}
\end{equation}
After multiplying by the total cross-section integrated over frequency,
we can obtain the final cross-section, given in Equation (\ref{eq1}),
as a function of frequency.

\section{Scattering Matrix for Dust Scattering}

\begin{table*}[t]
\caption{\label{tab:appB1}Best-fit parameters of the empirical formulae, Equation
(\ref{eq:appB1}), to the scattering matrix elements numerically-obtained
for the dust scattering of Ly\textgreek{a}. The albedo $a$ and the
asymmetry factor $g$ for the case of using only a single H-G phase
function are also shown.}
\medskip{}

\begin{centering}
\begin{tabular}{cccccccccccccc}
\hline 
 & $a$ & $g$ & $s_{1}$ & $g_{1}$ & $g_{2}$ & $p_{\ell}$ & $\theta_{0}$ & $s_{2}$ & $s_{3}$ & $s_{4}$ & $p_{c}$ & $s_{5}$ & $s_{6}$\tabularnewline
\hline 
\hline 
MW & 0.326 & 0.676 & 0.616 & 0.834 & 0.429 & 0.527 & 2.420 & 0.367 & 0.395 & 0.178 & 0.357 & 3.20 & 6.22\tabularnewline
LMC & 0.258 & 0.647 & 0.510 & 0.855 & 0.441 & 0.593 & $-0.332$ & 0.382 & 0.495 & 0.227 & 0.348 & 3.01 & 6.28\tabularnewline
SMC & 0.335 & 0.590 & 0.325 & 0.842 & 0.503 & 0.583 & 0.126 & 0.425 & 0.546 & 0.106 & 0.393 & 2.54 & 6.32\tabularnewline
\hline 
\end{tabular}
\par\end{centering}
\medskip{}
\end{table*}

\begin{figure*}[t]
\begin{centering}
\medskip{}
\includegraphics[clip,scale=0.9]{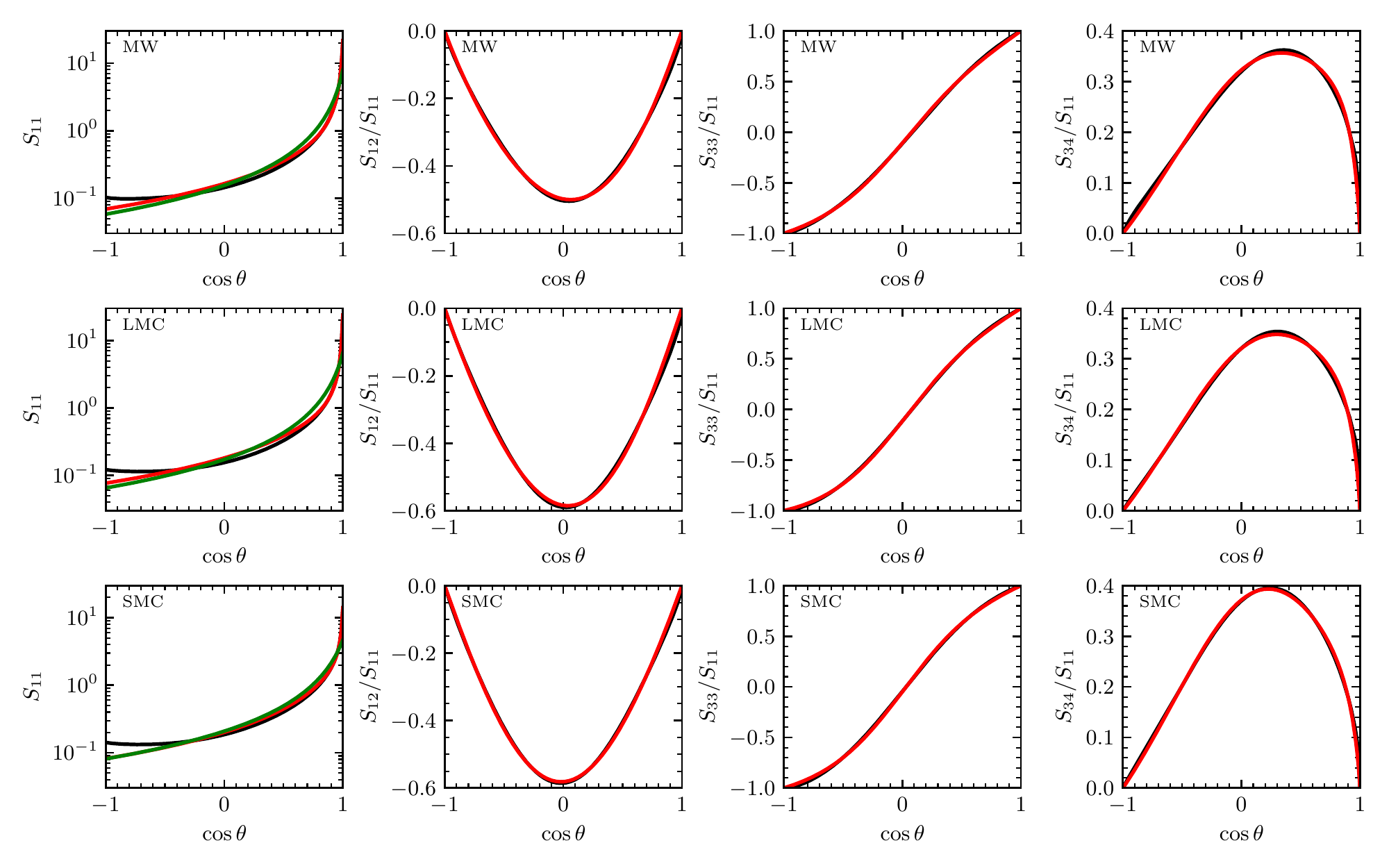}
\par\end{centering}
\begin{centering}
\medskip{}
\par\end{centering}
\caption{\label{fig_appB1}Scattering matrix elements for the MW, LMC, and
SMC dust models of \citet{2001ApJ...548..296W}, \citet{2003ARA&A..41..241D},
and \citet{2003ApJ...598.1017D}. Black curves denote numerically
obtained results as functions of $\cos\theta$. Red curves show the
best-fit curves given by Equation (\ref{eq:appB1}). Green curves
in the leftmost panels represent the best-fit results for the case
of using only one H-G phase function. The best-fit parameters are
shown in Table \ref{tab:appB1}.}
\medskip{}
\end{figure*}

\label{sec:scattering_matrix_for_dust_scattering}

This Appendix gives approximate formulae of the scattering matrix
elements for dust scattering, which are useful in performing Ly$\alpha$
RT simulation. We calculated all elements of the scattering matrix
for the MW, SMC, and LMC dust models of \citet{2001ApJ...548..296W},
\citet{2003ARA&A..41..241D}, and \citet{2003ApJ...598.1017D}, which
are composed of carbonaceous and silicate grains, using a Mie scattering
code.\footnote{https://www.astro.princeton.edu/\textasciitilde draine/scattering.html}
In the calculation, we also considered the contribution of free electrons
in the optical constants of graphite, as described in \citet{1984ApJ...285...89D}.
However, there was a typographical error in their Table 1. The effective
bulk scattering time to calculate the dielectric function for the
electric field vector parallel to the $c$-axis of graphite should
read as $\tau_{{\rm bulk}}=3.0\times10^{-14}$ in their Table 1 (B.
T. Draine, private communication).

\citet{1979ApJ...229..954W} proposed empirical equations to represent
the elements of the scattering matrix for the MW dust model of \citet{1977ApJ...217..425M}.
We thus attempted to reproduce our numerical results of the scattering
matrix elements using their equations. However, we found that their
equations are not adequate for the dust models of \citet{2001ApJ...548..296W}.
Instead, we slightly modified the equations and found that the following
formulae, obtained through the least-square fitting, reasonably well
reproduce the numerical results:

\begin{eqnarray}
S_{11} & = & s_{1}\phi_{{\rm HG}}\left(\cos\theta;g_{1}\right)+(1-s_{1})\phi_{{\rm HG}}\left(\cos\theta;g_{2}\right)\nonumber \\
\frac{S_{12}}{S_{11}} & = & -p_{{\rm \ell}}\frac{1-\cos^{2}\theta}{1+s_{2}^{2}\cos^{2}\left(\theta-\theta_{0}\right)}\nonumber \\
\frac{S_{33}}{S_{11}} & = & s_{3}\left(\frac{2\cos\theta}{1+\cos^{2}\theta}\right)+\left(1-s_{3}\right)\left\{ \cos\theta+s_{4}(\cos^{2}\theta-1)\right\} \nonumber \\
\frac{S_{34}}{S_{11}} & = & p_{c}\frac{1-\cos^{2}\left[\theta\left\{ 1+s_{5}\exp\left(-s_{6}\theta/\pi\right)\right\} \right]}{1+\cos^{2}\left[\theta\left\{ 1+s_{5}\exp\left(-s_{6}\theta/\pi\right)\right\} \right]}.\label{eq:appB1}
\end{eqnarray}
Here,

\begin{equation}
\phi_{{\rm {\rm HG}}}(\cos\theta;g)=\frac{1}{2}\frac{1-g^{2}}{\left(1+g^{2}-2g\cos\theta\right)^{3/2}}\label{eq:appB2}
\end{equation}
is the H-G phase funtion, $p_{\ell}$ the maximum linear polarization
level, and $p_{c}$ the maximum circular polarization level. The best-fit
parameters of the empirical equations are shown in Table \ref{tab:appB1}.
The table also shows the albedo, and the asymmetry factor $g=\left\langle \cos\theta\right\rangle $
for the case of using only one H-G function. Figure \ref{fig_appB1}
compares the numerical results and the best-fit curves of the scattering
matrix elements. The deviation between the approximate formulae and
the numerical values of $S_{11}$ at $\cos\theta\approx-1$ causes
no significant difference in RT results because the backward scattering
occurs rarely.

Note that all we need in the Monte Carlo RT simulation are not the
matrix elements themselves but $S_{11}$, $S_{12}/S_{11}$, $S_{33}/S_{11}$,
and $S_{34}/S_{11}$. Interestingly, the functional shapes of $S_{12}/S_{11}$
and $S_{33}/S_{11}$ are similar to those for the scattering case
by hydrogen atoms. However, $S_{11}$ for the scattering by dust is
totally different from that due to hydrogen atoms; the scattering
by dust is predominantly forward-directed, whereas that by hydrogen
is relatively more or less isotropic. In the wavelength range dealt
with in Ly$\alpha$ RT, the dust extinction and scattering properties
are not altered as the photon wavelength changes.

\section{Scattering Angle for Ly$\alpha$ Line Scattering}

\label{sec:App_scattering_angle}

The phase function for Ly$\alpha$ scattering by a hydrogen atom depends
on the frequency of incident Ly$\alpha$ photon \citep{1980A&A....84...68S,1994ASSL..189.....S}.
This appendix briefly describes an efficient algorithm, implemented
in LaRT following the approach described in \citet{2006PASJ...58..439S},
to randomly sample scattering angles from the frequency-dependent
phase function. The phase function for the angle $\theta$ between
the incident and scattered direction is:

\begin{equation}
P(\mu)=\left(\frac{3}{8}E_{1}\right)\mu^{2}+\left(\frac{4-E_{1}}{8}\right),\label{eq:appC1}
\end{equation}
where $\mu\equiv\cos\theta$. For $E_{1}=0$, this gives the isotropic
scattering, and the scattering angle can be sampled by $\mu=2\xi-1$
for a uniform random number $\xi$ between 0 and 1. In general, the
scattering angle can be obtained by inverting the cumulative integral
\begin{equation}
\int_{-1}^{\mu}P(\mu')d\mu'=\xi.\label{eq:appC2}
\end{equation}
 This equation results in the following cubic equation:
\begin{equation}
\mu^{3}+3p\mu=q,\label{eq:appC3}
\end{equation}
where
\begin{align}
p & \equiv\frac{4-E_{1}}{3E_{1}}\ \ {\rm and}\ \ q\equiv\frac{8\xi-4}{E_{1}}.\label{eq:appC4}
\end{align}
By setting 
\begin{equation}
\mu=2\left|p\right|^{1/2}y,\label{eq:appC5}
\end{equation}
we obtain the standard form to solve a cubic equation
\begin{equation}
4y^{3}+3\text{sgn}(p)y=\mathcal{Q},\label{eq:appC6}
\end{equation}
where 
\begin{equation}
\mathcal{Q}\equiv\frac{q}{2\left|p\right|^{3/2}}.\label{eq:appC7}
\end{equation}

If $p>0$ ($E_{1}>0$), we compare this equation with the identity
$4\sinh^{3}\Theta+3\sinh\Theta=\sinh(3\Theta)$ and obtain a real
solution:
\begin{align}
3\Theta & =\sinh^{-1}\mathcal{Q},\nonumber \\
y & =\sinh\Theta\nonumber \\
 & =\frac{1}{2}\left[\left(\mathcal{Q}+\sqrt{\mathcal{Q}^{2}+1}\right)^{1/3}-\left(\mathcal{Q}+\sqrt{\mathcal{Q}^{2}+1}\right)^{-1/3}\right].\label{eq:appC8}
\end{align}
Here, we utilized the identity $\sinh^{-1}\mathcal{Q}=\ln\left(\mathcal{Q}+\sqrt{\mathcal{Q}^{2}+1}\right)$.

We note that $\left|\mathcal{Q}\right|<1$ if $p<0$ $(-1/2\le E_{1}<0)$.
In this case, we compare Equation (\ref{eq:appC6}) with $4\cos^{3}\Theta-3\cos\Theta=\cos(3\Theta)$
and obtain ``three'' real solutions due to the periodicity of cosine
function:
\begin{align}
3\Theta & =\cos^{-1}\mathcal{Q}+2n\pi,\nonumber \\
y & =\cos\Theta\nonumber \\
 & =\cos\left(\frac{\cos^{-1}\mathcal{Q}+2n\pi}{3}\right),\label{eq:appC9}
\end{align}
where $n=0,1,2$. Among the three solutions, only the $n=2$ solution
is the appropriate one that satisfies the condition of $\left|\mu\right|\le1$
for $0\le\xi\le1$ and $-1/2\le E_{1}\le0$. The final result is summarized
in Equation (\ref{eq:32}).

\section{Density Matrix Formae in a Fixed Frame}

\begin{figure}[t]
\begin{centering}
\medskip{}
\includegraphics[clip,scale=0.53]{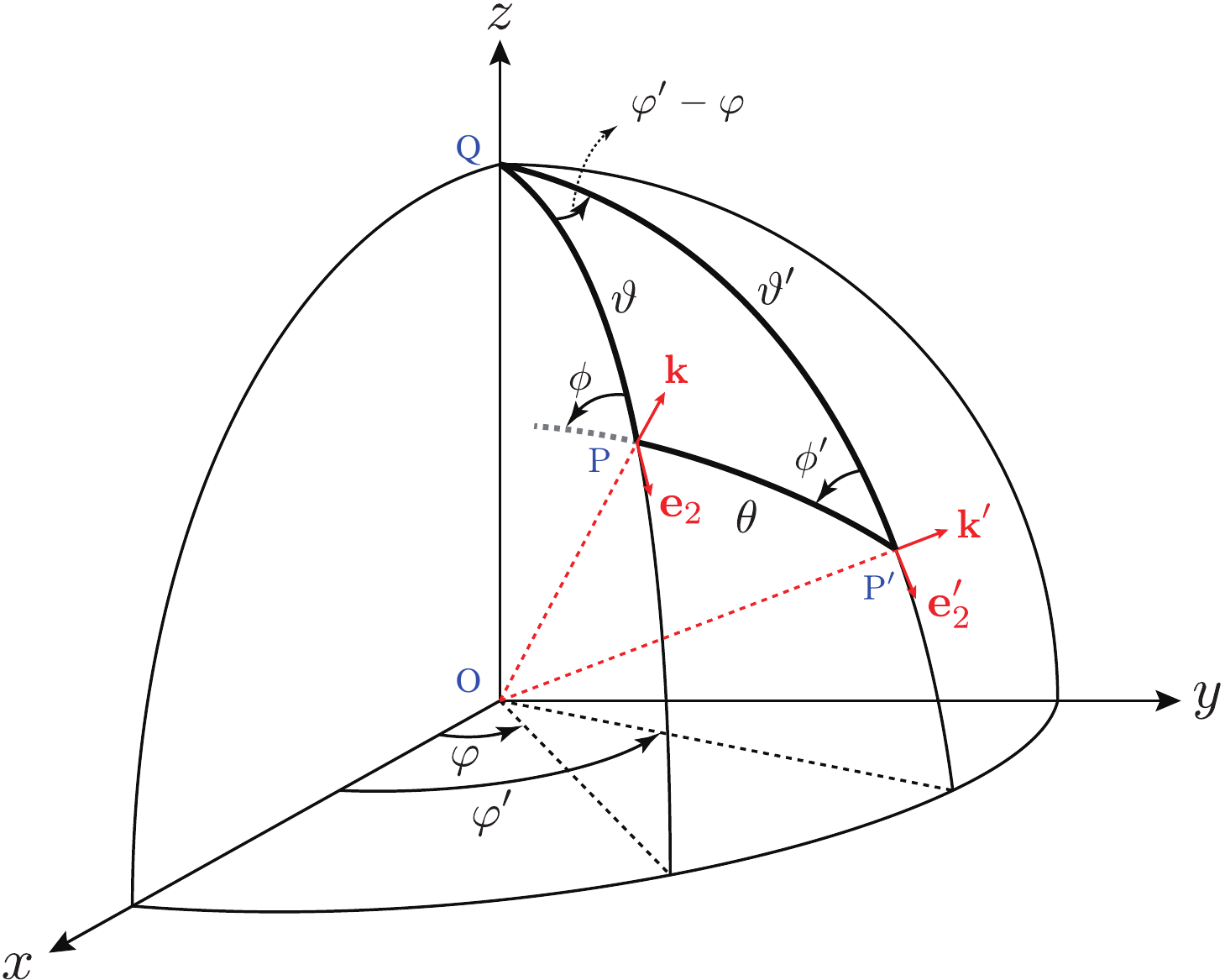}
\par\end{centering}
\begin{centering}
\medskip{}
\par\end{centering}
\caption{\label{fig_D1}Geometry for the density matrix formae in a fixed coordinate
system. The photon with the initial propagation direction $\mathbf{k}$
($OP$) is scattered in the direction $\mathbf{k}'$ ($OP'$). Note
that the orientation of polarization basis vectors is defined as left-handed
in the density matrix formalism; thus, the third basis vector is given
by $\mathbf{e}_{1}=\mathbf{k}\times\mathbf{e}_{2}$, but not shown
in the figure for clarity. The angle ($\phi$ or $\phi'$) between
the meridian plane ($OPQ$ or $OP'Q$) and the scattering plane ($OPP'$)
is measured in counterclockwise. Those with the prime symbol are for
the scattered photon, as oppose to the definition of \citet{1960ratr.book.....C}.}
\medskip{}
\end{figure}

\label{sec:density_matrix}

Instead of the Stokes parameters $I$, $Q$, $U$, and $V$, \citet{2002ApJ...567..922A},
\citet{2015JKAS...48..195A}, \citet{2017MNRAS.464.5018C}, \citet{2018ApJ...856..156E},
and \citet{2020JKAS_53_169C} use the density matrix. In their approach,
no circular polarization is assumed, i.e., $V=0$ and $\mathcal{E}_{1}\mathcal{E}_{2}^{*}=\mathcal{E}_{1}^{*}\mathcal{E}_{2}$,
where $\mathcal{E}_{1}$ and $\mathcal{E}_{2}$ are the electric field
components along two directions $\mathbf{e}_{1}$ and $\mathbf{e}_{2}$
at right angles to each other. The $2\times2$ Hermitian density matrix
$\mathbf{D}$ (or coherency matrix) is related to the Stokes parameters
by
\begin{align}
\mathbf{D} & \equiv\left(\begin{array}{cc}
\rho_{11} & \rho_{12}\\
\rho_{21} & \rho_{22}
\end{array}\right)=\left(\begin{array}{cc}
\mathcal{E}_{1}\mathcal{E}_{1}^{*} & \mathcal{E}_{1}\mathcal{E}_{2}^{*}\\
\mathcal{E}_{2}\mathcal{E}_{1}^{*} & \mathcal{E}_{2}\mathcal{E}_{2}^{*}
\end{array}\right)\nonumber \\
 & =\left(\begin{array}{cc}
I_{1} & U/2\\
U/2 & I_{2}
\end{array}\right),\label{eq:appD1}
\end{align}
where $I_{1}=\left|\mathcal{E}_{1}\right|^{2}$ and $I_{2}=\left|\mathcal{E}_{2}\right|^{2}$
are the intensities in the directions $\mathbf{e}_{1}$ and $\mathbf{e}_{2}$,
respectively. Therefore, the density matrix is in fact equivalent
to the Stokes vector. In this Appendix, we derive the density matrix
formulae, given by \citet{2002ApJ...567..922A} and others, starting
from ours.

The most critical difference between the density matrix approach and
our method is that they represent the polarization state in a ``fixed,''
left-handed coordinate system; however, this paper expresses it in
the photon's local frame. They represent the direction vector of the
scattered photon by the polar angles ($\vartheta$ and $\varphi$)
measured in a fixed reference frame (i.e., galaxy system). However,
in the present approach, we use the angles ($\theta$ and $\phi$)
measured in the incident photon's local frame. From this point of
view, we may refer to the aforementioned ``density matrix'' approch
as the ``fixed frame'' method and ours as the ``local frame'' method.

Coordinate systems for the fixed- and local frame methods are illustrated
in Figure \ref{fig_D1} (see also Figure 8 in \citealp{1960ratr.book.....C}).
In the figure, the polar angles of the incident and scattered photons
are $(\vartheta,\varphi)$ and $(\vartheta',\varphi')$, respectively.
The axes of the photon's local frame are given as follows:

\begin{align}
\mathbf{e}_{1} & =-\sin\varphi\mathbf{x}+\cos\varphi\mathbf{y}\nonumber \\
\mathbf{e}_{2} & =\cos\vartheta\cos\varphi\mathbf{x}+\cos\vartheta\sin\varphi\mathbf{y}-\sin\vartheta\mathbf{z}\nonumber \\
\mathbf{k} & =\sin\vartheta\cos\varphi\mathbf{x}+\sin\vartheta\sin\varphi\mathbf{y}+\cos\vartheta\mathbf{z},\label{eq:appD2}
\end{align}
where $\mathbf{x}$, $\mathbf{y}$, and $\mathbf{z}$ are axes of
the fixed frame. Here, we note that ($\mathbf{e}_{2}$, $\mathbf{e}_{1}$,
$\mathbf{k}$) corresponds to ($\mathbf{m}$, $\mathbf{n}$, $\mathbf{k}$).

To derive the formulae for the density matrix in a fixed frame, we
define a column vector:

\begin{align}
\overline{\mathbf{I}} & =\left(\begin{array}{c}
I_{2}\\
I_{1}\\
U\\
V
\end{array}\right)=\left(\begin{array}{c}
\rho_{22}\\
\rho_{11}\\
2\rho_{12}\\
0
\end{array}\right).\label{eq:appD3}
\end{align}
Here, $I_{2}$ and $I_{1}$ are intensities in two perpendicular directions
($\mathbf{e}_{2}$ and $\mathbf{e}_{1}$) in a left-handed coordinate
system, which correspond to the intensities $I_{m}=\left|E_{m}\right|^{2}$
and $I_{n}=\left|E_{n}\right|^{2}$, respectively, in our convention.
The scattering matrix for $\overline{\mathbf{I}}$ can readily be
obtained, using the scattering matrix (Equation (\ref{eq:18})) for
the Stokes vector, as follows:

\begin{align}
\overline{\mathbf{M}}(\theta) & =\begin{pmatrix}\frac{1}{2}\left(S_{11}+2S_{12}+S_{22}\right) & \frac{1}{2}\left(S_{11}-S_{22}\right) & 0 & 0\\
\frac{1}{2}\left(S_{11}-S_{22}\right) & \frac{1}{2}\left(S_{11}-2S_{12}+S_{22}\right) & 0 & 0\\
0 & 0 & S_{33} & 0\\
0 & 0 & 0 & S_{34}
\end{pmatrix}.\nonumber \\
\label{eq:appD4}
\end{align}
Substituting the matrix elements of Equation (\ref{eq:19}), the scattering
matrix for $\overline{\mathbf{I}}$ is given by
\begin{align}
\overline{\mathbf{M}}(\theta) & =\frac{3}{2}E_{1}\begin{pmatrix}\cos^{2}\theta & 0 & 0 & 0\\
0 & 1 & 0 & 0\\
0 & 0 & \cos\theta & 0\\
0 & 0 & 0 & 0
\end{pmatrix}+\frac{1}{2}E_{2}\begin{pmatrix}1 & 1 & 0 & 0\\
1 & 1 & 0 & 0\\
0 & 0 & 0 & 0\\
0 & 0 & 0 & 0
\end{pmatrix}\nonumber \\
 & \ \ +\frac{3}{2}E_{3}\begin{pmatrix}0 & 0 & 0 & 0\\
0 & 0 & 0 & 0\\
0 & 0 & 0 & 0\\
0 & 0 & 0 & \cos\theta
\end{pmatrix}.\label{eq:appD5}
\end{align}
The transformation matrix of $\overline{\mathbf{I}}$ for a rotation
of axes by an angle $\phi$ is found, from Equation (\ref{eq:14}),
to be

\begin{align}
\overline{\mathbf{L}}(\phi) & =\begin{pmatrix}\cos^{2}\phi & \sin^{2}\phi & \frac{1}{2}\sin2\phi & 0\\
\sin^{2}\phi & \cos^{2}\phi & -\frac{1}{2}\sin2\phi & 0\\
-\sin2\phi & \sin2\phi & \cos2\phi & 0\\
0 & 0 & 0 & 1
\end{pmatrix}\label{eq:appD6}
\end{align}
(see Equation (190) in \citealp{1960ratr.book.....C}).

The four intensity vector $\overline{\mathbf{I}}$ after a scattering
through the angles ($\theta,\phi$) are expressed by, in a fixed frame:
\begin{equation}
\overline{\mathbf{I}}'=\mathbf{\bar{\mathbf{P}}}\overline{\mathbf{I}},\label{eq:appD7}
\end{equation}
where the phase matrix can be written as follows:
\begin{align}
\bar{\mathbf{P}} & \equiv\overline{\mathbf{L}}(-\phi')\overline{\mathbf{M}}(\theta)\overline{\mathbf{L}}(\phi).\label{eq:appD8}
\end{align}
Here, the rotation angle $\phi$ denotes the angle between the meridian
plane $OPQ$ through and the plane of scattering $OPP'$, and $\phi'$
the angle between the planes $OP'Q$ and $OPP'$ in Figure \ref{fig_D1}.
The phase matrix can be decomposed into three terms:
\begin{align}
\bar{\mathbf{P}} & \equiv\bar{\mathbf{P}}_{1}+\bar{\mathbf{P}}_{2}+\bar{\mathbf{P}}_{3}\nonumber \\
\bar{\mathbf{P}}_{1} & =\frac{2}{3}E_{1}\times\nonumber \\
 & \ \ \ \ \ \begin{pmatrix}(mm)^{2} & (nm)^{2} & (mm)(nm) & 0\\
(mn)^{2} & (nn)^{2} & (mn)(nn) & 0\\
2(mm)(mn) & 2(nn)(nm) & (mm)(nn)+(nm)(mn) & 0\\
0 & 0 & 0 & 0
\end{pmatrix}\nonumber \\
\bar{\mathbf{P}}_{2} & =\frac{1}{2}E_{2}\begin{pmatrix}1 & 1 & 0 & 0\\
1 & 1 & 0 & 0\\
0 & 0 & 0 & 0\\
0 & 0 & 0 & 0
\end{pmatrix}\nonumber \\
\bar{\mathbf{P}}_{3} & =\frac{3}{2}E_{3}\begin{pmatrix}0 & 0 & 0 & 0\\
0 & 0 & 0 & 0\\
0 & 0 & 0 & 0\\
0 & 0 & 0 & \cos\theta
\end{pmatrix}.\label{eq:appD9}
\end{align}
Here, we introduced the abbreviations, similar to those of \citet{1960ratr.book.....C}
but for a right-handed system:
\begin{align}
(mm) & =\cos\theta\cos\phi'\cos\phi+\sin\phi'\sin\phi\nonumber \\
(nn) & =\cos\theta\sin\phi'\sin\phi+\cos\phi'\cos\phi\nonumber \\
(nm) & =\cos\theta\cos\phi'\sin\phi-\sin\phi'\cos\phi\nonumber \\
(mn) & =\cos\theta\sin\phi'\cos\phi-\cos\phi'\sin\phi.\label{eq:appD10}
\end{align}

Note that the above formulae are written using the angles $(\theta,\phi,\phi')$
in the photon's local frames. We now want to express the above equations
in terms of the angles $(\vartheta,\varphi,\vartheta',\varphi')$.
For the spherical triangle $QPP'$ in Figure \ref{fig_D1}, we have
the spherical cosine rules

\begin{align}
\cos\vartheta & =\cos\vartheta'\cos\theta+\sin\vartheta'\sin\theta\cos\phi'\nonumber \\
\cos\vartheta' & =\cos\vartheta\cos\theta+\sin\vartheta'\sin\theta\cos(\pi-\phi)\nonumber \\
\cos(\varphi'-\varphi) & =-\cos(\pi-\phi)\cos\phi'+\sin(\pi-\phi)\sin\phi'\cos\theta,\label{eq:appD11}
\end{align}
and the sine rule
\begin{equation}
\frac{\sin(\pi-\phi)}{\sin\vartheta'}=\frac{\sin\phi'}{\sin\vartheta}=\frac{\sin(\varphi'-\varphi)}{\sin\theta}.\label{eq:appD12}
\end{equation}
Using these expressions, we can readily show that

\begin{align}
(mm) & =\cos\Delta\varphi\cos\vartheta\cos\vartheta'+\sin\vartheta\sin\vartheta'\nonumber \\
(nn) & =\cos\Delta\varphi\nonumber \\
(nm) & =\sin\Delta\varphi\cos\vartheta'\nonumber \\
(mn) & =-\sin\Delta\varphi\cos\vartheta,\label{eq:appD13}
\end{align}
where $\Delta\varphi\equiv\varphi'-\varphi$.

Using the above expressions, we now obtain the transformation rules
for the density matrix elements:
\begin{equation}
\begin{array}{cll}
\rho_{22}' & = & \left[\frac{3}{2}E_{1}\left(\cos\Delta\varphi\cos\vartheta\cos\vartheta'+\sin\vartheta\sin\vartheta'\right)^{2}+\frac{1}{2}E_{2}\right]\rho_{22}\\
 &  & +\left[\frac{3}{2}E_{1}\left(\sin\Delta\varphi\cos\vartheta'\right)^{2}+\frac{1}{2}E_{2}\right]\rho_{11}\\
 &  & +\frac{3}{2}E_{1}\left(\cos\Delta\varphi\cos\vartheta\cos\vartheta'+\sin\vartheta\sin\vartheta'\right)\\
 &  & \ \ \ \ \ \ \ \ \times\left(\sin\Delta\varphi\cos\vartheta'\right)\left(2\rho_{12}\right),
\end{array}\label{eq:appD14}
\end{equation}
\begin{equation}
\begin{array}{cll}
\rho_{11}' & = & \left[\frac{3}{2}E_{1}\left(\sin\Delta\varphi\cos\vartheta\right)^{2}+\frac{1}{2}E_{2}\right]\rho_{22}\\
 &  & +\left[\frac{3}{2}E_{1}\left(\cos\Delta\varphi\right)^{2}+\frac{1}{2}E_{2}\right]\rho_{11}\\
 &  & -\frac{3}{2}E_{1}\left(\sin\Delta\varphi\cos\vartheta\right)\left(\cos\Delta\varphi\right)\left(2\rho_{12}\right),
\end{array}\label{eq:appD15}
\end{equation}
\begin{equation}
\begin{array}{cll}
2\rho_{12}' & = & -3E_{1}\left(\cos\Delta\varphi\cos\vartheta\cos\vartheta'+\sin\vartheta\sin\vartheta'\right)\\
 &  & \ \ \ \ \ \ \ \ \times\left(\sin\Delta\varphi\cos\vartheta\right)\rho_{22}\\
 &  & +3E_{1}\left(\cos\Delta\varphi\sin\Delta\varphi\cos\vartheta'\right)\rho_{11}\\
 &  & +\frac{3}{2}E_{1}\left(\cos2\Delta\varphi\cos\vartheta\cos\vartheta'+\cos\Delta\varphi\sin\vartheta\sin\vartheta'\right)\\
 &  & \ \ \ \ \ \ \ \ \times\left(2\rho_{12}\right).
\end{array}\label{eq:appD16}
\end{equation}
It can be easily verified that the above equations are equivalent
to Equation (11) of \citet{2015JKAS...48..195A} when $E_{1}=1/2$
(resonance transition $S_{1/2}-P_{3/2}$) and Equation (12) when $E_{1}=1$
(resonance transition $S_{1/2}-P_{1/2}$ and Rayleigh scattering),
except for the normalization factors and a typographical error.

We need to multiply our scattering matrix elements by a factor of
8 to obtain Equation (11) of \citet{2015JKAS...48..195A} and by a
factor of 2/3 for Equation (12). We also note that there was a typographical
error in the last term for $\rho_{22}'$ in Equation (5) of \citet{2002ApJ...567..922A},
Equation (11) of \citet{2015JKAS...48..195A}, and Equation (21) of
\citet{2018ApJ...856..156E}. The last term for $\rho_{22}'$ should
read
\begin{align}
6\left(\sin2\Delta\varphi\cos^{2}\vartheta'\cos\vartheta+2\sin\Delta\varphi\cos\vartheta'\sin\vartheta\sin\vartheta'\right)\rho_{12}.\nonumber \\
\label{eq:appD17}
\end{align}
The numeric factor ``6'' should come out in front of the parenthesis.
The typographical error was fixed in \citet{2020JKAS_53_169C}. The
above equations for the density matrix allow us to perform Monte-Carlo
simulations for an arbitrary $E_{1}$ using an appropriate rejection
method.

\section{Technique Using 100\%-Polarized Light}

\label{sec:App_100=000025-Polarized}

\citet{1999ApJ...520L..79R} describe a Monte-Carlo simulation technique,
initially developed by \citet{1969ApJ...158..219A} for Thomson scattering
in X-ray, to study the polarization state of Ly$\alpha$. In their
approach, every photon packet is assumed to be 100\% linearly polarized
along a unit vector perpendicular to its propagation direction. Light
is 100\% polarized if it is strictly monochromatic. However, it is
not generally so because, in actual situations, light is a superposition
of mutually-incoherent monochromatic light beams. Partial polarization
arises due to the incoherent superposition of mutually uncorrelated
photons with different polarization states in a statistical ensemble
of photons. The photons are mutually uncorrelated since the atomic
processes by which they are created are stochastically independent
of each other. Therefore, to obtain unpolarized, quasi-monochromatic
light by employing the 100\%-polarized photons, they let the photon's
polarization direction be randomly oriented. The algorithm was devised
for Rayleigh scattering (and additionally, the resonance transition
$S_{1/2}-P_{1/2}$; $E_{1}=1$). Then, it was later extended to the
case of the transition $S_{1/2}-P_{3/2}$ ($E_{1}=1/2$), which can
be regarded as a superposition of Rayleigh and isotropic scattering
\citep{2008MNRAS.386..492D,2016A&A...593A.122T}.

To obtain a random, new propagation direction after a scattering event,
\citet{1999ApJ...520L..79R} use the angle $\Psi$ between the polarization
and scattered directions, instead of the scattering angle $\theta$
defined by the incident and scattered directions. Suppose that $\beta$
is an angle between the incident photon's polarization vector and
the scattering plane, measured counterclockwise when viewed toward
the photon. The intensities parallel and perpendicular to the scattering
plane are, then, given by $I_{m}=I\cos^{2}\beta$ and $I_{n}=I\sin^{2}\beta$.
The phase function for 100\%-polarized photons can be expressed as
a function of $\Psi$, as follows:
\begin{align}
\mathcal{P}(\theta,\phi) & =\frac{I'}{I}\nonumber \\
 & =\frac{3}{4}E_{1}\left(\cos^{2}\theta+1\right)+E_{2}+\frac{3}{4}E_{1}\left(\cos^{2}\theta-1\right)\frac{Q}{I}\nonumber \\
 & =\frac{3}{2}E_{1}\left(\cos^{2}\theta\cos^{2}\beta+\sin^{2}\beta\right)+E_{2}\nonumber \\
 & =\frac{3}{2}E_{1}\sin^{2}\Psi+E_{2},\label{eq:appE1}
\end{align}
Here, we used the relations $Q=I_{m}-I_{n}$ and $\cos\Psi=\sin\theta\cos\beta$
to obtain the last equation. This indicates that the scattering event
can be regarded as a superposition of Rayleigh and isotropic scattering
only when $E_{1}\ge0$ and $E_{2}\ge0$. Otherwise, the phase function
cannot be interpreted as a superposition of two probability distribution
functions. The frequency range in which the superposition interpretation
is invalid ($E_{1}<0$) is
\begin{equation}
\nu_{{\rm H}}<\nu<\frac{1}{3}\left(2\nu_{{\rm K}}+\nu_{{\rm H}}\right)\ \ {\rm or}\ \ -\frac{\delta\nu_{{\rm KH}}}{2\Delta\nu_{{\rm D}}}<x<\frac{\delta\nu_{{\rm KH}}}{6\nu_{{\rm D}}},\label{eq:appE2}
\end{equation}
where $\delta\nu_{{\rm KH}}\equiv\nu_{{\rm K}}-\nu_{{\rm H}}$. This
frequency range corresponds to the wavelength range of $\lambda_{c}<\lambda<\lambda_{{\rm H}}$
in Figure \ref{fig02}.

We now provide a method to generate random angles $\Psi$ using the
inversion method, as described for the scattering angle $\theta$
in Appendix \ref{sec:App_scattering_angle}. Substituting $E_{2}=1-E_{1}$
into Equation (\ref{eq:appE1}), we obtain the distribution function
for $\Psi$:
\begin{align}
\mathcal{P}(\cos\Psi) & =\frac{E_{1}+2}{4}-\frac{3}{4}E_{1}\cos^{2}\Psi.\label{eq:appE3}
\end{align}
This function results in a cubic equation, which should be solved
for $\cos\Psi$, as follows:
\begin{equation}
\cos^{3}\Psi-\frac{E_{1}+2}{E_{1}}\cos\Psi=\frac{2(1-2\xi)}{E_{1}}\label{eq:appE4}
\end{equation}
As in Appendix \ref{sec:App_scattering_angle}, we obtain the following
solutions:
\begin{align}
\cos\Psi & =\begin{cases}
2\left|p\right|^{1/2}\cos\left[\left(\cos^{-1}\mathcal{Q}+4\pi\right)/3\right] & {\rm for}\ \;E_{1}>0\\
\\
\left|p\right|^{1/2}\left(W-1/W\right) & {\rm for}\ \;E_{1}<0\\
\\
2\xi-1 & {\rm for}\ \;E_{1}=0,
\end{cases}\label{eq:appE5}
\end{align}
where
\begin{align}
p & \equiv-\frac{E_{1}+2}{3E_{1}},\ \mathcal{Q}\equiv\frac{1-2\xi}{E_{1}\left|p\right|^{3/2}},\nonumber \\
W & \equiv\left(\mathcal{Q}+\sqrt{\mathcal{Q}^{2}+1}\right)^{1/3}.\label{eq:appE6}
\end{align}
In Equation (\ref{eq:appE5}), we, however, note that the solution
for $E_{1}<0$ should not be used in a Monte Carlo simulation using
100\%-polarized photons.

One may want to generate random angles $\Psi$ using the composition
method for a function composed of the Rayleigh function with a weight
$E_{1}$ and a uniform distribution function with a weight $1-E_{1}$.
In this case, random angles following Rayleigh function can be obtained
by setting $E_{1}=1$ in Equation (\ref{eq:appE5}).

A final note we want to mention is that the above technique using
100\%-polarized photons is valid only for $E_{1}\ge0$. In the case
of $E_{1}<0$, we may assume that $E_{1}=0$ because the frequency
range in which $E_{1}<0$ is very narrow (especially when the gas
temperature is as high as $T\gtrsim10^{3}$ K). If an accurate method
is desirable for $E_{1}<0$, the approach of using partially polarized
light, as adopted in LaRT, is required. Another but more elaborated
approach is to use the electric field $\mathbf{E}$ and the ``amplitude''
matrix, instead of using the Stokes parameters and the scattering
matrix.

\acknowledgements{This work was supported by a National Research Foundation of Korea
(NRF) grant funded by the Korea government (MSIP) (No. 2020R1A2C1005788).
Numerical simulations were partially performed by using a high performance
computing cluster at the Korea Astronomy and Space Science Institute. }

\end{document}